\documentclass[12pt]{article}
\usepackage[nosort]{cite}
\usepackage{amsmath,amsthm,amsfonts,amssymb,amscd,mathrsfs,slashed,graphicx,braket}

\newlength{\xtrawidth}
\newlength{\xtraheight}
\numberwithin{equation}{section}
\numberwithin{table}{section}
\numberwithin{figure}{section}
\newcommand{\kov}{\varkappa} 
\newcommand{\LR}{ {L/R} }

\def\G2{\operatorname{G}_2}
\def\dd{{\rm d}}
\def\adj{\mathbf{adj}}
\def\fun#1{\mathbf{#1}}
\def\dPint#1#2{\langle #1 , #2 \rangle_{dP_3}}
\newcommand{\be}{\begin{equation}}
\newcommand{\ee}{\end{equation}}

\newcommand{\sx}{\sigma}

\begin{document}
\begin{titlepage}
\begin{center}
\hfill BONN--TH--2017--01\\
\vskip 0.6in
{\LARGE\bf{Effective action from M-theory}}\\[2ex]
{\LARGE\bf{on twisted connected sum $G_2$-manifolds}}
\vskip 0.3in
{\bf
Thaisa C.~da~C.~Guio$^{1}$,
Hans Jockers$^{1}$,
Albrecht Klemm$^{1}$,
Hung-Yu~Yeh$^{1,2}$}
\vskip 0.15in
{\it $^{1}\,$Bethe Center for Theoretical Physics, Physikalisches Institut der Universit\"at Bonn,\\
Nussallee 12, D-53115 Bonn, Germany}
\vskip 0.1in
{\it $^{2}\,$Max Planck Institute for Mathematics,
Vivatsgasse 7, D-53111 Bonn, Germany}\\

\vskip 0.15in
{\tt tguio@th.physik.uni-bonn.de}\\
{\tt jockers@uni-bonn.de}\\
{\tt aklemm@th.physik.uni-bonn.de}\\
{\tt yeh@mpim-bonn.mpg.de}
\end{center}
\vskip 0.15in
\begin{center} {\bf Abstract} \end{center}

We study the four-dimensional low-energy effective $\mathcal{N}=1$ supergravity theory of the
dimensional reduction of M-theory on $G_2$-manifolds, 
which are constructed by Kovalev's twisted connected sum gluing 
suitable pairs of asymptotically cylindrical Calabi--Yau threefolds $X_{L/R}$ 
augmented with a circle $S^1$. In the Kovalev limit the Ricci-flat $G_2$-metrics 
are approximated by the Ricci-flat metrics on  $X_{L/R}$ and we identify 
the universal modulus --- the Kovalevton ---  that parametrizes this limit. We observe that 
the low-energy effective theory exhibits in this limit gauge theory 
sectors with extended supersymmetry. We determine the universal (semi-classical) 
K\"ahler potential of the effective $\mathcal{N}=1$ supergravity action as a function 
of the Kovalevton and the volume modulus of the $G_2$-manifold. This 
K\"ahler potential fulfills the no-scale inequality such that no anti-de-Sitter 
vacua are admitted. We describe geometric degenerations in $X_{L/R}$, which lead 
to non-Abelian gauge symmetries enhancements with various matter content.  
Studying the resulting gauge theory branches, we argue that they lead to transitions
compatible with the gluing construction and provide many new explicit examples
of $G_2$-manifolds.

\vfill
\noindent February, 2017
\end{titlepage}

\pagenumbering{roman}
\tableofcontents
\newpage
\section*{Glossary} \label{sec:glossary}
\def\widthone{12ex}
\def\widthtwo{69ex}
\def\widththree{10.5ex}
{\small
\hbox{
\vbox{
\offinterlineskip
\halign{\strut\vrule width1.2pt#&\hbox to \widthone{\hfil#\hfil}\vrule&\hbox to \widthtwo{~#\hfil}\vrule&\hbox to \widththree{~#\hfil}&#\vrule width1.2pt\cr 
\noalign{\hrule height 1.2pt}
&~Symbol\hfill&Description&~Pages\hfill&\cr
\noalign{\hrule height 1.2pt}
&\multispan3~Geometric spaces:\hfil&\cr
\noalign{\hrule}
&$M^{1,10}$&eleven-dimensional Lorentz manifold (M-theory space-time)&\pageref{eq:Splitting}&\cr
&$\mathbb{M}^{1,3}$&four-dimensional Minkowski space&\pageref{eq:Splitting},\pageref{eq:Metric}&\cr
&$Y$&compact seven-dimensional compactification manifold&\pageref{eq:Splitting},\pageref{eq:Metric},\pageref{eq:Yunion},\pageref{eq:K3fib}&\cr
&$Y^{\cdots}_{\cdots}$&twisted connected sum $G_2$-manifold from orthogonal gluing&\pageref{gl:Ydot},\pageref{eq:Yrank2rank3},\pageref{eq:Yintrank2}&\cr
&$\mathcal{M}$&moduli space of Ricci-flat metrics of $G_2$-metrics&\pageref{eq:ModSpace},\pageref{gl:Mc}&\cr
&$\mathcal{M}_\mathbb{C}$&semi-classical moduli space of M-theory on $G_2$-manifolds&\pageref{gl:Mc}&\cr
&$\Delta^\text{cyl}$&complex one-dimensional open cylinder $\left\{ x\in\mathbb{C}\,\middle|\,|z|>1 \right\}$&\pageref{gl:Dcyl}&\cr
&$X^\infty$&cylindrical Calabi--Yau threefold&\pageref{eq:metXoo}&\cr
&$X^\infty_\LR$&left/right cylindrical Calabi--Yau threefold&\pageref{eq:asymp}&\cr
&$X_\LR$&left/right asymptotically cylindrical Calabi--Yau threefold&\pageref{gl:XLR}&\cr
&$X_\LR(T)$&left/right truncated Calabi--Yau threefold $X_\LR$&\pageref{fig:twistedsum},\pageref{eq:seven}&\cr
&$K_\LR$&left/right compact complement of asymp. cyl. of $X_\LR(T)$&\pageref{gl:KLR},\pageref{eq:seven}&\cr
&$S^{*\,1}_\LR$&left/right circle of the asymptotic region of $X_\LR$&\pageref{fig:twistedsum}&\cr
&$S^1_\LR$&left/right circles in Kovalev's twisted connected sum&\pageref{gl:XLR},\pageref{fig:twistedsum}&\cr
&$Y_\LR^\infty$&left/right asymptotic seven-manifolds $X_\LR^\infty \times S^1_\LR$&\pageref{eq:asymp}&\cr
&$Y_\LR$&left/right non-compact seven-manifolds $X_\LR \times S^1_\LR$&\pageref{gl:XLR},\pageref{eq:K3fibLR}&\cr
&$Y_\LR(T)$&left/right truncated non-compact seven-manifolds $Y_\LR$&\pageref{fig:twistedsum},\pageref{eq:seven}&\cr
&$S$&K3 surface&\pageref{gl:S},\pageref{gl:Sii}&\cr
&$S_\LR$&left/right polarized K3 surface&\pageref{eq:asymp},\pageref{fig:twistedsum}&\cr
&$(Z,S)$&building block for asymptotically cylindrical Calabi--Yau threefold&\pageref{gl:bblock}&\cr
&$(Z_\text{sing},S)$&singular building block with $\mathcal{N}=2$ gauge theory sector&\pageref{eq:sing1},\pageref{eq:ZsingNonAb}&\cr
&$dP_\ell$&del Pezzo surface of degree $(9-\ell)$&\pageref{gl:dP3}&\cr
&$P$&weak Fano, semi-Fano or Fano threefold&\pageref{gl:Fano},\pageref{gl:tFano}&\cr
&$P_\Sigma$&toric weak Fano, toric semi-Fano or toric Fano threefold&\pageref{gl:tFano},\pageref{eq:K32Mcone}&\cr
&$\Sigma$&toric fan of a toric variety&\pageref{gl:tFano},\pageref{eq:K32Mcone}&\cr
&$\Delta/\Delta^*$&toric and dual toric polytopes&\pageref{gl:tFano},\pageref{eq:K32Mcone}&\cr
\noalign{\hrule}
&\multispan3~Forms and tensors:\hfil&\cr
\noalign{\hrule}
&$\eta_{\mu\nu}$&four-dimensional Minkowski metric&\pageref{eq:Metric}&\cr
&$g_{mn}$&Riemannian metric tensor of compactification manifold $Y$&\pageref{eq:Metric}&\cr
&$\hat g$&metric tensor of eleven-dimensional Lorentz manifold $M^{1,10}$&\pageref{eq:Metric}&\cr
&$\varphi$&$G_2$-structure (three-form)&\pageref{eq:Bpairing},\pageref{eq:ModSpace}&\cr
&$g_\varphi$&$G_2$-structure metric&\pageref{eq:G2met}&\cr
&$\omega_I^{(2)}$&basis of harmonic two-forms of $G_2$-manifold $Y$&\pageref{eq:ExpansionCfield},\pageref{eq:chiexp}&\cr
&$\rho^{(3)}_i$&basis of harmonic three-forms of $G_2$-manifold $Y$&\pageref{eq:SymAntiSym},\pageref{eq:ExpansionCfield}&\cr
&$\rho^\text{sym}_{i,(mn)}$&basis of zero modes of the Lichnerowicz Laplacian on $Y$&\pageref{eq:RicEq},\pageref{eq:chiexp}&\cr
&$\eta$&covariantly constant spinor of $G_2$-manifold $Y$&\pageref{gl:eta}&\cr
&$\omega^\infty$&K\"ahler form of cylindrical Calabi--Yau threefold $X^\infty$&\pageref{eq:ccyforms}&\cr
&$\Omega^\infty$&holomorphic three-form of cylindrical Calabi--Yau threefold $X^\infty$&\pageref{eq:ccyforms}&\cr
&$g_{X^\infty}$&Ricci-flat metric of cylindrical Calabi--Yau threefold $X^\infty$&\pageref{eq:metXoo}&\cr
&$g_\LR$&left/right Ricci-flat metric of Calabi--Yau threefold $X_\LR$&\pageref{eq:gLR}&\cr
\noalign{\hrule height 1.2pt}
}}}}
\newpage
{\small
\hbox{
\vbox{
\offinterlineskip
\halign{\strut\vrule width1.2pt#&\hbox to \widthone{\hfil#\hfil}\vrule&\hbox to \widthtwo{~#\hfil}\vrule&\hbox to \widththree{~#\hfil}&#\vrule width1.2pt\cr 
\noalign{\hrule height 1.2pt}
&~Symbol\hfill&Description&~Pages\hfill&\cr
\noalign{\hrule height 1.2pt}
&\multispan3~Forms and tensors \emph{(continued)}:\hfil&\cr
\noalign{\hrule}
&$\omega^{I,J,K}_\LR$&left/right triplet of hyper K\"ahler two-forms of K3 surface $S_\LR$&\pageref{gl:omegatripLR}&\cr
&$g_S$&Ricci-flat metric of K3 surface $S$&\pageref{eq:metXoo}&\cr
&$\varphi_0$&canonical $G_2$-structure of Calabi--Yau threefold times circle&\pageref{eq:G2prodform}&\cr
&$\varphi_{0\,L/R}^{\infty}$&left/right asymptotic torsion-free $G_2$-structure&\pageref{eq:phi0LR}&\cr
&$\widetilde{\varphi}_\LR(\gamma,T)$&left/right interpolating $G_2$-structure&\pageref{eq:phitilde}&\cr
&$\varphi(\gamma,T)$&torsion-free $G_2$-structure in Kovalev's twisted connected sum&\pageref{eq:PhiKov}&\cr
\noalign{\hrule}
&\multispan3~Coordinates and quantum fields:\hfil&\cr
\noalign{\hrule}
&$x^\mu$&coordinates of Minkowski space $\mathbb{M}^{1,3}$&\pageref{eq:Metric}&\cr
&$y^m$&local coordinates of compact seven-manifold $Y$&\pageref{eq:Metric}&\cr
&$S^i$&local coordinates of moduli space $\mathcal{M}$&\pageref{eq:ModSpace},\pageref{tab:N1multi}&\cr
&$P^i$&three-form scalar fields&\pageref{eq:ExpansionCfield},\pageref{tab:N1multi}&\cr
&$\phi^i$&local coordinates of moduli space $\mathcal{M}_\mathbb{C}$ \& chiral scalar fields&\pageref{eq:CStarget}&\cr
&$\chi^i_\alpha$&four-dimensional chiral fermions of $\mathcal{N}=1$ chiral multiplets&\pageref{tab:N1multi}&\cr
&$\Phi^i$&four-dimensional $\mathcal{N}=1$ chiral multiplets&\pageref{tab:N1multi},\pageref{eq:CStarget}&\cr
&$A^I_\mu$&four-dimensional vector bosons&\pageref{eq:ExpansionCfield},\pageref{tab:N1multi}&\cr
&$\lambda^I_\alpha$&four-dimensional gauginos&\pageref{tab:N1multi}&\cr
&$V^I$&four-dimensional $\mathcal{N}=1$ vector multiplets&\pageref{tab:N1multi}&\cr
&$\nu$&four-dimensional $\mathcal{N}=1$ chiral overall volume modulus&\pageref{eq:nukov},\pageref{eq:Kkov}&\cr
&$\kov$&Kovalevton (four-dimensional $\mathcal{N}=1$ chiral gluing modulus)&\pageref{eq:nukov},\pageref{eq:Kkov}&\cr
\noalign{\hrule}
&\multispan3~Parameters and coupling constants:\hfil&\cr
\noalign{\hrule}
&$\kappa_{4}$/$\kappa_{11}$&four-dimensional/eleven-dimensional Planck constant&\pageref{eq:kappa4}&\cr
&$V_{Y_0}$&constant reference volume of manifold $Y$&\pageref{eq:Vol},\pageref{gl:refVol}&\cr
&$V_Y$&moduli-dependent volume of manifold $Y$&\pageref{eq:Vol}&\cr
&$\lambda_0$&moduli-dependent dimensionless volume factor of manifold $Y$&\pageref{eq:Vol}&\cr
&$\gamma_0$&constant reference radius&\pageref{eq:defR}&\cr
&$\gamma$&moduli-dependent radius&\pageref{eq:G2prodform},\pageref{eq:defR}&\cr
&$\lambda$&inverse length scale of Calabi--Yau threefolds $X_\LR^\infty$&\pageref{eq:Clambda},\pageref{gl:invlength}&\cr
&$\tilde\lambda$&dimensionless inverse length scale of Calabi--Yau threefolds $X_\LR^\infty$&\pageref{gl:invlength}&\cr
&$\lambda_S$&inverse length scale of K3 surface $S$&\pageref{eq:Clambda}&\cr
&$R$&dimensionless volume modulus&\pageref{eq:defR},\pageref{eq:R(T)}&\cr
&$T$&dimensionless Kovalev parameter&\pageref{fig:twistedsum},\pageref{eq:seven},\pageref{eq:R(T)}&\cr
&$\tilde S$&dimensionless non-universal moduli&\pageref{gl:Stilde},\pageref{gl:Stildeii}&\cr
&$t_\LR$&left/right K\"ahler moduli of Calabi--Yau threefold $X_\LR$&\pageref{eq:gLR}&\cr
&$z_\LR$&left/right comp. struct. moduli of Calabi--Yau threefold $X_\LR$&\pageref{eq:gLR}&\cr
\noalign{\hrule}
&\multispan3~Cohomology groups and lattices:\hfil&\cr
\noalign{\hrule}
&$L$&two-form lattice of K3 surfaces $S_{L/R}$, $L=H^2(S_L,\mathbb{Z})=H^2(S_R,\mathbb{Z})$ &\pageref{eq:cohTwistedSum}&\cr
&$N_\LR$&left/right Picard lattices of polarized K3 surface $S_\LR$&\pageref{eq:cohTwistedSum}&\cr
&$T_\LR$&left/right transcendental lattices of polarized K3 surface $S_\LR$&\pageref{eq:cohTwistedSum}&\cr
&$k_\LR$&left/right kernels of two-form cohomology &\pageref{eq:cohTwistedSum}&\cr
&$W$&orthogonal pushout lattice&\pageref{eq:defW}&\cr
&$W_\LR$&left/right orthogonal complement to the intersection lattice $R$&\pageref{gl:WLR}&\cr
&$R$&intersection lattice&\pageref{eq:defW}&\cr
\noalign{\hrule height 1.2pt}
}}}}
\newpage
{\small
\hbox{
\vbox{
\offinterlineskip
\halign{\strut\vrule width1.2pt#&\hbox to \widthone{\hfil#\hfil}\vrule&\hbox to \widthtwo{~#\hfil}\vrule&\hbox to \widththree{~#\hfil}&#\vrule width1.2pt\cr 
\noalign{\hrule height 1.2pt}
&~Symbol\hfill&Description&~Pages\hfill&\cr
\noalign{\hrule height 1.2pt}
&\multispan3~Cohomology groups and lattices \emph{(continued)}:\hfil&\cr
\noalign{\hrule}
&$N_L\!\perp_R\!N_R$&orthogonal pushout of lattices $N_\LR$ at the intersection lattice $R$&\pageref{eq:Wperp}&\cr
&$N_L \perp N_R$&perpendicular gluing of lattices $N_\LR$&\pageref{eq:Worth}&\cr
&$\langle \cdot , \cdot \rangle_\cdot$&lattice intersection pairing&\pageref{eq:indpairing}&\cr
&$\kappa$&lattice intersection matrix&\pageref{tab:g2:f22},\pageref{tab:Zrank2e2=-4Blocks},\pageref{tab:34Blocks}&\cr
&$\Delta^\kappa$&discriminant of the lattice intersection matrix&\pageref{tab:g2:f22},\pageref{tab:Zrank2e2=-4Blocks},\pageref{tab:34Blocks}&\cr
\noalign{\hrule}
&\multispan3~Gauge theory data:\hfil&\cr
\noalign{\hrule}
&$(\cdot)^\flat$&superscript for Higgs branch quantity&\pageref{eq:dimhAb},\pageref{eq:HnonAb}&\cr
&$(\cdot)^\sharp$&superscript for Coulomb branch quantity&\pageref{eq:dimcAb},\pageref{eq:CnonAb}&\cr
&$H^\flat$&$\mathcal{N}=2$ Higgs branch&\pageref{eq:dimhAb},\pageref{eq:HnonAb}&\cr
&$C^\sharp$&$\mathcal{N}=2$ Coulomb branch&\pageref{eq:dimcAb},\pageref{eq:CnonAb}&\cr
&$h^\flat$&complex dimension of the Higgs branch&\pageref{eq:dimhAb},\pageref{eq:HnonAb}&\cr
&$c^\sharp$&complex dimension of the Coulomb branch&\pageref{eq:dimcAb},\pageref{eq:CnonAb}&\cr
&$G$&gauge group&\pageref{eq:GAb},\pageref{eq:GnonAb}&\cr
&$\adj$&adjoint representation&\pageref{tab:SpecNonAbelian},\pageref{tab:tabP3},\pageref{tab:tabW6}&\cr
&$\fun{k}$&fundamental representation of $SU(k)$&\pageref{tab:SpecNonAbelian},\pageref{tab:tabP3},\pageref{tab:tabW6}&\cr
\noalign{\hrule}
&\multispan3~Miscellaneous:\hfil&\cr
\noalign{\hrule}
&$\Delta$&Laplacian of compact seven-manifold $Y$&\pageref{eq:OpSymAntiSym},\pageref{eq:Da}&\cr
&$\Delta_L$&Lichnerowicz Laplacian of compact seven-manifold $Y$&\pageref{eq:RicEq},\pageref{eq:Da}&\cr
&$\slashed{D}$&Dirac operator of compact seven-manifold $Y$&\pageref{eq:fermzeromodes},\pageref{gl:DRS}&\cr
&$\slashed{D}^\text{RS}$&Rarita--Schwinger operator of compact seven-manifold $Y$&\pageref{eq:fermzeromodes},\pageref{gl:DRS}&\cr
&$K$&K\"ahler potential of $\mathcal{N}=1$ supergravity action&\pageref{eq:KahlerPotential},\pageref{eq:Kkov},\pageref{eq:Kkovcor}&\cr
&$f_{IJ}$&gauge kinetic coupling functions of $\mathcal{N}=1$ supergravity action&\pageref{eq:GaugeKineticCoupling}&\cr
&$W$&superpotential of $\mathcal{N}=1$ supergravity action&\pageref{eq:Superpotential}&\cr
&$(\cdot)_\LR$&left/right subscript in Kovalev's twisted connected sum&\pageref{gl:XLR},\pageref{fig:twistedsum}&\cr 
&$F_\Lambda$&gluing diffeomorphism in Kovalev's twisted connected sum&\pageref{eq:gluingdiff}&\cr
&$r$&hyper K\"ahler rotation mapping polarized K3 surface $S_L$ to $S_R$&\pageref{gl:r}&\cr
&MM\#$_\rho$&reference to rank $\rho$ Fano threefold in the Mori--Mukai classification&\pageref{gl:MMKlist}&\cr
&K\#&reference to toric semi-Fano threefolds in the Kasprzyk list\label{gl:lastentry}&\pageref{gl:MMKlist}&\cr
\noalign{\hrule height 1.2pt}
}}}}
\newpage
\pagenumbering{arabic}

\section{Introduction}
M-theory compactifications on seven-dimensional manifolds with $G_2$ 
holonomy offer the opportunity to geometrically study the properties of
$\mathcal{N}=1$ effective theories in a setting that is non-perturbative from the superstring 
point of view \cite{Candelas:1984yd,deWit:1986mwo,Acharya:1996ci,Acharya:1998pm}.
As M-theory is conjectured to be the non-perturbative extension of type~IIA theory, 
it is natural to compare it to  F-theory compactifications on elliptically-fibered 
Calabi--Yau fourfolds. It leads to effective $\mathcal{N}=1$ theories in four dimensions and 
is the geometrization of non-perturbative type IIB compactifications on the complex 
three-dimensional base of the fibration. It also includes a varying axio-dilaton 
background due to space-time-filling seven-branes.
 
While standard techniques of complex algebraic geometry provide immediately  
hundreds of thousands of elliptically-fibered Calabi--Yau fourfolds --- for instance realized
as hypersurfaces  and complete intersections in weighted projective spaces
or more generally in toric ambient spaces~\cite{Klemm:1996ts,Kreuzer:1997zg}  --- for a long time there were only about 
a hundred examples of $G_2$-manifolds constructed by the resolution of special 
orbifolds of seven-dimensional torus $T^7$~\cite{MR1424428}. Likewise, the holomorphic terms in the
four-dimensional low-energy effective $\mathcal{N}=1$ supergravity action obtained from Calabi--Yau fourfolds
are computable in the underlying algebraic setting. In particular, the flux-induced superpotential is essentially
determined by the integral periods of the Calabi--Yau fourfolds \cite{Gukov:1999ya},
which for compact fourfolds have systematically been determined in refs.~\cite{Bizet:2014uua,Gerhardus:2016iot}. 
Furthermore, the holomorphic gauge kinetic coupling functions are in principle accessible in this setting 
as well as threshold corrections to the gauge kinetic terms~\cite{Donagi:2008kj,Blumenhagen:2008aw}.\footnote{The
gauge kinetic coupling functions are holomorphic and one loop exact. However, physical gauge kinetic terms receive
further corrections in $\mathcal{N}=1$ effective theories, see for instance ref.~\cite{Ibanez:2012zz} for a thorough discussion.} 

M-theory, however, has the clear advantage that --- at least in the supergravity
limit --- we expect an explicit description in terms of eleven-dimensional supergravity,
for which a unique eleven-dimensional supergravity action exists~\cite{Cremmer:1978km}.\footnote{The maximal dimension that 
allows this supersymmetry representation was previously pointed out in ref.~\cite{Nahm:1977tg}.}
Therefore, by simply studying Kaluza--Klein reductions of this action on compact seven-dimensional
manifolds one obtains four-dimensional low-energy effective theories that capture already
many of the physical properties of the associated M-theory compactifications \cite{Candelas:1984yd,deWit:1986mwo,Acharya:2000ps,Beasley:2002db,Lukas:2003dn,Lukas:2003rr,House:2004pm}.
The resulting semi-classical four-dimensional effective action is then further corrected by 
non-perturbative effects specific to M-theory, such as M2- and M5-brane instantons wrapping internal cycles of
the seven-dimensional compactification manifold \cite{Witten:1996bn,Harvey:1999as}.

More recently, a new construction of $G_2$-manifolds has been proposed by
Kovalev~\cite{MR2024648}, which we loosely refer to as Kovalev's twisted connected
sum. The essential idea is to consider suitable pairs of non-compact asymptotically
cylindrical Calabi--Yau threefolds times a circle --- referred to as building blocks --- 
that are glued in their asymptotic region to compact seven-dimensional manifolds admitting a Ricci-flat metric of $G_2$~holonomy. 
Further explorations of this construction show that Kovalev's twisted connected sum
offers again a large number of explicit examples of $G_2$-manifolds~\cite{MR3369307,Halverson:2014tya,Braun:2016igl}.
In fact, the origin of this multitude is similarly
based on toric geometry as the asymptotically cylindrical Calabi--Yau threefolds --- which furnish
the distinct summands in Kovalev's twisted connected sum --- can for instance be constructed  
using the toric weak Fano threefolds, defined by the $4\,319$ reflexive polytopes in three 
dimensions, by blowing up a suitable curve and removing the anti-canonical 
class~\cite{MR3109862}. Currently, there is no systematic gluing prescription for the whole 
class available, but admissible gluing conditions can be straightforwardly established for 
building blocks constructed from $899$ toric varieties of the semi-Fano type defined in~\cite{MR3369307}. 
They admit in general different K\"ahler cones and --- as discussed further in this work --- 
they can be degenerated and resolved to yield different building blocks realizing distinct
branches of both Abelian and non-Abelian gauge theories. An estimate --- even accounting for
the possibility that the homeomorphism class of the constructed examples occur multiple times ---
yields nevertheless a factor of ten for each admissible building block. This leads to an estimated 
number of $10^8 \times m_g$ different $G_2$ examples, where $m_g\ge 1$ is the multiplicity due to the different 
gluings. A rough estimate for non-homeomorphic elliptically-fibered Calabi--Yau fourfolds 
yields alone $10^{18}$ hypersurfaces in toric varieties.\footnote{This is based on 
the observation that the number of $d$-dimensional reflexive polytopes grows at least exponentially in $d$.
Mark Gross's proof for the finiteness of elliptically-fibered Calabi--Yau threefolds in ref.~~\cite{MR1272978}
suggests that the estimated number of elliptically-fibered Calabi--Yau fourfolds is again finite and the order of
magnitude independent of the construction.}  The still huge difference in the orders 
of magnitude is maybe due to the fact that Kovalev's twisted connected sum only realizes a 
particular class of $G_2$-manifolds. Namely, Crowley and Nordstr\"om define a (non-trivial) $\mathbb{Z}_{48}$-valued 
homotopy invariant for $G_2$-manifolds that takes the value~$24$ for any twisted connected 
sum $G_2$-manifold \cite{MR3416118}.

One goal of this paper is to study the four-dimensional $\mathcal{N}=1$ low-energy
effective action that arises from M-theory compactifications on $G_2$-manifolds that are of the twisted
connected sum type. In order to determine the defining data of the resulting $\mathcal{N}=1$
supergravity theory --- such as the K\"ahler potential, the gauge kinetic coupling functions, and
the superpotential --- an important question is to which extent the harmonic analysis of the
asymptotically cylindrical Calabi--Yau threefold summands with their Ricci-flat Calabi--Yau metrics
approximates the one of the Ricci-flat $G_2$-metric of the resulting compact $G_2$-manifold. 
In a certain limit --- to be referred to as the Kovalev limit in the following --- the corrections to
the $G_2$-metric expressed in terms of the Calabi--Yau data become exponentially
suppressed \cite{MR2024648,MR3369307,MR3399097,MR3109862}. Thus, it is the Kovalev limit
that allows us to reliably deduce from the geometry of the Calabi--Yau summands the resulting
low-energy effective action. To some extent we can think of the Kovalev limit of M-theory
on $G_2$-manifolds as the analog of the large volume limit of Calabi--Yau compactifications of
type~II string theories.

The topological analysis of the Kovalev's twisted connected sum gives a rather detailed description on how the cohomology of the resulting $G_2$-manifold is constructed from the (relative) cohomology of the asymptotically cylindrical Calabi--Yau threefold summands \cite{MR2024648,MR3369307}. This cohomological data determines the $\mathcal{N}=1$ vector and chiral multiplets of the resulting four-dimensional theory. In the Kovalev limit we find that the $\mathcal{N}=1$ vector multiplets furnish gauge theory sectors of extended supersymmetry. Specifically, $\mathcal{N}=1$ vector multiplets attributed to the interior of the asymptotically Calabi--Yau threefolds combine with $\mathcal{N}=1$ chiral multiplets to $\mathcal{N}=2$ vector multiplets realizing $\mathcal{N}=2$ gauge theory sectors, whereas $\mathcal{N}=1$ vector multiplets associated to the mutual asymptotic region of the Calabi--Yau summands enhance to $\mathcal{N}=4$ gauge theory sectors.

In the spectrum of twisted connected sum $G_2$-manifolds, we identify two universal $\mathcal{N}=1$ chiral fields $\nu$ and $\kov$. The real part of the chiral field $\nu$ furnishes the overall volume modulus of the $G_2$-manifold. The chiral field $\kov$ is specific to Kovalev's twisted connected sum construction, as its real part parametrizes the Kovalev limit. In the sequel we refer to this multiplet as the Kovalevton $\kov$. Restricting the dynamics of the two universal chiral moduli fields in the vicinity of the Kovalev limit, we arrive at the universal expression for the K\"ahler potential of the four-dimensional effective supergravity theory
\begin{equation}
  K \,=\, - \log\left[ (\nu + \bar\nu)^4 (\kov+\bar\kov)^3 \right] \ .
\end{equation}  
This simple semi-classical K\"ahler potential --- only capturing the dynamics of the two universal chiral multiplets --- fulfills the no-scale inequality implying a manifest non-negative F-term scalar potential, such that no (supersymmetric) anti-de-Sitter vacua can occur. 

From a physics point of view, the most interesting question is whether 
the twisted sum construction can accommodate singularities that  
in the four-dimensional effective theory lead to enhanced non-Abelian gauge symmetries, to 
geometrically engineerable matter content, to a chiral spectrum and to  
transitions within the class ${\cal N}=1$ effective theories connecting 
topologically inequivalent $G_2$-manifolds. The first three questions 
have been addressed in refs.~\cite{Acharya:2000gb,Witten:2001uq,Acharya:2001gy,Berglund:2002hw,Halverson:2015vta} 
and the last one in refs.~\cite{Atiyah:2000zz,Atiyah:2001qf,Halverson:2014tya}, however, mainly in the 
context of local models. The Kovalev construction and the above described 
decoupling into sectors with different amount of supersymmetry allow us
to discuss some of these questions in particular the first two and the 
last one in the context of global $G_2$-manifolds. More explicitly, we 
find that, by degenerating certain algebraic equations in the description of the 
building  blocks and blowing up the corresponding singularities, we 
can achieve various Abelian and non-Abelian gauge symmetries --- for example including the standard model 
gauge group --- as well as matter in the adjoint, bi-fundamental and fundamental representations. Moreover,
following refs.~\cite{Greene:1995hu,Strominger:1995cz,Klemm:1996kv,Katz:1996ht,Berglund:1996uy,Katz:1996xe} we analyze
the Higgs and Coulomb branches of the $\mathcal{N}=2$ gauge theory sectors 
to realize transitions in the building blocks of twisted connected sum $G_2$-manifolds.
The remarkable fact is now that the predicted spectra in the various gauge
theory branches agree with the changes among the $\mathcal{N}=1$ supergravity spectra
of the corresponding compact $G_2$-manifolds. This leads us to propose
geometric transitions among $G_2$-manifolds that are physically connected via 
branches of $\mathcal{N}=2$ gauge theory sectors.

The paper is organized as follows:
in Section~\ref{sec:Mcompactification} we review 
the geometry of the $G_2$-manifolds and the Kaluza--Klein reduction of eleven-dimensional
supergravity on these spaces. We focus on the moduli space of such $G_2$-compactifications, 
on the four-dimensional low-energy effective spectrum and action, and
on the resulting four-dimensional $\mathcal{N}=1$ supergravity description in terms of
the K\"ahler potential, the (flux-induced) superpotential, and the gauge kinetic coupling functions.
Section~\ref{sec:Kovalev} reviews Kovalev's twisted connected sum construction. Firstly, we introduce
the asymptotically cylindrical Calabi--Yau threefolds, and secondly we summarize the twisted connected
sum constructed from a suitable pair of such Calabi--Yau threefolds. Due to the importance to our analysis,
a particular emphasis is put on the Kovalev limit. 
In Section~\ref{sec:M-theoryonTCS} we describe M-theory compactifications on twisted connected sum $G_2$-manifolds. We start with a description of the low-energy effective $\mathcal{N}=1$
spectrum as deduced from the cohomology of the Calabi--Yau summands. We analyze the universal properties
of the low-energy effective theory attributed to Kovalev's twisted connected sum.
In Section~\ref{sec:N=4sectors} we apply the method of orthogonal gluing to explicitly construct novel examples of twisted connected $G_2$-manifolds. We argue that in the Kovalev limit these examples directly relate to Abelian $\mathcal{N}=4$ gauge theory sectors.  
In Section~\ref{sec:N=2sectors} we study the emergence of Abelian and non-Abelian $\mathcal{N}=2$ gauge theory sectors. For both the Abelian and
non-Abelian $\mathcal{N}=2$ gauge theory sectors we establish a correspondence between $\mathcal{N}=2$ Higgs and Coulomb
branches of the gauge theory and the associated phases of twisted connected $G_2$-manifolds. We illustrate the different
physical aspects of the proposed correspondence with explicit examples of $G_2$-manifolds.
Our conclusions are presented in Section~\ref{sec:conclusions}. For the convenience of the reader, we collect in the glossary on pages~\pageref{sec:glossary} to \pageref{gl:lastentry} our notational conventions of the used mathematical symbols together with a reference to their appearence in the main text. In Appendix~\ref{app:KKFermions} we give further technical details on the $G_2$-compactifications of fermionic terms, supplementing the material presented in Section~\ref{sec:Mcompactification}.

\section{M-theory on $G_2$-manifolds}
\label{sec:Mcompactification}
An eleven-dimensional Lorentz manifold $M^{1,10}$\label{gl:M110} together with a four-form flux $G$ of an anti-symmetric three-form tensor field $\hat{C}$ describe the geometry of the low-energy effective action of M-theory. Firstly --- due to fermionic degrees of freedom in M-theory --- the Lorentz manifold ${M}^{1,10}$ must be spin. Secondly --- consistency 
of the effective action at one loop --- imposes the cohomological flux quantization condition \cite{Witten:1996md}
\begin{equation} \label{eq:Gquant}
  \frac{G}{2\pi}  - \frac{\lambda}2 \, \in \, H^4(M^{1,10},\mathbb{Z}) \ , \qquad \lambda = \frac{p_1(M^{1,10})}{2} \ .
\end{equation}  
The class $\lambda$ is integral since the first Pontryagin class $p_1$ is even for seven-dimensional spin manifolds.

In this work we study compactifications of M-theory to four-dimensional Minkowski space $\mathbb{M}^{1,3}$ with $\mathcal{N}=1$ supersymmetry. That is to say that, for the eleven-dimensional Lorentz manifold ${M}^{1,10}$, we consider the compactification ansatz
\begin{equation} \label{eq:Splitting}
  {M}^{1,10} \,=\, \mathbb{M}^{1,3} \times Y \ 
\end{equation}
with the seven-dimensional compact smooth manifold~$Y$. In the absence of background fluxes such a 
four-dimensional $\mathcal{N}=1$ Minkowski vacuum implies that the internal space $Y$ must be a $G_2$-manifold~\cite{Candelas:1984yd,Acharya:1996ci,Acharya:1998pm}.

A $G_2$-manifold $Y$ is a seven-dimensional Ricci-flat Riemannian manifold with $G_2$ holonomy and not a proper 
subgroup thereof. Furthermore, the manifold $Y$ is spin with a single globally defined covariantly constant 
spinor~\cite{MR696037}. Note that for $G_2$-manifolds the characteristic class $\lambda$ is 
always even~\cite{Harvey:1999as,Beasley:2002db}, which is consistent with the quantization 
condition~\eqref{eq:Gquant} for the considered compactification ansatz of vanishing background flux.

Let us briefly recall some relevant aspects of $G_2$-manifolds. First of all, the exceptional Lie group $G_2$ 
is a fourteen-dimensional simply connected subgroup of $SO(7)$, and we can think of $G_2$ in the following way. 
A three-form $\varphi$ on $\mathbb{R}^7$ gives rise to a canonical symmetric bilinear form on $\mathbb{R}^7$
\begin{equation} \label{eq:Bpairing}
  B_\varphi(X,Y) = -\frac16 (X \,\lrcorner\, \varphi) \wedge (Y \,\lrcorner\, \varphi) \wedge \varphi \ 
\end{equation}
with values in $\Lambda^7(\mathbb{R}^7)^*$. For a generic three-form $\varphi$, the bilinear form $B_\varphi$ yields a non-degenerate pairing of some signature $(p,q)$ with $p+q=7$ 
(with respect to an oriented volume form on $\mathbb{R}^7$)~\cite{MR916718,Hitchin:2000jd}. 
In particular, there is an open set $\Lambda^3_+(\mathbb{R}^7)^*$ in the space of three-forms 
$\Lambda^3(\mathbb{R}^7)^*$ such that $B_\varphi$ is a positive definite bilinear form 
for $\varphi\in\Lambda^3_+(\mathbb{R}^7)^*$. Then $\operatorname{GL}(7,\mathbb{R})$ acts on 
the three-form $\varphi$ and $G_2$ is its fourteen-dimensional stabilizer subgroup. 
Since $G_2$ leaves the positive definite pairing $B_\varphi$ invariant, it is actually a 
subgroup of $SO(7)$. 

Since the Lie group $G_2$ is the stabilizer group of the described three-form, a seven-dimensional oriented manifold $Y$ together with 
the three-form $\varphi$ in $\Omega^3_+(Y)$ --- which is the space of smooth three-forms oriented-isomorphic 
to $\Lambda^3T^*_pY \simeq \Lambda^3_+(\mathbb{R}^7)^*$ for any $p\in M$ --- becomes a $G_2$-structure 
manifold.\footnote{Note that $\Lambda_+^3(\mathbb{R}^7)^*$ is a convex open set in $\Lambda^3(\mathbb{R}^7)^*$. 
Thus --- with a partition of unity --- we can construct a $G_2$-structure on any smooth paracompact seven-dimensional manifold $Y$.}
Thus we call  $\varphi$ a $G_2$-structure on $Y$. Furthermore, the positive definite 
pairing~\eqref{eq:Bpairing} defines a Riemannian metric $g_\varphi$ on $Y$. 
Namely, at any point $p\in M$ and for any basis $\partial_1|_p, \ldots, \partial_7|_p$ at $T_pY$, we obtain the positive definite inner product 
\begin{equation} \label{eq:G2met}
\begin{aligned}
  &g_\varphi(X_p,Y_p) \,=\, \frac{B_\varphi(X_p,Y_p)(\partial_1|_p, \ldots, \partial_7|_p)}{\operatorname{vol}_\varphi(\partial_1|_p, \ldots, \partial_7|_p)} \ , \\
  &{\operatorname{vol}_\varphi(\partial_1|_p, \ldots, \partial_7|_p)}^9 \,=\, \det \left[ B_\varphi(\partial_i|_p,\partial_j|_p)(\partial_1|_p, \ldots, \partial_7|_p) \right] \ ,
\end{aligned}
\end{equation}  
for vectors $X_p$ and $Y_p$ in the tangent space $T_pY$.

The remarkable  and important theorem for the following is that a $G_2$-structure manifold has a subgroup of $G_2$ as its 
holonomy group if and only if the three-form $\varphi$ is harmonic with respect to the constructed $G_2$-metric $g_\varphi$~\cite{MR696037}, i.e., 
\begin{equation} \label{eq:tfree}
  d \varphi = 0 \ , \qquad d *_{g_\varphi} \varphi = 0 \ ,
\end{equation}
in terms of the Hodge star $*_{g_\varphi}$ of the metric $g_\varphi$. Such a harmonic three-form in $\Omega^3_+(Y)$ is called torsion-free. Requiring in addition a finite fundamental group $\pi_1(Y)$ ensures that $Y$ has $G_2$ holonomy and not a proper subgroup thereof. Thus these manifolds --- that is to say with finite fundamental group $\pi_1(Y)$ and torsion-free $G_2$-structure --- are referred to as $G_2$-manifolds.

The system of partial differential equations~\eqref{eq:tfree} for torsion-free $G_2$-structures are highly non-linear due to 
relation~\eqref{eq:G2met} between the metric $g_\varphi$ and the $G_2$-structure $\varphi$. Nevertheless, given a 
torsion-free $G_2$-structure of a $G_2$-manifold, the local structure of the moduli space $\mathcal{M}$ of 
$G_2$-manifolds is known due to Joyce~\cite{MR1424428}.  In particular the Betti number $b_3(Y)$ is the 
dimension of $\mathcal{M}$. We will discuss these  aspects of the local structure in the next section.

\subsection{Kaluza--Klein reduction on $G_2$-manifolds}
Eleven-dimensional $\mathcal{N}=1$ supergravity compactified on a seven-dimensional manifold $Y$ without four-form background fluxes 
to four-dimensional supergravity has been first discussed in ref.~\cite{Candelas:1984yd}. A possible warp factor in this 
compactification has been considered in ref.~\cite{deWit:1986mwo}, where it was shown that the warping breaks 
supersymmetry.\footnote{While these authors give the criteria for the compactification manifold $Y$ to yield four-dimensional $\mathcal{N}=1$ supergravity, they do not refer to the $G_2$-manifolds in Berger's classification of special holonomy manifolds~\cite{MR0079806}, likely because compact examples of $G_2$-manifolds were only found much later in ref.~\cite{MR1424428}.} The structure of the massless four-dimensional 
$\mathcal{N}=1$ multiplets that arise from such compactification was shown in~ref.~\cite{Acharya:1998pm} to possess $b_2(Y)$ abelian $U(1)$ 
vector fields and $b_3(Y)$ neutral chiral fields $\Phi^i$ (see also~ref.~\cite{Acharya:1996ci}). The inclusion of background fluxes $G$ ---  
generating a superpotential $W$ for the neutral chiral fields $\Phi^i$ and thereby generically breaking supersymmetry ---  
has been analyzed in refs.~\cite{Acharya:2000ps,Beasley:2002db}.

Let us now review the Kaluza--Klein reduction of eleven-dimensional $\mathcal{N}=1$ supergravity, which furnishes the 
low-energy effective description of M-theory. The massless spectrum of this maximally supersymmetric supergravity theory consists only of the eleven-dimensional gravity 
multiplet. Its bosonic massless field content is given by the eleven-dimensional space-time metric tensor $\hat{g}_{MN}$, 
the three-form tensor $\hat{C}_{[MNP]}$, whereas the fermionic massless field content is given by the eleven-dimensional 
gravitino $\hat{\Psi}^{\alpha}_M$. The degrees of freedom of the massless component fields in the gravity 
multiplet transform in the following irreducible representations of the little group $SO(9)$: 
\begin{itemize}
 \item The metric  $\hat{g}_{MN}$ in the traceless symmetric representation $\mathbf{44}$. 
 \item The three-form $\hat{C}_{[MNP]}$ in the anti-symmetric three-tensor representation $\mathbf{84}$.
 \item The gravitino $\hat{\Psi}^{\alpha}_M$ in the spinorial representation $\mathbf{128}_s$.
\end{itemize}

We now perform the Kaluza--Klein reduction to four-dimensional Minkowski space $\mathbb{M}^{1,3}$ with the compactification ansatz~\eqref{eq:Splitting}. To solve Einstein's equations in the absence of background fluxes, we consider the block diagonal metric\footnote{In the presence of background fluxes in the internal space~$Y$, the ansatz for the metric is generalized to a warped product $\hat{g}(x,y) = e^{2A(y)}\eta_{\mu\nu} dx^{\mu} dx^{\nu} + e^{-2A(y)}g_{mn} dy^{m} dy^{n}$ in terms of the function $A(y)$ on $Y$ called the warped factor \cite{Acharya:2000ps}, which generically breaks four-dimensional $\mathcal{N}=1$ supersymmetry~\cite{deWit:1986mwo,Acharya:2000ps}.}
\begin{equation} \label{eq:Metric}
   \hat{g}(x,y) = \eta_{\mu\nu}dx^{\mu} dx^{\nu} + g_{mn}(y)dy^m dy^n \ ,
\end{equation}
where $x^\mu$ and $y^m$ furnish (local) coordinates of the four-dimensional Minkowski space $\mathbb{M}^{1,3}$ with the flat space-time metric $\eta_{\mu\nu}$ and the seven-dimensional $G_2$-manifold $Y$ with the Ricci-flat Riemannian metric $g_{mn}$, respectively. Notice that we use upper-case latin letters for eleven-dimensional indices, lower-case latin letters for seven-dimensional indices, and greek letters for four-dimensional indices.

The first task is to deduce the massless spectrum of the effective four-dimensional low-energy theory. We start with 
the gravitational degrees of freedom, which infinitesimally describe the fluctuations of the metric background~\eqref{eq:Metric}, 
i.e., $\hat g \to \hat g + \delta\hat g$. Firstly, we obtain the four-dimensional metric fluctuations $\delta g_{\mu\nu}$, which corresponds to the gravitational degrees of the four-dimensional low-energy effective theory. Secondly, since the fundamental group of $G_2$-manifolds is finite, there are no massless gravitational Kaluza--Klein vectors. Finally, we determine the gravitational Kaluza--Klein scalars $S^{i}$, which furnish coordinates on the moduli space of $G_2$-metrics. At a given point $S^i$ in the moduli space we fix a reference metric, and consider its  infinitesimal deformation under $\delta S^i$, i.e., 
\begin{equation} \label{eq:Metdeform}
   g_{mn}(S^i) dy^m dy^n \to g_{mn}(S^i) dy^m dy^n + \sum_i \delta S^{i} \,\rho^{\text{sym}}_{i,(mn)}(S^i) dy^m dy^n \ .
\end{equation}
Then, solving Einstein's equations to linear order in the {\sl symmetric metric fluctuations} $\rho^{\text{sym}}_{i,(mn)}$, we obtain
\begin{equation} \label{eq:RicEq}
  \operatorname{Ric}\left(g + \sum \delta S^{i} \,\rho^{\text{sym}}_i \right) \,=\, 0 
  \quad \Rightarrow \quad \Delta_L \,\rho^{\text{sym}}_i \,=\, 0 \ ,
\end{equation}  
in terms of the Lichnerowicz Laplacian $\Delta_L$ for the symmetric tensor fields. Using the $G_2$-structure $\varphi$ on $Y$, 
we construct the anti-symmetric three-form tensors
\begin{equation} \label{eq:SymAntiSym}
  \rho^{(3)}_{i,[mnp]} \,=\,g^{rs} \rho^{\text{sym}}_{i,r[m} \varphi_{np]s}\ .
 \end{equation}  
On $G_2$-manifolds the symmetric tensor $\rho^{\text{sym}}_i$ is a zero mode of the Lichnerowicz Laplacian 
operator if and only if the above constructed three-form $\rho^{(3)}_i$ is harmonic~\cite{MR2733250}, namely
\begin{equation} \label{eq:OpSymAntiSym}
  \Delta_L \,\rho^{\text{sym}}_i \,=\,0  \quad \Leftrightarrow \quad \Delta \rho^{(3)}_i  \,=\,0\ .
\end{equation}    
Thus, the massless Kaluza--Klein scalars $S^{i}$ arise from harmonic three-forms $\rho^{(3)}_{i}$, 
which represent a basis for the vector space $H^3(Y)$ of dimension $b_3(Y)$. 
According to eq.~\eqref{eq:tfree} the harmonic three-forms $\rho^{(3)}_i$ are the first order 
deformations to the torsion-free $G_2$-structure 
\begin{equation}
   \varphi(S^i) \to \varphi(S^i) + \sum_i \delta S^{i} \rho^{(3)}_i(S) \ .
\end{equation}   

At a given point $S^i$ in moduli space the harmonic three-forms $\rho_i^{(3)}$ of $Y$ fall into representations of the structure 
group $G_2$, and $H^3(Y)$ splits as~\cite{MR1424428}
\begin{equation}
  H^3(Y) =H^3_{\bf 1}(Y) \oplus H^3_{\bf 27}(Y) \ , \quad
  \dim H^3_{\bf 1}(Y) =1 \ , \quad \dim H^3_{\bf 27}(y) = b_3(Y) - 1 \ ,
\end{equation}
where the three-form representatives transform in the representations ${\bf 1}$ and ${\bf 27}$ of $G_2$, respectively. 
The harmonic torsion-free $G_2$-structure $\varphi$ corresponds to the unique singlet, and the associated deformation simply 
rescales the volume of the $G_2$-manifold~$Y$. The remaining harmonic forms in the representation $\bf 27$ infinitesimally 
deform the torsion-free $G_2$-structure such that the volume of $Y$ remains constant at first order approximation. Analogously, the symmetric tensors~$\rho_i^{\text{sym}}$ solving the Lichnerowicz Laplacian $\Delta_L$ split into a 
unique singlet --- given by the metric tensors $g$ --- and $b_3(Y)-1$ traceless symmetric tensors in the representation $\bf 27$ 
of the $G_2$-structure group.

Above we have seen that the infinitesimal deformations can be identified with harmonic three-forms. In ref.~\cite{MR1424428} Joyce shows that these infinitesimal deformations are actually unobstructed to all orders. That is to say that the vicinity $U_{\varphi(S^i)} \subset \mathcal{M}$ of a given torsion-free $G_2$-structure $\varphi(S^i) \in \mathcal{M}$ --- at a given point $S^i$, $i=1,\ldots,b_3(Y)$, in the moduli space --- is locally diffeomorphic to the de Rham cohomology $H^3(Y)$,\footnote{We assume that the scalars $S^i$ describe a generic point in $\mathcal{M}$. First of all, the associated $G_2$-manifold $Y$ should be smooth. Furthermore, it should not be a special symmetric point corresponding to an orbifold singularity in $\mathcal{M}$.} i.e., 
\begin{equation} \label{eq:ModSpace}
  \mathcal{P}_{\varphi(S^i)}: U_{\varphi(S^i)}\subset\mathcal{M} \to H^3(Y) \ , \ \varphi \mapsto [\varphi] \ .
\end{equation}
Hence, the Betti number $b_3(Y)$ is indeed the dimension of $\mathcal{M}$, and the scalar fields $S^i$ furnish local coordinates on $\mathcal{M}$ with the infinitesimal deformations $\delta S^i$ spanning the tangent space $T_{S^i}\mathcal{M}$. This local structure implies that the massless infinitesimally metric deformations $\rho^{\text{sym}}_i$ --- or alternatively the first order deformations $\rho^{(3)}_i$ to the torsion-free $G_2$-structure --- extend order-by-order to unobstructed finite deformations, which therefore describe locally the moduli space $\mathcal{M}$ of $G_2$-manifolds. While the harmonic three-forms $\rho_i^{(3)}$ themselves depend (non-linearly) on the moduli space coordinates $S^i$, we can --  due to eq.~\eqref{eq:ModSpace} -- locally  expand the cohomology class $[\varphi]$ of the torsion-free $G_2$-structure $\varphi$ as
\begin{equation} \label{eq:cohexpand}
  [\varphi(S^i)] \,=\, \sum_i S^i \, [\rho^{(3)}_i] \ ,
\end{equation}  
which is a useful local description of the moduli space of $Y$.\footnote{Note, however, that a given cohomology class $[\varphi(S^i)]$ is not necessarily represented by a unique torsion-free $G_2$-structure $\varphi(S^i)$. Mathematically, not much is known about the global structure of the moduli space $\mathcal{M}$ and, in particular, about the global map $\mathcal{P}: \mathcal{M} \to H^3(Y)$.}  

Massless four-dimensional modes arise from the coefficients in the decomposition of the eleven-dimensional anti-symmetric 
three-form tensor $\hat C$ as
\begin{equation} \label{eq:ExpansionCfield}
	\hat{C}(x,y)\
	\,=\, \sum_I A^{I} (x) \wedge \omega^{(2)}_I (y) + \sum_i P^{i}(x) \wedge \rho^{(3)}_i(y) \ ,
\end{equation}
in terms of the harmonic two-forms $\omega^{(2)}_I$ and three-forms $\rho^{(3)}_i$ identified with non-trivial cohomology representatives of 
$H^2(Y)$ and $H^3(Y)$ of dimension $b_2(Y)$ and $b_3(Y)$, respectively. Thus, as there are no dynamical degrees of freedom in four-dimensional 
anti-symmetric three-form tensor fields and due to the absence of harmonic one-forms on the internal $G_2$-manifolds, the four-dimensional vectors 
$A^{I}$, $I=1,\ldots,b_2(Y)$, and the four-dimensional scalars $P^{i}$, $i=1,\ldots,b_3(Y)$, are the only massless modes obtained from the dimensional 
reduction of the eleven-dimensional anti-symmetric three-form tensor field $\hat C$. 

Let us now turn to the dimensional reduction of the eleven-dimensional gravitino $\hat\Psi$, which geometrically is a section of $T^*M^{1,10} \otimes S M^{1,10}$, where $SM^{1,10}$ denotes a spin bundle of $M^{1,10}$. Upon dimensional reduction the gravitino $\hat\Psi$ enjoys the expansion 
\begin{equation} \label{eq:ExpansionGravitino}
   \hat\Psi(x,y) \,=\, \left( \psi_\mu(x) dx^\mu + \psi^{\ast}_\mu(x) dx^\mu \right) \zeta(y)
   + \left( \chi(x) + \chi^{\ast}(x) \right) \zeta^{(1)}_n(y) dy^n \ .
\end{equation}
Here $(\psi_\mu,\psi^{\ast}_\mu)$  and $(\chi, \chi^{\ast})$ 
are four-dimensional Rarita--Schwinger and four-dimensional spinor fields of both chiralities in 
$\mathbb{M}^{1,3}$.\footnote{In our conventions the fermionic fields $\psi_\mu$, $\chi$ and $\psi^{\ast}_\mu$, $\chi^{\ast}$ are chiral and anti-chiral,  respectively, such that $\psi_\mu+\psi^{\ast}_\mu$ and $\chi+\chi^{\ast}$ become Majorana fermions.} $\zeta$ is a section of the (real) spin bundle $SY$ of the compact $G_2$-manifold $Y$.\footnote{As the $G_2$-structure group of $Y$  --- a subgroup of $SO(7)$ --- is simply connected, it defines a canonical spin structure on $Y$.} Furthermore, $\zeta^{(1)}$ is a section of the (real) Rarita--Schwinger bundle $T^*Y \otimes SY$, which locally takes the form $\theta^{(1)} \otimes \tilde\zeta$ in terms of the local one-form $\theta^{(1)}$ and the spinorial section $\tilde\zeta$.

On the $G_2$-manifold the spin bundle splits as $SY\simeq T^*Y \oplus \mathbb{R}$ \cite{MR1424428} such that the section $\zeta$ decomposes accordingly 
\begin{equation} \label{eq:zeta}
  \zeta = \sum_m a_m(y) \gamma^m \eta + b(y) \eta \ .
\end{equation}
Here, $\eta$\label{gl:eta} is the covariantly constant Majorana spinor of the $G_2$-manifold and $\gamma^m$ are the seven-dimensional gamma matrices. Similarly, we analyze the Rarita--Schwinger section $\zeta^{(1)}$ of $T^*Y \otimes SY$. It decomposes as 
\begin{equation} \label{eq:zeta1}
  \zeta^{(1)} \,=\, \sum_{n,m} a^{28}_{(nm)}(y)  dy^n \otimes  \gamma^m \eta
   + \sum_{n,m} a^{14}_{[nm]}(y) dy^n \otimes \gamma^m \eta 
   +  \sum_n b_n^7(y)  dy^n \otimes \eta  \ .
\end{equation}
The superscripts in the symmetric tensor $a^{28}_{(nm)}(y)$, the anti-symmetric tensor $a^{14}_{[nm]}(y)$, 
and the vector $b_n^7(y)$ indicate the dimension of their respective representations with respect to the structure group $G_2$. While the anti-symmetric 
tensor $a^{14}_{[nm]}(y)$ and the vector $b_n^7(y)$ transform in the irreducible representations $\bf 14$ and $\bf 7$, the symmetric tensor $a^{28}_{(nm)}(y)$ 
further decomposes into the trace and the traceless symmetric part, which respectively correspond to the irreducible representations $\bf 1$ and $\bf 27$.

The massless four-dimensional fermionic spectrum results from the zero modes of the seven-dimensional Dirac operator $\slashed{D}$ and Rarita--Schwinger operator $\slashed{D}^\text{RS}$, i.e,
\begin{equation} \label{eq:fermzeromodes}
  \slashed{D} \zeta \,=\, 0 \ , \qquad \slashed{D}^\text{RS} \zeta^{(1)} \,=\, 0 \ .
\end{equation}   
The zero modes of these operators on $G_2$-manifolds are discussed in Appendix~\ref{sec:ZeroModes} and are also determined in ref.~\cite{Font:2010sj}. For the spinorial section $\zeta$, the covariantly constant spinor $\eta$ --- i.e., $b(y) \equiv 1$ --- is the only zero mode of the Dirac operator. In the Rarita--Schwinger section $\zeta^{(1)}$, the one-form tensor $b^7(y)=b_n^7(y)dy^n$ does not contribute any zero modes. All zero modes 
arise from the zero modes of the Lichnerowicz Laplacian and the two-form Laplacian acting respectively on the symmetric tensors $a^{28}(y) = a^{28}_{(nm)}(y) dy^n \otimes dy^m$ and 
the anti-symmetric tensors $a^{14}(y) = a^{14}_{[nm]}(y) dy^n \wedge dy^m$, i.e.,
\begin{equation} \label{eq:Da}
   \Delta_L a^{28}(y) \,=\, 0 \ , \qquad \Delta a^{14}(y) \,=\, 0 \ .
\end{equation}   
The zero modes of the Lichnerowicz Laplacian on $G_2$-manifolds are again identified with harmonic three-forms according to eqs.~\eqref{eq:SymAntiSym} and \eqref{eq:OpSymAntiSym} --- with a single zero mode and $b_3(Y)-1$ traceless symmetric zero modes transforming in the $G_2$-representations $\bf 1$ and $\bf 27$, respectively. Therefore, the zero modes of the Rarita--Schwinger bundle on $Y$ are in one-to-one correspondence with non-trivial cohomology elements of both $H^3(Y)$ and $H^2(Y)$,\footnote{A priori, the constructed zero modes furnish elements of $H^3_\mathbf{1}(Y)$, $H^3_\mathbf{27}(Y)$ and $H^2_\mathbf{14}(Y)$ that transform in the specified representations of the $G_2$-structure group. However, on $G_2$-manifolds all non-trivial three- and two-form cohomology elements can respectively be represented in the representations $\mathbf{1}$, $\mathbf{27}$, and $\mathbf{14}$, which justifies the identification of zero modes with $H^3(Y)$ and $H^2(Y)$ \cite{MR1424428}, cf. also Appendix~\ref{app:KKFermions}.} and we arrive at the expansion of the four-dimensional chiral fermions
\begin{equation} \label{eq:chiexp}
   \chi(x) \zeta^{(1)}(y) \,=\,  \sum_{i=1}^{b_3(Y)}\chi^i(x) \rho^{\text{sym}}_{i,(nm)} dy^n \otimes\gamma^m\eta 
   + \sum_{I=1}^{b_2(Y)} \lambda^I \omega_{I[nm]}^{(2)} dy^n\otimes\gamma^m\eta \ ,
\end{equation}
in terms of the bases of zero modes $\rho^{\text{sym}}_i$ of the Lichnerowicz Laplacian and of the harmonic two-forms $\omega_I^{(2)}$. 

\begin{table}[t]
\hfil
\hbox{
\vbox{
\offinterlineskip
\halign{\vrule height2.7ex depth1.2ex width1.2pt\hfil~#\hfil\vrule width0.8pt&~#\hfil&\vrule~#\hfil&\vrule width0.8pt~#\hfil\vrule width1.2pt\cr 
\noalign{\hrule height 1.2pt}
Multiplicity &\multispan2\hfil Massless 4d component fields \hfil& \hfil ~Massless 4d \hfil \cr
& bosonic fields & fermionic fields & \hfil ~$\mathcal{N}=1$ multiplets \hfil \cr
\noalign{\hrule height 1.2pt}
$1$ & metric $g_{\mu\nu}$ & gravitino $\psi_\mu,\psi_\mu^*$ & gravity multiplet \cr
\noalign{\hrule}
$i=1,\ldots,b_3(Y)$ & scalars $(S^i,P^i)$ & spinors $\chi^i,\chi^{*\,i}$ & chiral multiplets $\Phi^i$ \cr
\noalign{\hrule}
$I=1,\ldots,b_2(Y)$ & vectors $A^I_\mu$ & gauginos $\lambda^I_\alpha$ & vector multiplets $V^I$ \cr
\noalign{\hrule height 1.2pt}
}
}}
\hfil
\caption{This table summarizes the massless four-dimensional low-energy effective $\mathcal{N}=1$ supergravity spectrum that is obtained from the dimensional reduction of M-theory  --- or rather of eleven-dimensional supergravity --- on a smooth $G_2$-manifold~$Y$.} 
\label{tab:N1multi}
\end{table}

Now, we can spell out the massless four-dimensional spectrum in terms of $\mathcal{N}=1$ supergravity multiplets as obtained from the 
dimensional reduction of M-theory upon the $G_2$-manifolds~$Y$. It consists of the four-dimensional supergravity multiplet, 
$b_3(Y)$ (neutral) chiral multiplets $\Phi^i$, and $b_2(Y)$ (Abelian) vector multiplets $V^I$, as summarized in detail in Table~\ref{tab:N1multi}.

To specify the four-dimensional low-energy effective $\mathcal{N}=1$ supergravity action for the determined spectrum 
of the massless fields, we insert the mode expansions for the metric~\eqref{eq:Metdeform}, the anti-symmetric three-form 
tensor~\eqref{eq:ExpansionCfield}, and the gravitino~\eqref{eq:ExpansionGravitino} into the eleven-dimensional 
supergravity action~\cite{Cremmer:1978km}, which in terms of the eleven-dimensional Hodge star $*_{11}$ and the eleven-dimensional gamma matrices $\hat\Gamma^M$ reads 
\begin{multline} \label{eq:11dSUGRA}
   S_{11d} \,=\, 
   \frac{1}{2\kappa^2_{11}}\int \left( *_{11} \hat R_S - \frac12 d\hat C \wedge *_{11} d\hat C 
   - *_{11} i \, \bar{\hat\Psi}_M \hat\Gamma^{MNP} \hat D_N \hat \Psi_P  \right) \\
   -\frac{1}{192\kappa_{11}^2} \int *_{11} \bar{\hat\Psi}_M \hat\Gamma^{MNPQRS}  \hat \Psi_N 
   (d\hat C)_{[PQRS]} -\frac{1}{2\kappa^2_{11}} \int d\hat{C} \wedge\ast_{11} \hat{F}\\
   - \frac1{12\kappa^2_{11}} \int d\hat C \wedge d\hat C \wedge \hat C + \ldots \ ,
\end{multline}
where we denote $\hat{F}_{[MNPQ]} = 3\bar{\hat{\Psi}}_{[M}\hat{\Gamma}_{NP}\hat{\Psi}_{Q]}$.
The first line contains the kinetic terms of the eleven-dimensional supergravity multiplet, i.e., the Einstein--Hilbert term in terms of the Ricci scalar $\hat R_S$, 
the kinetic term for the anti-symmetric three-form tensor $\hat C$, and the Rarita--Schwinger kinetic term for the gravitino $\hat\Psi$. 
The second line comprises the interaction terms and the third line is the Chern--Simons term of the eleven-dimensional supergravity action. 
There are additional four-fermion interactions denoted by `$\ldots$'  \cite{Cremmer:1978km}. The coupling constant $\kappa_{11}$ relates to the eleven-dimensional 
Newton constant $\hat G_N$, the eleven-dimensional Planck length $\hat\ell_P$ and Planck mass $\hat M_P$ according to
\begin{equation}
   \kappa^2_{11} \,=\, 8\pi \hat G_N \,=\,\frac{(2\pi)^8\,\hat \ell_P^9}{2} \,=\, \frac{(2\pi)^8}{2\,\hat M_P^9} \ .
\end{equation}

To perform the Kaluza--Klein reduction let us introduce the moduli-dependent volume $V_Y(S^i)$ of the $G_2$-manifold~$Y$ given by
\begin{equation}
  V_Y(S^i) \,=\, \int_Y d^7y \sqrt{ \det g(S^i)_{mn} }  \ .
\end{equation}  
Furthermore, we introduce a reference $G_2$-manifold $Y_0$ with respect to some background expectation values 
$S^i_0= \langle S^i \rangle$, upon which we carry out the dimensional reduction. This allows us to introduce the 
dimensionless (but yet moduli-dependent) volume factor
\begin{equation} \label{eq:Vol}
  \lambda_0(S^i) \,=\, \frac{V_Y(S^i)}{V_{Y_0}} \,=\, \frac17 \int_Y \varphi \wedge *_{g_\varphi} \varphi\ , 
\end{equation}
in terms of the reference volume $V_{Y_0}=V_Y(S^i_0)$. Here the choice of $Y_0$ fixes via the resulting volume factor $V_{Y_0}$ the normalization of the three-form $\varphi$.

Then --- using eqs.~\eqref{eq:Metdeform} and \eqref{eq:ExpansionCfield} --- the dimensional reduction of the Einstein--Hilbert term and the three-form tensor $\hat C$ yields the four-dimensional action \cite{Beasley:2002db} 
\begin{equation} \label{eq:BosAct}
\begin{aligned} 
     S_{4d}^\text{bos} \,=\, \frac{1}{2\kappa_4^2} 
       \int &\Big[  *_4 R_S + \frac{\kappa_{IJk}}{2} \left(S^k F^I \wedge *_4 F^J - P^k F^I \wedge F^J \right)  \\
        &\quad - \frac{1}{2 \lambda_0} \int_Y \rho^{(3)}_i \wedge \ast_{g_\varphi} \rho^{(3)}_j \left(dP^i\wedge *_4 dP^j + dS^i \wedge *_4 dS^j \right)  \Big] \ .      
\end{aligned}
\end{equation}
in terms of the four-dimensional Hodge star $*_4$, the Ricci scalar $R_S$ with respect to the metric $g_{\mu\nu}$, the reference volume $V_{Y_0}$, 
and the seven-dimensional Hodge star $*_7$. Here we have performed the Weyl rescaling of the four-dimensional metric according to 
\begin{equation}
     g_{\mu\nu} \,\rightarrow\, \frac{g_{\mu\nu}}{\lambda_0(S^i)} \ ,
\end{equation}
such that the four-dimensional coupling constant $\kappa_4$ --- relating the four-dimensional Newton constant $G_N$, the four-dimensional Planck length $\ell_P$ and the Planck mass $M_P$ --- becomes
\begin{equation} \label{eq:kappa4}
  \kappa_4^2 \,=\, \frac{\kappa_{11}^{2}}{V_{Y_0}} \ , \qquad
  \kappa_4^2 \,=\, 8\pi G_N \,=\, 8\pi \ell_P^2 \,=\, \frac{8\pi}{M_P^2} \ .
\end{equation}
Furthermore, the couplings $\kappa_{IJk}$ arise from the topological intersection numbers
\begin{equation}
     \kappa_{IJk} \,=\, \int_Y \omega^{(2)}_I \wedge \omega^{(2)}_J \wedge \rho^{(3)}_k \ .
\end{equation}

We can now bring the (bosonic) action~\eqref{eq:BosAct} into the conventional form of four-dimensional $\mathcal{N}=1$
supergravity \cite{Wess:1992cp}.\footnote{The structure of the four-dimensional effective $\mathcal{N}=1$ action from type~II and M-theory dimensional reduction including Kaluza--Klein modes
--- which we do not consider here --- has recently been discussed in refs.~\cite{Becker:2016xgv,Becker:2016edk}.}
To identify the chiral multiplets --- that is to say, to identify the complex structure of the K\"ahler target space --- we observe that --- at least to the leading order --- 
the action of the membrane instantons generating non-perturbative superpotential interactions is given by \cite{Harvey:1999as}
\begin{equation} \label{eq:CStarget}
  \phi^i = -P^i + i S^i \ .
\end{equation}
Hence, due to holomorphy of the $\mathcal{N}=1$ superpotential, the complex fields $\phi^i$ furnish complex coordinates of the K\"ahler target space 
and thus represent the complex scalar fields in the $\mathcal{N}=1$ chiral multiplets $\Phi^i$ in Table~\ref{tab:N1multi}. This allows us to quickly read off 
from the action~\eqref{eq:BosAct} the K\"ahler potential and the gauge kinetic coupling functions \cite{Beasley:2002db, Lukas:2003dn}
\begin{align}
    K (\phi, \bar{\phi})&\,=\, -3\,\text{log} \left(\frac{1}{7} \int_Y \varphi \wedge*_{g_{\varphi}} ~\varphi\right) \ , \label{eq:KahlerPotential} \\
    f_{IJ}(\phi)&\,=\, \frac{i}{2}\sum_k  \phi^{k}\int_Y \omega^{(2)}_I \wedge \omega^{(2)}_J \wedge \rho^{(3)}_k 
    \,=\, \frac{i}{2} \sum_k \kappa_{IJk}\phi^k .\label{eq:GaugeKineticCoupling}
\end{align}
Note that the holomorphy of the gauge kinetic coupling functions is in accordance with the complex chiral coordinates \eqref{eq:CStarget}. The moduli space metric is then given by
\begin{equation}
  g_{i\bar{j}} = \partial_i \partial_{\bar{j}} K 
  = \frac{1}{4\lambda_0} \int_Y \rho^{(3)}_i \wedge \ast_{g_\varphi} \rho^{(3)}_j \ .
\end{equation}

Thus, we see that in the physical theory the real scalar fields $S^i$ and $P^i$ combine to the complex chiral scalars $\phi^i$ according to eq.~\eqref{eq:CStarget}. These complex scalar fields parametrize locally the (semi-classical) M-theory moduli space $\mathcal{M}_\mathbb{C}$\label{gl:Mc} of the $G_2$-compactification on $Y$ of complex dimension $b_3(Y)$, where the real subspace $\operatorname{Re}(\phi^i)=0$ of real dimension $b_3(Y)$ is the geometric moduli space $\mathcal{M}$ of $G_2$-metrics on $Y$.\footnote{For corrections to the semi-classical moduli space $\mathcal{M}_\mathbb{C}$ see ref.~\cite{Becker:2014rea}.} Note, however, that the derived moduli space $\mathcal{M}_\mathbb{C}$ merely arises from the semi-classical dimensional reduction of eleven-dimensional supergravity on the $G_2$-manifold $Y$. For the resulting four-dimensional $\mathcal{N}=1$ supersymmetric theory, one expects on general grounds that the flat directions of $\mathcal{M}_{\mathbb{C}}$ are lifted at the quantum level due to non-perturbative effects in M-theory \cite{Harvey:1999as} --- even in the absence of background fluxes. 

Finally, let us remark that the presence of non-trivial four-form background fluxes~$G$ of anti-symmetric three-form tensor fields $\hat{C}$ supported on the $G_2$-manifold $Y$ generates a flux-induced superpotential~\cite{Acharya:2000ps,Beasley:2002db,House:2004pm}. While the 
superpotential enters quadratically in the bosonic action, it appears linearly in the fermionic action generating a gravitino 
mass term $M_\psi$ \cite{Wess:1992cp}
\begin{equation} \label{eq:gravimass}
  \mathcal{L}^{M_\psi}_{4d} \,=\,  \frac{1}{2\kappa_4^2} e^{K/2} 
    \left(\bar{W}(\bar\phi)\,\psi^T_{\mu} \gamma^{\mu\nu} \psi_{\nu} + W(\phi)\,\bar{\psi}_{\mu}\gamma^{\mu\nu}\psi^{*}_{\nu} \right) \ .
\end{equation}
This linear dependence on $W$ allows us to directly derive the superpotential from the dimensional reduction of the gravitino terms, as carried in detail in Appendix~\ref{app:super}, where we determine the holomorphic superpotential to be\footnote{Non-vanishing four-form fluxes induce a gravitational back-reaction to the eleven-dimensional metric. This requires a warped metric ansatz \cite{Acharya:2000ps} that breaks supersymmetry \cite{deWit:1986mwo,Acharya:2000ps}. As the presented derivation neglects such back-reactions, the resulting effective action becomes more accurate the smaller the effect of warping. Similarly as argued in ref.~\cite{House:2004pm} this is the case for a small number of four-form flux quanta.}
\begin{equation} \label{eq:Superpotential}
  W(\phi^i) \,=\, \frac{1}{4} \int_Y G \wedge \left(-\frac12\hat C + i\varphi\right) \ .
\end{equation}   
Our result yields the flux-induced superpotential in refs.~\cite{Acharya:2000ps,Beasley:2002db,House:2004pm,Becker:2014rea}. As explained in ref.~\cite{Beasley:2002db}, in order to obtain the chiral combination $-\delta\hat C + i \delta \varphi$ in the variation $\delta W$ of the superpotential $W$, it is necessary to introduce the relative factor $\frac12$ between $\hat C$ and $\varphi$ in formula~\eqref{eq:Superpotential}. Note that --- both in the presence and in the absence of background fluxes $G$ --- we expect generically additional non-perturbative superpotential contributions arising from membrane instanton effects~\cite{Acharya:1998pm,Harvey:1999as}.

\section{Kovalev's construction of $G_2$-manifolds} \label{sec:Kovalev} 
In this section we focus on the construction of $G_2$-manifolds as put forward by Kovalev~\cite{MR2024648} and further developed by Corti et al. \cite{MR3109862,MR3369307}. These $G_2$-manifolds are obtained from a certain twisted connected sum of two asymptotically cylindrical Calabi--Yau threefolds times an additional circle $S^1$. In the Kovalev limit the Ricci-flat metric of the obtained $G_2$-manifold can be approximated by the metrics of the two Calabi--Yau summands. In order to set the stage for the derivation of the low-energy effective action for M-theory compactified on such $G_2$-manifolds, we identify the Kovalev limit in the moduli space of the constructed $G_2$-manifolds.

\subsection{Asymptotically cylindrical Calabi--Yau threefold} \label{sec:ccy}
A (complex) three-dimensional Calabi--Yau cylinder $X^\infty$ is the product of a compact Calabi--Yau twofold --- which we take to be a compact K3 surface $S$\label{gl:S} --- with an open cylinder $\Delta^{\rm cyl}$, here given as the complement of the unit disk in the complex plane $\mathbb{C}$, i.e., $\Delta^{\rm cyl} = \left\{ z\in\mathbb{C} \,\middle|\, |z| > 1 \right\}$\label{gl:Dcyl}. The K\"ahler form $\omega^\infty$ and the holomorphic three-form $\Omega^\infty$ of $X^\infty$ read
\begin{equation} \label{eq:ccyforms} 
\begin{aligned}
  \omega^{\infty} & \,=\, {\gamma^*}^2 \frac{i {\dd} z \wedge {\dd} \bar z}{2 z \bar z} + \omega_S\,=\,{\gamma^*}^2\,{\dd} t \wedge {\dd} \theta^* +\omega_S \ ,  \\   
  \Omega^{\infty} & \,=\, -{\gamma^*}\frac{i {\dd} z }{z} \wedge\Omega_S\,=\,{\gamma^*}( {\dd \theta^*} - i {\dd} t)\wedge \Omega_S\ ,
\end{aligned}
\end{equation}
in terms of the K\"ahler form $\omega_S$ and the holomorphic two-form $\Omega_S$ of the K3 surface~$S$, the complex coordinate $z = e^{t+ i \theta^*}$ and the length scale $\gamma^*$ of the cylinder $\Delta^{\rm cyl}$.\footnote{With the conventional mutual normalization $(-1)^{\frac{n(n-1)}{2}} \left(\frac{i}{2}\right)^n \Omega \wedge \bar \Omega=\frac{\omega^n}{n!}$ between the K\"ahler form $\omega$ and the holomorphic $n$-form $\Omega$ of Calabi--Yau $n$-folds, we note that --- assuming this normalization for $\omega_S$ and $\Omega_S$ of the K3 surface $S$ --- the K\"ahler form $\omega^\infty$ and the holomorphic three-form $\Omega^\infty$ of $X^\infty$ are conventionally normalized.} Then --- with the metric $g_S$ of the K3 surface --- the product metric $g_{X^\infty}$ of the Calabi--Yau cylinder $X^\infty$ becomes
\begin{equation} \label{eq:metXoo}
  g_{X^\infty} \,=\, {\gamma^*}^2  \left(\dd t^2 + \dd{\theta^*}^2 \right) + g_S \ .
\end{equation}
The length scale $\gamma^*$ furnishes the radius of the cylindrical metric on $\Delta^{\rm cyl}$, whereas the map $\xi: X^\infty \to \mathbb{R}^+$ with $\xi=\log|z|$ projects on the longitudinal direction of the cylinder such that $\xi^{-1}(\mathbb{R}^+) = X^\infty$. 

As defined in refs.~\cite{MR2024648,MR3109862,MR3399097}, an asymptotically cylindrical Calabi--Yau threefold~$X$ is a non-compact Calabi--Yau threefold with $SU(3)$ holonomy and a complete Calabi--Yau metric $g_X$ with the following properties. There exists a compact subspace $K\subset X$ such that the complement $X \setminus K$ is diffeomorphic to a three-dimensional Calabi--Yau cylinder $X^\infty$\label{gl:KLR} and such that the K\"ahler and the holomorphic three-form of $X$ approach fast enough $\omega^\infty$ and $\Omega^\infty$ of the cylindrical Calabi--Yau threefold  $X^\infty$, as given in eqs.~\eqref{eq:ccyforms}. More precisely, given the diffeomorphism~$\eta: X^\infty \to X\setminus K$, we require that in the limit $\xi\to+\infty$ and for any positive integer $k$ \cite{MR2024648,MR3109862,MR3399097}
\begin{equation}  \label{eq:limit}
\begin{aligned}
   \eta^* \omega - \omega^\infty\,&=\,\dd \mu\  &\text{with} &&|\nabla^k \mu| \,=\,O(e^{- \lambda \gamma^* \xi }) \ , \\
   \eta^* \Omega - \Omega^\infty\,&=\,\dd \nu\  &\text{with} &&|\nabla^k \nu| \,=\,O(e^{- \lambda \gamma^* \xi })\ ,
\end{aligned}   
\end{equation} 
for certain choices of $\mu$ and $\nu$ with the norm $|\,\cdot\,|$ and Levi--Civita connection~$\nabla$ of the metric $g_{X^\infty}$. The scale $\lambda$ has inverse length dimension and is determined by the (inverse) length scale of the asymptotic region $X^\infty$. To be precise \cite{MR2024648}
\begin{equation} \label{eq:Clambda}
  \lambda \,=\, \min\left\{ \tfrac1{\gamma^*}, \lambda_S \right\} \ ,
\end{equation}
where $\lambda_S$ is the square root of the smallest positive eigenvalue of the Laplacian of the K3~surface~$S$ in the asymptotic Calabi--Yau cylinder $X^\infty$.

\subsection{Kovalev's twisted connected sum}  \label{sec:generalideaofkovalev} 
From any Calabi--Yau threefold $X$ with $SU(3)$ holonomy we can always construct the seven-manifold $X \times S^1$ with the torsion-free $G_2$-structure
\begin{equation} \label{eq:G2prodform}
  \varphi_0\,=\,\gamma\, \dd \theta \wedge \omega + {\rm Re} (\Omega)\ ,
  \qquad  *\varphi_0\,=\,\frac{1}{2} \omega^2 - \gamma \,\dd \theta \wedge {\rm Im}(\Omega) \ ,        
\end{equation}
in terms of the K\"ahler form $\omega$, the holomorphic three-form $\Omega$ of $X$ and the coordinate $\theta$ of $S^1$ with radius $\gamma$. However, the resulting seven-dimensional manifold with $\varphi_0$ still has $SU(3)$ holonomy, which is a subgroup of $G_2$.

For Kovalev's construction of the $G_2$-manifolds $Y$~\cite{MR2024648}, the essential idea is to first construct two suitable asymptotically cylindrical Calabi--Yau threefolds $X_\LR$ (referred to as left and right).\label{gl:XLR} Then for each of them one takes a direct product with $S^1_\LR$ in order to obtain two seven-manifolds $Y_\LR$ with torsion-free $G_2$-structures $\varphi_{0\, L/R}$ and $SU(3)$ holonomy as above. To obtain a genuine compact $G_2$-manifold~$Y$, the asymptotic regions of type $Y^\infty_\LR=X^\infty_\LR\times S_\LR^1$ are glued together in such a way that the obtained manifold $Y$ admits a torsion-free $G_2$-structure resulting in $G_2$~holonomy and not a subgroup thereof.

To obtain the full $G_2$ holonomy, it is necessary to reduce the infinite fundamental groups $\pi_1(Y_\LR)$ to a finite fundamental group $\pi_1(Y)$ by gluing their asymptotic regions appropriately. For suitable choices of $Y_\LR$ this can be achieved by Kovalev's twisted connected sum construction \cite{MR2024648}. Recall that the asymptotic regions of $Y_\LR$ are given by
\begin{equation} \label{eq:asymp}
  Y^\infty_\LR\,=\, X^\infty_\LR\times S_\LR^1 \,=\, S_\LR \times \Delta^{\rm cyl}_\LR \times  S_\LR^1 \ ,
\end{equation}
with the K3 surfaces $S_\LR$ and the cylinders $\Delta^{\rm cyl}_\LR$ (cf. Section~\ref{sec:ccy}). Let us denote by $\omega^I_\LR,\omega^J_\LR$ and $ \omega^K_\LR$\label{gl:omegatripLR} the triplets of mutually orthogonal hyper~K\"ahler two-forms, satisfying the relations $(\omega^I_\LR)^2= (\omega^J_\LR)^2= (\omega^K_\LR)^2$. Then the $SU(2)$ structures of the asymptotic polarized K3 surfaces $S_\LR$ determine their K\"ahler two-forms $\omega_{S_\LR}^{\infty}$ and their holomorphic two-forms $\Omega_{S_\LR}^{\infty}$ according to
\begin{equation}
    \omega_{S_\LR}^{\infty}\,=\,\omega^I_\LR \ , \qquad 
    \Omega_{S_\LR}^{\infty}\,=\, \omega^J_\LR+ i \, \omega^K_\LR \ .
\end{equation}
With eqs.~\eqref{eq:ccyforms} and \eqref{eq:G2prodform}  this explicitly specifies the asymptotic torsion-free $G_2$-structures
\begin{multline} \label{eq:phi0LR}
   \varphi_{0\,L/R}^\infty\,=\,\gamma_\LR  \dd \theta_\LR \wedge 
   \left( \gamma_\LR^{*\,2} \dd t_\LR \wedge \dd \theta^*_\LR+ \omega_{S_\LR}^\infty \right) \\
   + \gamma_\LR^* \dd \theta^*_\LR \wedge {\rm Re} (\Omega_{S_\LR}^\infty)  + \gamma_\LR^* \dd t_\LR \wedge {\rm Im} (\Omega_{S_\LR}^\infty) \ .
\end{multline}
Following Kovalev \cite{MR2024648}, let us now assume that the asymptotically cylindrical Calabi--Yau threefolds $X_\LR$ are chosen such that the resulting asymptotic polarized K3~surfaces $S_\LR$ are mutually isometric with respect to a hyper~K\"ahler rotation $r: S_L \to S_R$\label{gl:r} obeying\footnote{These conditions impose rather non-trivial constraints on the pair of asymptotically cylindrical Calabi--Yau threefolds $X_\LR$, which --- at least for certain classes of pairs --- have been studied systematically in ref.~\cite{MR3369307}. In this section we assume that these conditions on $X_\LR$ are met, and we come back to this issue in Section~\ref{sec:examples}, where we explicitly construct asymptotically cylindrical Calabi--Yau threefold $X_\LR$ fulfilling these constraints.}
\begin{equation} \label{eq:HyperKaehler}
   r^* \omega_R^I=\omega_L^J\ ,\quad r^* \omega_R^J=\omega_L^I \ ,\quad  r^* \omega_R^K=-\omega_L^K \ .     
\end{equation}
Then there is a family of diffeomorphisms $F_\Lambda: Y^\infty_L \to Y^\infty_R$ with constant $\Lambda \in \mathbb{R}$ given by
\begin{equation} \label{eq:gluingdiff} 
  F_\Lambda:\ (\theta_L^*,t_L,u_L^\alpha,\theta_L) \mapsto 
  (\theta_R^*,t_R,u_R^\alpha,\theta_R) = (\theta_L,\Lambda-t_L,r(u^\alpha_L),\theta_L^*) \ ,
\end{equation}
in terms of the local coordinates $(\theta^*_\LR,t_\LR)$ of $\Delta^{\rm cyl}_\LR$, $u^\alpha_\LR$ of $S_\LR$, and $\theta_\LR$ of $S^1_\LR$. Now it is straightforward to check that if and only if the radii are equal 
\begin{equation} \label{eq:radii}
  \gamma := \gamma_L = \gamma_R = \gamma^*_L = \gamma^*_R \ ,
\end{equation}  
this asymptotic diffeomorphism is also an asymptotic isometry because it leaves the asymptotic $G_2$-structures $\varphi_{0\,L/R}^\infty$ --- and hence the asymptotic metric --- invariant, i.e., 
\begin{equation} \label{eq:Fsym}
  F^*_\Lambda \varphi_{0\,R} \,=\, \varphi_{0\,L} \ .
\end{equation}

\begin{figure}[t]
\center
\includegraphics[width=13cm]{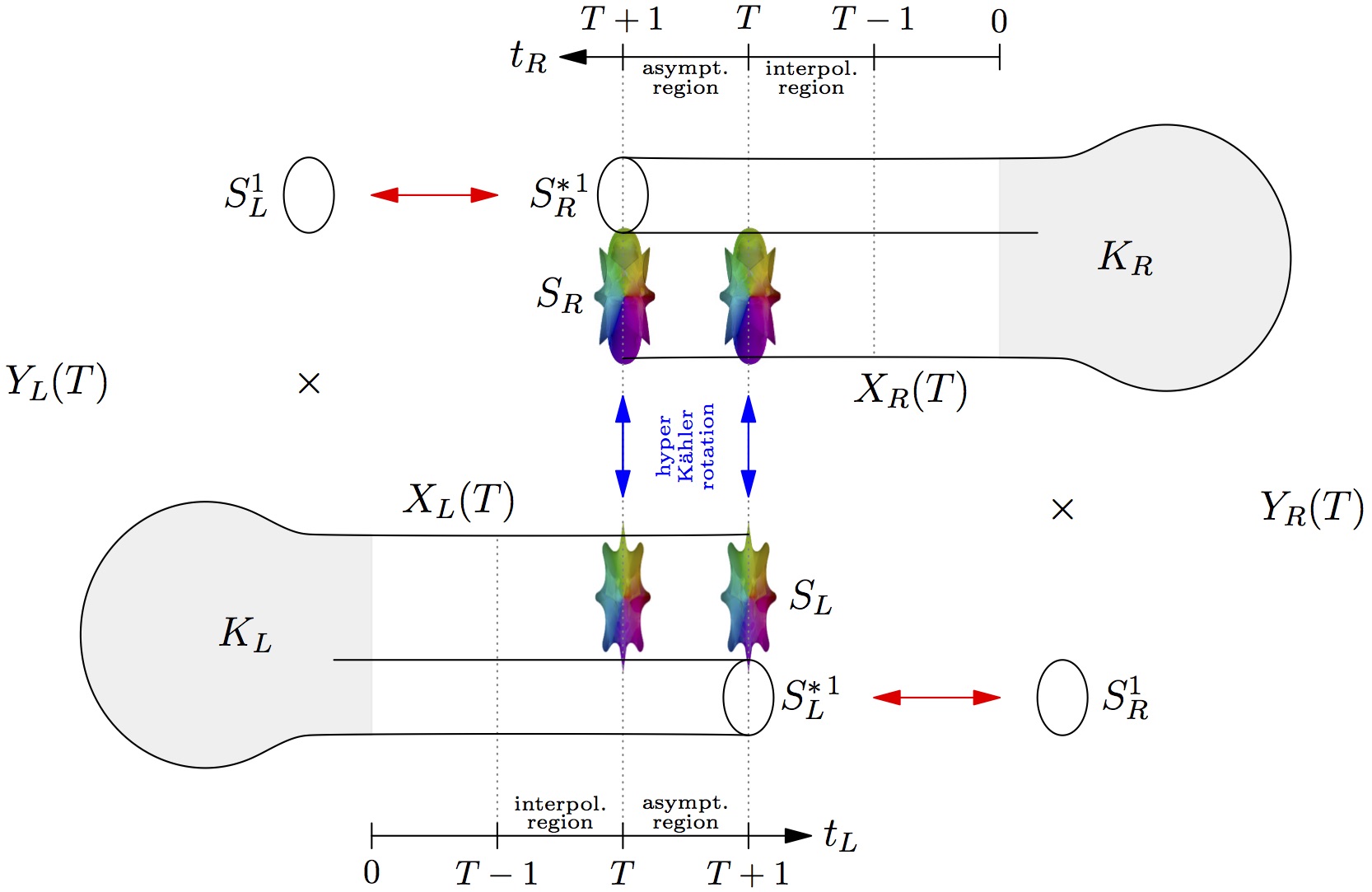}
\caption{This picture illustrates Kovalev's twisted connected sum. $X_\LR(T)$ are the truncated asymptotically cylindrical Calabi--Yau threefolds together with their compact subspaces $K_\LR$. Their Cartesian products with the circles $S^1_\LR$ yield the two seven-dimensional components $Y_\LR(T)$ combined to form the $G_2$-manifold $Y$. There are two essential aspects in the gluing procedure. Firstly --- as indicated by the red horizontal arrows --- the circles $S_\LR^1$ are identified with the asymptotic circles of $X_\LR(T)$ here denoted by $S_{R/L}^{*\,1}$. Secondly --- as depicted by the blue vertical arrows --- the asymptotic polarized K3 surfaces $S_\LR$ must be matched with a certain hyper~K\"ahler rotation. Finally, the diagram highlights the interpolating regions, $t_\LR \in (T-1,T]$, and the asymptotic gluing regions, $t_\LR \in (T,T+1)$, important for the construction of the $G_2$-structure $\varphi(\gamma,T)$ of $Y$.} \label{fig:twistedsum}
\end{figure}

As schematically depicted in Figure~\ref{fig:twistedsum}, the $G_2$-manifold $Y$ is now obtained by gluing the asymptotic ends of $Y_\LR$ with the help of the diffeomorphism $F_\Lambda$. Let $X_\LR(T)$ and $Y_\LR(T)$ be the truncated asymptotically cylindrical Calabi--Yau threefolds and the truncated seven-manifolds given by cutting off their asymptotic regions at $t_\LR=T+1$ for some (large) $T$, i.e., 
\begin{equation} \label{eq:seven}
  X_\LR(T) \,=\, K_\LR \cup \eta_\LR(\mathbb{R}_{<T+1}) \ , \quad Y_\LR(T) \,=\, X_\LR(T) \times S^1_\LR  \ .
\end{equation}
Here the diffeomorphisms $\eta_\LR$ and the compact subspaces $K_\LR$ are defined in Section~\ref{sec:ccy}. Then, using the (restricted) diffeomorphism $F_{2T+1}$ --- which maps the coordinate $t_L \in (T,T+1)$ to $t_R = -t_L + 2T+1 \in (T,T+1)$ --- we glue the two seven-manifolds $Y_\LR(T)$ at the overlap $t_\LR \in (T,T+1)$ to arrive at the compact seven-manifold
\begin{equation} \label{eq:Yunion}
  Y \,=\, Y_L(T) \cup_{F_{2T+1}} Y_R(T)  \ .
\end{equation}
Finally, to construct a $G_2$-structure $\varphi$ on $Y$, we first introduce interpolating $G_2$-structures on the two pieces $Y_\LR(T)$. Let $\alpha: \mathbb{R} \to [0,1]$ be a smooth function interpolating between $0$ and $1$ within the interval $(-1,0)$, namely $\alpha(s)=0$ for $s\le-1$ and $\alpha(s)=1$ for $s\ge0$. Then we endow the truncated asymptotically cylindrical Calabi--Yau threefolds $X_\LR$ with the forms \cite{MR2024648,MR3109862}
\begin{equation}
\begin{aligned}
   \widetilde{\omega}^T_\LR \,&=\, \omega_\LR -d\left(\alpha(t-T) \mu_\LR \right)  \ , \\
   \widetilde{\Omega}^T_\LR \,&=\, \Omega_\LR - d \left(\alpha(t-T) \nu_\LR \right)  \ ,
\end{aligned}
\end{equation}
in terms of the forms $\mu_\LR$ and $\nu_\LR$ of eqs.~\eqref{eq:limit}. By construction the forms $\widetilde{\omega}^T_\LR$ and $\widetilde{\Omega}^T_\LR$ smoothly interpolate between the corresponding Calabi--Yau cylinder forms~\eqref{eq:ccyforms} and the asymptotically cylindrical Calabi--Yau forms~\eqref{eq:limit}. At the interpolating regions $t_\LR \in (T-1,T)$ the symplectic forms $\omega^{T}_\LR$ fail to induce a Ricci-flat metric  and the three-forms $\Omega^T_\LR$ cease to be holomorphic. Analogously to eq.~\eqref{eq:G2prodform}, the interpolating $G_2$-structures $\widetilde{\varphi}_\LR(\gamma,T)$ on $Y_\LR$ read
\begin{equation} \label{eq:phitilde}
    \widetilde{\varphi}_\LR(\gamma,T) 
    \,=\, \gamma\, \dd \theta \wedge \widetilde{\omega}^T_\LR + {\rm Re} (\widetilde{\Omega}^T_\LR) \ ,
\end{equation}
which according to eq.~\eqref{eq:Fsym} glue together to a well-defined $G_2$-structure $\widetilde{\varphi}(\gamma,T)$ on the seven-manifold $Y$.

Note that the constructed $G_2$-structure $\widetilde{\varphi}(\gamma,T)$ is closed but not torsion-free. The torsion of $\varphi(\gamma,T)$ is measured by $\dd\!*\!\widetilde{\varphi}(\gamma,T)$. It is only non-vanishing at the interpolating regions $t_\LR\in(T-1,T)$, where it is of order $O(e^{-\gamma\lambda T})$ due to eq.~\eqref{eq:limit} \cite{MR2024648}.\footnote{Note that --- due to relation~\eqref{eq:radii} and the hyper~K\"ahler isometry~\eqref{eq:Fsym} --- we have the identifications $\lambda=\lambda_L=\lambda_R$ among inverse length scales.} Hence, it is plausible that we can view $\widetilde{\varphi}(\gamma,T)$ as an order $O(e^{-\gamma \lambda T})$ approximation to a torsion-free $G_2$-structure $\varphi(\gamma,T)$, which equips the seven-manifold $Y$ with a Ricci-flat metric. Indeed, Kovalev shows that, for sufficiently large $T$, there exists in the same three-form cohomology class of $\widetilde{\varphi}(\gamma,T)$ a torsion-free $G_2$-structure $\varphi(\gamma,T)$ such that, for any positive integer $k$, \cite{MR2024648}
\begin{equation} \label{eq:PhiKov}
  \varphi(\gamma,T) = \widetilde{\varphi}(\gamma,T) + \dd \rho(\gamma,T) \quad
  \text{with} \quad \left|\nabla^k\rho(\gamma,T)\right| \,=\, O(e^{-\gamma \lambda T} ) \ ,
\end{equation}
in terms of the norm $|\,\cdot\,|$ and the Levi--Civita connection $\nabla$ of the metric induced from the asymptotic $G_2$-structure \eqref{eq:Fsym}.

Finally, the relationship $\pi_1(Y) = \pi_1(X_L)\times \pi_1(X_R)$ among the fundamental groups in the twisted connected sum implies that  the torsion-free $G_2$-structure $\varphi(\gamma,T)$ indeed gives rise to a genuine $G_2$-manifold $Y$ with $G_2$ holonomy \cite{MR2024648}. 

This summarizes the main result of ref.~\cite{MR2024648} --- clarified and further developed in refs.~\cite{MR3369307,MR3399097,MR3109862} --- namely 
the analysis proof that the $G_2$-structure $\widetilde{\varphi}(\gamma,T)$ in Kovalev's twisted connected sum construction furnishes an approximation to the 
torsion-free $G_2$-structure $\varphi(\gamma,T)$ in the same three-form cohomology class, which gives rise to a compact seven-dimensional Ricci-flat Riemannian manifold~$Y$ with $G_2$ 
holonomy.

\subsection{The Kovalev limit of $G_2$-manifolds}  
\label{sec:Kovalevlimit}
As discussed in the last section, the torsion-free $G_2$-structure $\varphi(\gamma,T)$ in Kovalev's twisted connected sum is approximated via eq.~\eqref{eq:G2prodform} in terms of the holomorphic forms~$\Omega_\LR$ and the K\"ahler forms~$\omega_\LR$ of the asymptotically cylindrical Calabi--Yau threefolds~$X_\LR$. According to eqs.~\eqref{eq:limit} and \eqref{eq:PhiKov} this approximation is of order $O(e^{-\gamma\lambda T})$. 

We now discuss the torsion-free $G_2$-structure $\varphi(\gamma,T)$ as a function of the parameters $\gamma$ and $T$. Except for the overall volume modulus, we keep all other moduli of the asymptotically cylindrical Calabi--Yau threefolds $X_\LR$ fixed. That is to say, we consider the moduli dependence of the two metrics $g_\LR$ of $X_\LR$ as 
\begin{equation} \label{eq:gLR}
  g_\LR(z_\LR,t_\LR) \,=\, \gamma_0^2\,R^2\,\tilde g(z_\LR,\tilde t_\LR) \ ,
\end{equation}
where $z_\LR$ and $t_\LR$ are the (dimensionless) complex structure moduli and the K\"ahler moduli of $X_\LR$, respectively. The constant $\gamma_0$ has dimension of length such that the metrics $\tilde g_\LR$ become dimensionless. We split the K\"ahler moduli $t_\LR$ further into the overall volume modulus $R$ and the remaining K\"ahler moduli $\tilde t_\LR$ --- measuring ratios of volumes of subvarieties in $X_\LR$ --- such that\footnote{For an asymptotically cylindrical Calabi--Yau threefold with a single K\"ahler modulus $t$ the volume modulus $R$ relates to this K\"ahler modulus as $R^2=t$ without the presence of any further moduli $\hat t$.}
\begin{equation}
   t_\LR^{a} \,=\, \begin{cases} R^2 & a = 1 \\ R^2\,\tilde t_\LR^a & a\ne 1 \end{cases} \ .
\end{equation}
In order to obtain the $G_2$-manifold $Y$ from the seven-dimensional building blocks~$Y_\LR$, the hyper~K\"ahler compatibility condition \eqref{eq:HyperKaehler} constrains the explicit values of the moduli $z_\LR$ and $\tilde t_\LR$. Furthermore, the required identification~\eqref{eq:radii} of the radii of all circles in the asymptotic region of $Y_\LR$  determines the volume modulus $R$ as the dimensionless ratio
\begin{equation} \label{eq:defR}
   R \,=\, \frac{\gamma}{\gamma_0} \ ,
\end{equation}
and it justifies to introduce a mutual volume modulus $R$ for both threefolds $X_\LR$.  

In the Kovalev limit of large $RT$, the volume $V_Y$ of the constructed $G_2$-manifold $Y$ becomes
\begin{multline}
   V_Y(\tilde S,R,T) \,=\, 
      V_{Y_L(T)}(z_L,t_L) + V_{Y_R(T)}(z_R,t_R)  \\
    - (2\pi)^2 \gamma_0^3 R^3 V_{S}(\tilde\rho_S,R) +  O(e^{-\tilde\lambda R T}) \ .
\end{multline}
The resulting volume depends on the moduli $R$ and $T$ and the remaining moduli of the $G_2$-manifold $Y$ --- collective denoted by $\tilde S$ singling out those moduli fields $\tilde\rho_S$ deforming the K3 surface~$S$. The volumes $V_{Y_\LR(T)}$ of the truncated building blocks $Y_\LR(T)$ are calculated with the metrics of the (truncated) asymptotically cylindrical Calabi--Yau threefolds $X_\LR(T)$ using the expressions \eqref{eq:ccyforms} and the volume formula~\eqref{eq:Vol} (with the dimensionful reference volume~$V_{Y_0} = \gamma_0^7$\label{gl:refVol}). As the sum of the first two terms counts the volume of the overlapping region twice, we need to subtract this contribution again once. It is given by the product of the volumes of the overlapping interval, the asymptotic two-torus $S_L^1 \times S_{L}^{*\,1}\equiv S_R^1 \times S_{R}^{*\,1}$, and the asymptotic K3-surface $S_L\equiv S_R \equiv S$\label{gl:Sii} according to
\begin{equation} \label{eq:VolY}
   V_{Y_L(T) \cap Y_R(T)} \,=\,  (2\pi)^2 \gamma_0^3 R^3 V_{S}(\tilde\rho_S,R) 
   \,=\, (2\pi)^2 \gamma_0^7 R^7 V_{S}^{\tilde g}(\tilde\rho_S) \ .
\end{equation} 
In the last equality, the volume of the K3-surface $S$ is expressed in terms of the dimensionless (asymptotic) metric $\tilde g$. Due to approximation \eqref{eq:limit} of the metrics of the building blocks $Y_\LR(T)$ by the limiting metrics of $Y_\LR^\infty(T)$ in the interpolation region $t_\LR \in (T-1,T)$ and due to the overall correction~\eqref{eq:PhiKov} to the torsion-free $G_2$-structure $\varphi(\gamma,T)$, the computed volume of the $G_2$-manifold $Y$ receives exponentially suppressed corrections in $\tilde\lambda RT$ for large $RT$, where --- because of eq.~\eqref{eq:Clambda} --- the dimensionless constant $\tilde\lambda$ reads $\tilde\lambda=\lambda\gamma_0$\label{gl:invlength}. 

Note that --- up to exponentially correction terms suppressed for large $RT$ --- the volume~\eqref{eq:VolY} is entirely determined by the (relative) periods and the K\"ahler forms of the asympotically cylindrical Calabi--Yau threefolds $X_\LR$. However, due to non-compactness of $X_\LR$, the relative periods and the K\"ahler volumes of non-compact cycles diverge linearly in $T$. Therefore, in order to obtain the finite periods and finite volumes of the truncated asymptotically cylindrical Calabi-Yau threefolds $X_{L/R}(T)$, a suitable regularization scheme must be employed to extract the required geometric data from the diverging periods and infinite volumes. This analysis is beyond the scope of this work, but we plan to get back to this issue elsewhere.

Thus, instead of deriving the entire moduli dependence of the volume of the $G_2$-manifold~$Y$, we focus on the moduli dependence of the two fields $R$ and $T$ --- viewing the remaining moduli fields $\tilde S$\label{gl:Stilde} as parameters. First, we compute the volumes $V_{Y_\LR(T)}$ in eq.~\eqref{eq:VolY} as
\begin{equation}
\begin{aligned}
  V_{Y_\LR(T)} \,&=\, \frac{\gamma_0^7}7 \int_{Y_\LR(T)} \varphi_\LR \wedge *\varphi_\LR \\
  \,&=\, 
   \frac{\gamma_0^7}7 \int_{Y_\LR(T) \setminus K_\LR} \varphi_\LR^\infty \wedge *\varphi_\LR^\infty
   + V_{K_\LR} +  (2\pi)^2 \gamma_0^7 R^7 \Delta_\LR(\tilde S,T) \\
  \,&=\, (2\pi)^2 \gamma_0^7 R^7 \left[ V_{S}^{\tilde g}(\tilde\rho_S) \left( (T+1) + D_\LR(\tilde S) \right)
  + \Delta_\LR(\tilde S,T) \right] \ .
\end{aligned}
\end{equation}
Here we split the integration by performing the integral over the compact parts $K_\LR$ and the asymptotic regions $Y_\LR(T) \setminus K_\LR$. The former part factors into the volume of the K3 surface $S$ and a contribution $D_\LR(\tilde S)$ to be discussed in greater detail momentarily. The latter part is evaluated with respect to the asymptotic $G_2$-structure $\varphi_\LR^\infty$, which introduces the correction term $\Delta_\LR(\tilde S,T)$ such that 
\begin{equation}
 \Delta_\LR(\tilde S,T) \,=\, R \int_0^{T+1} dt \, f_\LR(\tilde S,Rt) e^{-\tilde\lambda R t} \,=\, C_\LR(\tilde S) + O(e^{-\tilde\lambda R T}) \ ,
\end{equation}  
in terms of the function $f_\LR(\tilde S,Rt)$ determined by eq.~\eqref{eq:limit}. Thus, taking the correction terms $\Delta_\LR$ into account, we arrive at
\begin{equation} \label{eq:VolYII}
  V_{Y}(\tilde S,R,T) \,=\,  (2\pi)^2 \gamma_0^7 R^7 V_{S}^{\tilde g}(\tilde\rho_S)
  \,\left(2 T +\alpha(\tilde S)\right) +  O(e^{-\tilde\lambda R T})  \ ,
\end{equation}
with 
\begin{equation}
    \alpha(\tilde S) = \left( D_L(\tilde S) + C_L(\tilde S)\right) +\left( D_R(\tilde S)+ C_R(\tilde S)\right) + 1 \ .
\end{equation}  
As discussed, the moduli-dependent contributions $D_\LR(\tilde S)+C_\LR(\tilde S)$ --- and hence the moduli dependent function $\alpha(\tilde S)$ --- are in principle computable from the (relative) periods and the K\"ahler forms of the asymptotically cylindrical Calabi--Yau threefolds~$X_\LR$. 

In the context of M-theory compactification on the $G_2$-manifold~$Y$, the volume $V_Y$ determines the four-dimensional Planck constant $\kappa_{4}$ according to eq.~\eqref{eq:kappa4}. We refer to the Kovalev limit as the approximation in which the four-dimensional Planck constant $\kappa_4$ --- and hence the volume $V_Y$ --- remains constant, while the exponential correction terms become sufficiently small. Namely, for fixed moduli $\tilde S$ we require that the dimensionless quantity $\chi$, that is given by
\begin{equation}
    \chi^7
    \, = \,  R^7 (2 T +\alpha)  \ ,
\end{equation}
remains constant. Requiring a constant four-dimensional Planck constant yields the functional dependence for the modulus $R$
\begin{equation} \label{eq:R(T)}
   R(T) \,=\, \frac{\chi}{\sqrt[7]{2T+\alpha}} \ ,
\end{equation}
such that corrections terms scale as
\begin{equation}
  O(e^{-\tilde\lambda R T}) \,=\, O(e^{- \frac{\tilde\lambda\chi T}{\sqrt[7]{2T+\alpha}}} ) \ .
\end{equation}
Thus, for large $T$, with $R=R(T)$ --- as in eq.~\eqref{eq:R(T)} --- the corrections for the volume $V_Y$ in eq.~\eqref{eq:VolYII} are exponentially suppressed. Furthermore, for $T\gg \tilde\lambda\chi$ --- loosely referred to as the Kovalev limit --- the torsion-free $G_2$-structure of $Y$ is well approximated in terms of the geometric data of the asymptotically cylindrical Calabi--Yau threefolds $X_\LR$ for a given four-dimensional Planck constant $\kappa_4$. However, taking literally the limit $T\to \infty$ with $R=R(T)$ does not yield a limiting Riemannian manifold but instead yields only a Hausdorff limit in the sense of Gromov--Hausdorff convergence of compact metric spaces. In the context of M-theory, the Kovalev limit implies that --- while for large $T$ with $R=R(T)$ the Ricci-flat $G_2$-metric gets more and more accurately approximated in terms of the Ricci-flat Calabi--Yau metrics of $X_\LR$ --- the discussed semi-classical dimensional reduction on the $G_2$-manifold $Y$ in terms of the Kaluza--Klein zero modes becomes less accurate due to the emergence of both light Kaluza--Klein modes and substantial non-perturbative membrane and M5-brane instanton corrections.

\section{M-theory on twisted connected sums} \label{sec:M-theoryonTCS} 
In this section we analyze the four-dimensional low-energy effective action of M-theory compactifications on $G_2$-manifolds that are of the twisted connected sum type. We focus on the Kovalev limit, in which the asymptotically cylindrical Calabi--Yau metrics of the summands furnish a good approximation to the $G_2$-metric. Our first task is to analyze the four-dimensional $\mathcal{N}=1$ spectrum of such compactifications, which --- according to Table~\ref{tab:N1multi} --- amounts to expressing the de~Rham cohomology groups of the resulting $G_2$-manifolds in terms of the cohomology groups of the Calabi--Yau summands. In particular, we explicitly identify the chiral modulus governing the Kovalev limit, referred to as the Kovalevton $\kov$, and we discuss the effective action and its physical properties in this limit.

\subsection{Spectrum and cohomology} \label{sec:spectrumandcohomology}
To deduce the four-dimensional $\mathcal{N}=1$ supersymmetric spectrum of M-theory compactified on a $G_2$-manifold of the twisted connected sum type, we should analyze the de Rham cohomology of $Y$ as arising from the cohomology of the asymptotically cylindrical Calabi--Yau summands. A partial answer to this question has already been given in Kovalev's paper \cite{MR2024648}. In ref.~\cite{MR3369307} Corti et al. have presented a systematic analysis of the cohomology of $Y$, which we summarize and use here.

Following refs.~\cite{MR3109862,MR3369307}, let us first introduce the notion of a building block $(Z,S)$\label{gl:bblock}, which allows us to construct an asymptotically cylindrical Calabi--Yau threefold $X$. In this work a building block $(Z,S)$ is a pair consisting of a smooth K3 fibration $\pi: Z \rightarrow \mathbb{P}^1$ together with a smooth K3 fiber $S=\pi^{-1}(p)$ for some $p\in\mathbb{P}^1$. We require that the anti-canonical class $-K_Z$ is primitive and that $S$ is linearly equivalent to the anti-canonical class, i.e., $S \sim -K_Z$. Due to the fibrational structure, the self-intersection of $S$ is trivial, which implies that the manifold $X = Z \setminus S$ has the topology of an asymptotically cylindrical Calabi--Yau threefold that admits a Ricci-flat K\"ahler metric \cite{MR2024648,MR3109862,MR3369307}.\footnote{This can readily be seen as follows: The trivial normal bundle of $S$ in $X$ defines a tubular neighborhood $T_\epsilon(S)\subset X$. By construction $T^*_\epsilon(S):=T_\epsilon(S) \setminus S$ is homeomorphic to $\Delta^\text{cyl} \times S$, which has the topology of a three-dimensional Calabi--Yau cylinder $X^\infty$ to be viewed as the asymptotic region of the asymptotically cylindrical Calabi--Yau threefold $X$, as discussed in Section~\ref{sec:ccy}.}  As in refs.~\cite{MR3109862,MR3369307}, we further impose two technical assumptions on the building block~$(Z,S)$. Namely, we demand that $H^3(Z,\mathbb{Z})$ is torsion-free and that the integral two-form cohomology $H^2(X,\mathbb{Z})$ embeds primitively into the K3 lattice~$L=H^2(S,\mathbb{Z})$ via the pullback map of the inclusion $\rho: S \hookrightarrow X$ (which is well-defined up to homotopy). 

To construct a $G_2$-manifold $Y$ we now consider a pair of building blocks $(Z_\LR,S_\LR)$ such that the polarized K3 surfaces $S_\LR$ are isometric and fulfill the hyper K\"ahler matching condition $r: S_L \to S_R$. From the asymptotically cylindrical Calabi--Yau threefolds $X_\LR = Z_\LR \setminus S_\LR$ we can then construct the $G_2$-manifold $Y$ with Kovalev's twisted connected sum construction as detailed in Section~\ref{sec:generalideaofkovalev}. Under the assumptions in the definition of the building blocks $(Z_\LR,S_\LR)$, Corti et al. derive the cohomology of the $G_2$-manifold $Y$ from the building blocks \cite{MR3369307}
\begin{equation} \label{eq:cohTwistedSum}
\begin{aligned}
 &\pi_1(Y) \, = \, H^1(Y,\mathbb{Z}) = 0 \ , \\
 &H^2(Y,\mathbb{Z}) \,\simeq\, \left( k_L \oplus k_R\right) \oplus \left( N_L \cap N_R \right)\ ,\\
 &H^3(Y,\mathbb{Z}) \,\simeq\, H^3(Z_L,\mathbb{Z}) \oplus H^3(Z_R,\mathbb{Z}) 
    \oplus k_L \oplus k_R \oplus N_L \cap T_R\, \oplus \, N_R \cap T_L \\
 &\qquad\qquad\qquad\qquad    \oplus \mathbb Z[S] \oplus L/\left(N_L+ N_R\right)\ .
\end{aligned}
\end{equation}
Here $[S]$ is the Poincar\'e dual three-form of a K3 fiber $S$ in the building blocks $(Z_\LR,S_\LR)$, and $L$ denotes the K3 lattice $L\simeq H^2(S_L,\mathbb{Z})\simeq H^2(S_R,\mathbb{Z})$. Furthermore, the inclusion maps $\rho_\LR: S_\LR \hookrightarrow X_\LR$ induce the maps $\rho^*_\LR: H^2(X_\LR,\mathbb{Z}) \to L$, which define the kernels ${k}_\LR := \operatorname{ker} \rho_\LR^*$, the images $N_\LR:=\operatorname{Im} \rho_\LR^*$, and the transcendental lattices
\begin{equation} \label{eq:Nperp}
  T_\LR =N_\LR^{\bot} = \left\{l\in L\ \middle\vert\ \langle l, N_\LR \rangle = 0 \right\} \ .
\end{equation}
Note that --- by the assumptions imposed on the cohomological properties of the buildings blocks --- the images $N_\LR$ are primitive sublattices of the K3 lattice $L$. We further assume that the sum $N_L + N_R$ embeds primitively into the K3 lattice $L$. As a consequence the quotient $L / (N_L + N_R)$ is torsion-free, and --- due to the assumed torsion-freeness of $H^3(Z_\LR,\mathbb{Z})$ --- all the integral cohomology groups~\eqref{eq:cohTwistedSum} are torsion-free as well.

Thus, the properties of the building blocks determine the cohomology~\eqref{eq:cohTwistedSum} and therefore the effective four-dimensional $\mathcal{N}=1$ supersymmetric spectrum of the M-theory compactification according to Table~\ref{tab:N1multi}. Let us now in particular focus on the neutral chiral moduli multiplets $\Phi^i$ associated to the three-form cohomology $H^3(Y,\mathbb{Z})$. The derivation of the cohomology in ref.~\cite{MR3369307} is essentially based upon the Mayer--Vietoris sequence applied to the union \eqref{eq:Yunion} of the overlapping non-compact seven-manifolds $Y_\LR$ defined in eq.~\eqref{eq:seven} in terms of the asymptotically cylindrical Calabi--Yau threefolds $X_\LR$ such that
\begin{equation}
\begin{aligned}
    H^3(Y,\mathbb{Z}) \,&=\, 
    \operatorname{ker}\left( H^3(Y_L,\mathbb{Z}) \oplus H^3(Y_R,\mathbb{Z})
    \xrightarrow{(\iota_L^*,-\iota_R^*)}
       H^3(T^2 \times S,\mathbb{Z}) \right) \\ 
      &\qquad\qquad \oplus
      \operatorname{coker}\left( H^2(Y_L,\mathbb{Z}) \oplus H^2(Y_R,\mathbb{Z})
      \xrightarrow{(\iota_L^*,-\iota_R^*)} 
       H^2(T^2 \times S,\mathbb{Z})\right) \ ,
\end{aligned}       
\end{equation}
where $Y_R \cap Y_L$ deformation retracts to $T^2 \times S$ and the maps are induced from the standard inclusion maps $\iota_\LR: T^2 \times S\hookrightarrow Y_\LR$. The three-form cohomolology in \eqref{eq:cohTwistedSum} is distributed among these two summands in the following way \cite{MR3369307}
\begin{equation} \label{eq:split1}
\begin{aligned}
  \operatorname{ker}&\left( H^3(Y_L,\mathbb{Z}) \oplus H^3(Y_R,\mathbb{Z})
   \to H^3(T^2 \times S,\mathbb{Z}) \right) \\ 
    &\qquad\quad \,=\, H^3(Z_L,\mathbb{Z}) \oplus H^3(Z_R,\mathbb{Z}) 
    \oplus k_L \oplus k_R \oplus N_L \cap T_R\, \oplus \, N_R \cap T_L \ , \\[1ex]
    \operatorname{coker}&\left( H^2(Y_L,\mathbb{Z}) \oplus H^2(Y_R,\mathbb{Z})
      \to 
       H^2(T^2 \times S,\mathbb{Z})\right) \,=\, 
       \mathbb Z[S] \oplus L/\left(N_L+ N_R\right) \ .
\end{aligned}
\end{equation}
Expanding the torsion-free $G_2$-structure $\varphi$ of $Y$ in terms of the above assembled three-form cohomology basis as in eq.~\eqref{eq:cohexpand}, the coefficients of the individual cohomology elements capture (in the Kovalev limit) particular geometric moduli of the twisted connected sum and their summands. 

First of all, the kernel contributions in \eqref{eq:split1} describe the moduli of the asymptotically cylindrical Calabi--Yau manifolds $X_\LR$. In particular, the coefficients of $H^3(Z_\LR)$ and $k_\LR$ realize the complex structure moduli and the K\"ahler moduli of the asymptotically cylindrical Calabi--Yau manifolds $X_\LR$, respectively. Furthermore, $N_L \cap T_R$ captures mutual K\"ahler moduli of $X_L$ and complex structure moduli of $X_R$, which are interlinked in this way due to the non-trivial gluing with the hyper K\"ahler rotation~\eqref{eq:HyperKaehler} --- exchanging $X_L$ and $X_R$. The intersection $N_R \cap T_L$ enjoys an analog interpretation. 

Second of all, we analyze the cokernel contribution in eq.~\eqref{eq:split1}. To get a better geometric picture for these moduli, we first observe that --- due to the K3~fibrations $Z_\LR \to \mathbb{P}^1$ --- the Calabi--Yau threefolds $X_\LR$ are K3~fibrations over a disk $D_\LR$. As a result $Y_\LR$ become $K3$~fibrations over solid tori $T_\LR \equiv S^1_\LR \times D_\LR$, namely
\begin{equation} \label{eq:K3fibLR}
\begin{CD}
  S_\LR @>>> Y_\LR  \\
  @. @VV\pi V \\
  @. \ T_\LR\ .
\end{CD} 
\end{equation}
The gluing diffeomorphism~\eqref{eq:gluingdiff} in the twisted connected sum identifies the boundary of the disk $D_L$ with the circle $S^1_R$ and the circle $S^1_L$ with the boundary of the disk $D_R$ such that the two solid tori $T_\LR$ are glued together to a three-sphere $S^3$. Thus, the resulting $G_2$-manifold~$Y$ is a topological K3 fibration\footnote{The topological K3 fibration in the context of twisted connected sums has also been discussed in ref.~\cite{Braun:2016igl}.}
\begin{equation} \label{eq:K3fib}
\begin{CD}
  S @>>> Y  \\
  @. @VV\pi V \\
  @. \ S^3 \ .
\end{CD} 
\end{equation}
The cohomology three-forms of the cokernel~\eqref{eq:split1} describe moduli of the asymptotic boundary of $\partial Y_L \simeq \partial Y_R \simeq T^2 \times S$. Their dual homology three-cycles restrict to relative three-cycles in the summands $Y_L$ and $Y_R$, and hence the associated moduli are sensitive to the overlapping gluing regions $Y_\LR(T)\setminus K_\LR$, cf. Figure~\ref{fig:twistedsum}. In particular, the three-form generator~$[S]$ is Poincar\'e dual to a K3 fiber $S$, and hence its dual homology three-cycle is the $S^3$ base of the fibration~\eqref{eq:K3fib}. As a consequence the modulus associated to $[S]$ measures the volume of the $S^3$ base. Similarly, the remaining cokernel moduli measure volumes of three-cycles that project under the map $\pi: Y \to S^3$ to paths in the $S^3$ base connecting the disjoint compact subsets  $\pi(K_L)$ and $\pi(K_R)$ of $S^3$. Note that $2T+1$ is the distance between these two compact subsets in terms of the parameter $T$ introduced in Section~\ref{sec:Kovalevlimit}. Therefore, it now follows that --- in the Kovalev limit --- all cokernel moduli depend linearly on the parameter $T$, and geometrically $T$ enjoys the interpretation of a squashing parameter for the $S^3$ base of the K3 fibration~\eqref{eq:K3fib}. 

The split into two types of moduli fields in eq.~\eqref{eq:split1} motivates us to introduce two universal geometric moduli $v$ and $b$. For any $G_2$-manifold there is the universal volume modulus $v$ that is associated to the singlet $H^3_{\mathbf{1}}(Y,\mathbb{Z})$ of the three-form cohomology. It simply rescales the torsion-free $G_2$-structure $\varphi$. In the twisted connected sum we additionally identify the squashing modulus $b$ of the $S^3$ base in the fibration~\eqref{eq:K3fib}. Note that $b \to +\infty$ describes the Kovalev limit discussed in Section~\ref{sec:Kovalevlimit}. According to eq.~\eqref{eq:cohexpand}, the torsion-free $G_2$-structure $\varphi$ depends on these two moduli as
\begin{equation} \label{eq:KovG2}
  \varphi(v,b,\tilde S) \,=\, v \left[ \left(\rho^\text{ker}_0  + \sum_{\hat\imath} \tilde S^{\hat\imath} \rho^\text{ker}_{\hat\imath} \right)
  + b \left( [S] + \sum_{\tilde\imath} \tilde S^{\tilde\imath} \rho^\text{coker}_{\tilde\imath} \right) \right] \ .
\end{equation}
Here $[S]$ is the harmonic three-form that is Poincar\'e dual to the K3 fiber $S$. Furthermore, $(\rho^\text{ker}_0,\rho^\text{ker}_{\hat\imath})$ and $\rho^\text{coker}_{\tilde\imath}$ form a basis of harmonic three-forms arising from the kernel contributions and the cokernel part $L/\left(N_L+ N_R\right)$ in \eqref{eq:split1}, respectively. $\tilde S^{\hat\imath}$ and $\tilde S^{\tilde\imath}$ are the respective associated geometric real moduli fields.\label{gl:Stildeii}\footnote{Note that the kernel contribution~\eqref{eq:split1} is at least one-dimensional, such that we can always choose a basis element $\rho_0^{ker}$ \cite{MR3369307}.}

Thus, the description of the torsion-free $G_2$-structure $\varphi(v,b,\tilde S)$ gives rise to two universal $\mathcal{N}=1$ neutral chiral moduli multiplets $\nu$ and $\kov$ given by
\begin{equation} \label{eq:nukov}
  \operatorname{Re}(\nu) \,=\, v \ , \qquad \operatorname{Re}(\kov) \,=\, v b \ .
\end{equation}
In particular, we refer to the chiral multiplet $\kov$ as the Kovalevton since it describes in the limit $\operatorname{Re}(\kov) \to +\infty$ --- while keeping $\operatorname{Re}(\nu)$ constant ---  the Kovalev limit discussed in Section~\ref{sec:Kovalevlimit}.

The remaining real moduli fields are not universal and relate to the non-universal neutral chiral multiplets as
\begin{equation} \label{eq:chiralvar}
   \operatorname{Re}(\phi^{\hat\imath}) \,=\, v \tilde S^{\hat\imath} \ , \qquad
   \operatorname{Re}(\phi^{\tilde\imath}) \,=\, v b \tilde S^{\tilde\imath} \ .
\end{equation}
They depend on the topological details of the building blocks $(Z_\LR,S_\LR)$ and the choice of gluing diffeomorphism \eqref{eq:gluingdiff}.

Finally, the two-form cohomology $H^2(Y,\mathbb{Z})$ for (smooth) $G_2$-manifolds yields four-dimensional massless abelian $\mathcal{N}=1$ vector multiplets, cf. Table~\ref{tab:N1multi}. In Kovalev's twisted connected sum we get two types of $\mathcal{N}=1$ vector multiplets according to eq.~\eqref{eq:cohTwistedSum}, which we discuss in the sequel.

Firstly, the kernel contributions $k_L$ and $k_R$ associate to zero modes of the two summands $Y_L$ and $Y_R$. Thus, we can view $Y_\LR$ as the local geometries governing these gauge theory degrees of freedom. As the individual summands $Y_\LR=S^1_\LR \times X_\LR$ have $SU(3)$ holonomy in the Kovalev limit, we expect that the two gauge theory sectors of the kernels $k_L$ and $k_R$ exhibit $\mathcal{N}=2$ supersymmetry. Indeed --- in addition to the abelian $\mathcal{N}=1$ vector multiplets --- the kernels $k_\LR$ of the local geometries $Y_\LR$ also contribute to the three-form cohomology $H^3(Y,\mathbb{Z})$ resulting in $\mathcal{N}=1$ neutral chiral multiplets. Thus, the abelian $\mathcal{N}=1$ vector and the neutral $\mathcal{N}=1$ chiral multiplets associated to the $k_\LR$ readily combine into four-dimensional $\mathcal{N}=2$ vector multiplets.

Secondly, in the Kovalev limit the abelian vector multiplets obtained from the intersection $N_L \cap N_R$ can be attributed to the local geometry of the asymptotic regions $Y_L(T) \cap Y_R(T) \simeq T^2 \times S\times (0,1)$, which has $SU(2)$ holonomy. Thus, we expect that these vector multiplets give rise to a four-dimensional abelian $\mathcal{N}=4$ gauge theory sector, which can be seen as follows. To any two-form $\omega^{(2)}$ in $N_L \cap N_R$ of the K3 surface $S$, we attribute the three-forms
\begin{equation} \label{eq:3forms}
  \omega^{(2)} \wedge h(t) d\theta_L \ , \qquad \omega^{(2)} \wedge h(t) d\theta_R \ , \qquad
  \omega^{(2)} \wedge h(t)dt \ ,
\end{equation}
in terms of the coordinates $\theta_\LR$ of $S^1_L\times S^1_R \simeq T^2$ and the smooth bump function $h(t)$ in the coordinate $t$ of the interval~$(0,1)$.\footnote{The bump function $h(t)$ is given by a smooth non-negative function $h(t): (0,1) \to \mathbb{R}$ with compact support, which is normalized such that $\int_0^1 h(t)dt = 1$.} These three-forms yield geometrically non-trivial cohomology elements of compact support in $H^3_{c}(T^2 \times S\times (0,1),\mathbb{Z})$, which give rise to normalizable scalar fields. They combine with three scalar deformations of the hyper K\"ahler metric of the K3 surface $S$ to three complex scalar moduli fields, which furnish three neutral four-dimensional $\mathcal{N}=1$ chiral multiplets. It is these three $\mathcal{N}=1$ chiral multiplets that combine with the $\mathcal{N}=1$ vector multiplet of $\omega^{(2)}$ to one $\mathcal{N}=4$ vector multiplet.\footnote{Alternatively, we can consider the five-dimensional theory obtained from M-theory on $T^2 \times S$ with $SU(2)$ holonomy. Then the two-form cohomology element $\omega^{(2)}$ is accompanied by the two three-form cohomology elements $\omega^{(2)}\wedge d\theta_\LR$. Combined with the mentioned three hyper K\"ahler metric deformations these cohomology elements provide the zero modes of five scalar fields, which --- together with the vector field and the superpartners --- assemble into a five-dimensional $\mathcal{N}=2$ vector multiplet for each harmonic two-form $\omega^{(2)}$. Upon dimensional reduction to four dimensions, we arrive at four-dimensional $\mathcal{N}=4$ vector multiplets.} Note that the three-forms \eqref{eq:3forms} canonically extend to Kovalev's $G_2$-manifold $Y$. However, they become trivial in cohomology because $N_L \cap N_R$ is not an element of $H^3(Y,\mathbb{Z})$ according to eq.~\eqref{eq:cohTwistedSum}. Nevertheless, we can Fourier expand any of these three-forms into eigenforms with respect to the three-form Laplacian $\Delta$ of the $G_2$-manifold $Y$. By a simple scaling argument we find that the eigenvalues of these three-form Fourier modes scale with $T^{-1}$, i.e., they are inversely proportional to the parameter $T$ realizing the Kovalev limit. Therefore, we argue that the normalizable zero modes associated to the three-forms~\eqref{eq:3forms} acquire a mass term $m^2 \simeq O(T^{-1})$, which vanishes in the Kovalev limit. Furthermore, we expect that the scalar fields associated to the hyper K\"ahler metric deformations are generically obstructed at first order by a mass term that also vanishes in the Kovalev limit. As a consequence, we deduce that the massless four-dimensional abelian $\mathcal{N}=4$ vector multiplets of the asymptotic region decomposes into a massless four-dimensional abelian $\mathcal{N}=1$ vector multiplet and three massive four-dimensional $\mathcal{N}=1$ chiral multiplets with masses of order $O(T^{-1/2})$. Thus, we expect that the four-dimensional $\mathcal{N}=4$ gauge theory sector is only realized in the strict Kovalev limit $T \to +\infty$.

\begin{table}[t]
\hfil
\hbox{
\vbox{
\offinterlineskip
\halign{\vrule height2.7ex depth1.2ex width1.2pt~\hfil#\hfil\vrule width0.8pt&~\hfil#\hfil&\vrule~\hfil#\hfil&\vrule width0.8pt#&~\hfil#\hfil&\vrule~\hfil#\hfil&\vrule width1.2pt#\cr 
\noalign{\hrule height 1.2pt}
local geometry &\multispan2\hfil~multiplicity of $\mathcal{N}=1$ multiplets \hfil&& \multispan2\hfil ~$U(1)$ vector multiplets \hfil &\cr
(Kovalev limit) & $U(1)$ vectors & chirals && multiplicity & supersym. &\cr
\noalign{\hrule height 1.2pt}
$Y_L = S^1_L \times X_L$ & $\dim k_L$ & $\dim k_L$ && $\dim k_L$  & $\mathcal{N}=2$ &\cr
$SU(3)$ holonomy~&&&&&&\cr
\noalign{\hrule}
$Y_R = S^1_R \times X_R$ & $\dim k_R$ & $\dim k_R$ && $\dim k_R$  & $\mathcal{N}=2$ &\cr
$SU(3)$ holonomy~&&&&&&\cr
\noalign{\hrule}
$T^2\times S \times (0,1)$ & $\dim N_L\cap N_R$ & $3\cdot \dim N_L\cap N_R$ &&  $\dim N_L\cap N_R$ & $\mathcal{N}=4$ &\cr
$SU(2)$ holonomy~&&&&&&\cr
\noalign{\hrule height 1.2pt}
}
}}
\hfil
\caption{Shown are the abelian gauge theory sectors of the local geometries appearing in twisted connected sum $G_2$-manifolds in the Kovalev limit $T\to+\infty$. The left column specifies the Ricci-flat local geometries with their holonomies. The middle column lists the $\mathcal{N}=1$ chiral multiplets that assemble in the right column to vector multiplets of extended supersymmetry.}
\label{tab:locSpec}
\end{table}

The discussed local abelian gauge theory sectors are summarized in Table~\ref{tab:locSpec}. In particular, we find that in the Kovalev limit --- at least in the absence of background four-form fluxes and for smooth $G_2$-manifolds $Y$ --- the spectrum of all abelian gauge theory sectors exhibit extended supersymmetry. The observed extended supersymmetries of the local geometries appearing in Kovalev's twisted connected sum become relevant in Sections~\ref{sec:N=4sectors} and~\ref{sec:N=2sectors} because they impose strong constraints on the non-Abelian gauge theory sectors with charged matter fields.

\subsection{The K\"ahler potential} \label{sec:kaehlerpotential}  
The aim of this subsection is to describe the universal properties of the four-dimensional low-energy effective action in terms of the universal chiral multiplets $\nu$ and $\kov$. We first establish that --- while keeping the ratio $\operatorname{Re}(\nu)/\operatorname{Re}(\kov)$ constant --- the chiral multiplet~$\nu$ directly relates to the (dimensionless) volume modulus $R$ of Section~\ref{sec:Kovalevlimit} as
\begin{equation}
    \operatorname{Re}(\nu) \,=\, R^3 \ .
\end{equation}    
This relation comes about because the $\operatorname{Re}(\nu)$ measures (dimensionless) volumes of three-cycles while $R$ measures (dimensionless) length scales in the $G_2$-manifold $Y$. Apart from the overall volume dependence, the Kovalevton $\kov$ measures the squashed volume of the $S^3$ base. Therefore, from expression~\eqref{eq:VolYII} of the volume $V_Y(\tilde S,R,T)$ we arrive at the relation
\begin{equation}
  \operatorname{Re}(\kov) \,=\, (2\pi)^2 R^3 (2 T +\alpha(\tilde S)) \ ,
\end{equation}
where $\tilde S$ denotes collectively the remaining geometric moduli fields $\tilde S^{\tilde\imath}$ and $\tilde S^{\hat\imath}$. 

Thus --- using eqs.~\eqref{eq:KahlerPotential} and \eqref{eq:VolYII} --- we find that the universal structure of the four-dimensional low-energy effective $\mathcal{N}=1$ supergravity action is governed by the K\"ahler potential
\begin{equation} \label{eq:Kkov}
    K(\nu,\bar\nu,\kov,\bar\kov) \,=\, 
    -4 \log(\nu+\bar\nu) - 3 \log(\kov + \bar\kov) - 3 \log \left( V_S^{\tilde g}(\tilde S) \right) \ .
\end{equation}
Note that this K\"ahler potential is only a valid approximation both in the large volume regime and in the Kovalev regime, where quantum corrections and metric corrections of the asymptotically cylindrical Calab--Yau threefolds are suppressed.  The semi-classical large volume limit arises when both $\operatorname{Re}(\nu)$ and $\operatorname{Re}(\kov)$ are taken sufficiently large, and when $\operatorname{Re}(\kov)$ is (parametrically) larger than $\operatorname{Re}(\nu)$ --- cf. the discussion at the end of Section~\ref{sec:Kovalevlimit} --- while the corrections to the $G_2$-metric in the twisted connected sum are suppressed.

Let us discuss some basic properties of the derived K\"ahler potential. First of all, the structure of the K\"ahler potential is reminiscent of the Kovalev limit, in which the volume of the $G_2$-manifold is dominated by the cylindrical region $S \times T^2 \times I$ in terms of the interval $I$ of size $2T+1$. That is to say that, in this limit, the individual summands in eq.~\eqref{eq:Kkov} reflect the volume of the K3 surface $S$, the squashed volume of the $S^3$ base dominated by $T^2 \times  I$, and the moduli dependence of the K3 fiber $S$ on the non-universal moduli $\tilde S$. As long as we treat the non-universal moduli fields $\tilde S$ as constants, the K\"ahler geometry for the universal K\"ahler moduli $\nu$ and $\kov$ factorizes into two (complex) one-dimensional parts with a block diagonal K\"ahler metric. However, this block structure in the K\"ahler metric vanishes as soon as we treat the non-universal moduli fields $\tilde S$ dynamically, because relation~\eqref{eq:chiralvar} implies that the real geometric moduli $\tilde S$ also depend non-trivially on the chiral fields $\nu$ and $\kov$ as
\begin{equation}
  \tilde S^{\hat\imath} = \frac{\phi^{\hat\imath}+\bar\phi^{\hat\imath}}{\nu+\bar\nu} \ , \qquad
  \tilde S^{\tilde\imath} = \frac{\phi^{\tilde\imath}+\bar\phi^{\tilde\imath}}{\kov+\bar\kov} \ .
\end{equation}

We now briefly discuss the K\"ahler potential~\eqref{eq:Kkov} with the non-universal moduli fields $\tilde S$ treated as constants. We first observe that 
\begin{equation}
   g^{i\bar\jmath} \partial_i K \partial_{\bar\jmath} K - 3 \,=\, 4 \, \ge \, 0 \ , \qquad i\in\{\nu,\kov\}, \ \bar\jmath\in\{\bar\nu,\bar\kov\} \ ,
\end{equation}
in terms of the inverse K\"ahler metric $g^{i\bar\jmath}$. This implies that the no-scale inequality $g^{i\bar\jmath} \partial_i K \partial_{\bar\jmath} K - 3 \ge 0$ is fulfilled (but not saturated). The no-scale inequality is a property of the K\"ahler potential only and it guarantees that the scalar potential of the described four-dimensional $\mathcal{N}=1$ supergravity theory is positive semi-definite for any non-vanishing superpotential \cite{Kramer:2005Hamburg}. 
As a result the analyzed K\"ahler potential (of the two chiral fields $\nu$ and $\kov$ only) does not admit a negative cosmological constant and hence no (supersymmetric) anti-de-Sitter vacua.

Finally, we record the K\"ahler potential including the leading order correction to the Kovalev limit, which according to eq.~\eqref{eq:VolYII} takes the form
\begin{equation} \label{eq:Kkovcor}
  K =  
  - \log\left[ \left(V_S^{\tilde g}(\tilde S)\right)^3 (\nu +\bar\nu)^4 (\kov +\bar\kov)^3 +
  A(\tilde S,\nu+\bar\nu,\kov+\bar\kov) e^{-\lambda \frac{\kov+\bar\kov}{(\nu+\bar\nu)^{1/3}}}   \right] \ ,
\end{equation}
where the coefficient of the exponentially suppressed correction is expected to generically depend on both universal and non-universal geometric moduli fields. A detailed analysis of this class of K\"ahler potential may exhibit interesting phenomenological properties, which, is, however, beyond the scope of this work.

\section{Twisted connected sums from orthogonal gluing}  \label{sec:N=4sectors}
In this section we discuss the explicit construction of twisted connected sum $G_2$-manifolds by the method of orthogonal gluing \cite{MR3369307}. This construction offers a systematic way to fulfill the matching condition~\eqref{eq:HyperKaehler} for a pair of asymptotically cylindrical Calabi--Yau threefolds $X_\LR$ that are obtained from building blocks $(Z_\LR,S_\LR)$ associated to semi-Fano threefolds $P_\LR$. We focus on building blocks $(Z_\LR,S_\LR)$ of polarized K3~surfaces $S_\LR$ with Picard lattices of low rank and generate a list of new examples in order to get an impression of the multitude of possibilities to realize twisted connected sum $G_2$-manifolds in terms of orthogonal gluing. 

Our motivation for studying the method of orthogonal gluing is also to pan out the possibilities to obtain gauge theory sectors in twisted connected sum $G_2$-manifolds. In Section~\ref{sec:M-theoryonTCS} we have established that in the Kovalev limit this spectrum of vector multiplets assembles itself into $\mathcal{N}=2$  and $\mathcal{N}=4$ sectors as summarized in Table~\ref{tab:locSpec}. While we postpone the analysis of the $\mathcal{N}=2$ sectors to the next section, the focus in this section is on the $\mathcal{N}=4$ gauge theory sectors.

In the context of the orthogonal gluing construction, a certain intersection lattice~$R$ determines the $\mathcal{N}=4$ gauge theory sector. In particular, the rank of this intersection lattice becomes the rank of the gauge group. We further argue that in the $\mathcal{N}=4$ gauge theory sector the method of orthogonal gluing does not admit enhancements to non-Abelian gauge groups. As a result, we therefore arrive in the Kovalev limit at four-dimensional Abelian $\mathcal{N}=4$ gauge theory sectors with gauge group $U(1)^{\operatorname{rk} R}$. 

\subsection{Asymptotically cylindrical Calabi--Yau examples} \label{sec:ACYex}
Following ref.~\cite{MR3109862}, we obtain from toric weak Fano threefolds $P$\label{gl:Fano} a rich class of building blocks~$(Z,S)$, which in turn give rise to asymptotically cylindrical Calabi--Yau threefolds as discussed in Section~\ref{sec:spectrumandcohomology}. A projective smooth threefold $P$ is weak Fano if its anti-canonical divisor $-K_P$ is nef and big, which means that the intersections obey $-K_P C \ge 0$ for any algebraic curve $C$ in $P$ and $(-K_P)^3>0$.\footnote{A smooth projective variety is Fano if its anti-canonical divisor is ample, i.e., $-K_P C > 0$ for any algebraic curve $C$ in $P$.} Assuming further that two global sections $s_0$ and $s_1$ of the anti-canonical divisor $-K_P$ intersect transversely in a smooth reduced curve $\mathcal{C}=\{ s_0 =0 \} \cap \{ s_1 =0 \}\subset P$ and that $S=\left\{ \alpha_0 s_0 + \alpha_1 s_1 =0\right\}\subset P$ is a smooth K3~surface for some choice of $[\alpha_0:\alpha_1] \in \mathbb{P}^1$, a building block~$(Z,S)$ is obtained from the blow up $\pi_\mathcal{C}: Z \to P$ along $\mathcal{C}$, i.e.,
\begin{equation} \label{eq:ZBlowup}
    Z \,=\, \operatorname{Bl}_\mathcal{C}\!P \,=\, \left\{ (x,z) \in P \times \mathbb{P}^1 \,\middle|\, z_0 s_0 + z_1 s_1 = 0 \right\} \ ,
\end{equation}
together with the proper transform~$S$ of the smooth anti-canonical divisor $S$ on $P$ \cite{MR3109862}. Note that, for ease of notation, we use the same symbol $S$ for both the K3 surface in the toric weak Fano threefold $P$ and its proper transform in the blow-up $Z$. Then the K3 fibration $\pi: Z \to \mathbb{P}^1$ becomes, cf. Section~\ref{sec:spectrumandcohomology},
\begin{equation}
  \pi: Z \to \mathbb{P}^1,\, (x,z) \mapsto z \ ,
\end{equation}  
and $S=\pi^{-1}([\alpha_0,\alpha_1])$ is the K3~surface of the building block $(Z,S)$. 
Moreover, the three-form Betti number $b_3(Z)$ of the blown-up threefold $Z$ becomes
\begin{equation}
     b_3(Z) \,=\, b_3(P) + 2 g(\mathcal{C}) \,=\, b_3(P) + (-K_P)^3 + 2 \ .
\end{equation}
Here $g(\mathcal{C})$ denotes the genus of the curve $\mathcal{C}$ and the last equality follows from the adjunction formula.

For twisted connected sum $G_2$-manifolds, however, it is essential to find a pair of asymptotically cylindrical Calabi--Yau threefolds that fulfill the matching condition~\eqref{eq:HyperKaehler}. A common strategy is to first focus only on the moduli spaces of the polarized $K3$~surfaces $S_\LR$, ignoring their origin from the building blocks $(Z_\LR,S_\LR)$ \cite{MR3369307,MR3109862,Crowley:2014ij}. Once a matching pair of polarized $K3$~surfaces $S_\LR$ is found, it is necessary to check if these particular $K3$~surfaces $S_\LR$ arise as zero sections of anti-canonical divisors in $Z_\LR$. Beauville's theorem guarantees that indeed any general $K3$~surface polarized by the anti-canonical divisor of a Fano threefold can be realized as the zero locus of a global anti-canonical section for a suitable choice of the Fano threefold in its moduli space \cite{MR2112574}. Thus, for a pair of building blocks obtained from Fano threefolds it suffices to construct a matching pair of polarized $K3$~surfaces $S_\LR$ to ensure the existence of a pair of matching building blocks $(Z_\LR,S_\LR)$ in the moduli spaces of these building blocks.

However, for building blocks $(Z_\LR,S_\LR)$ obtained from weak Fano threefolds the procedure of simply matching their polarized $K3$~surfaces $S_\LR$ may not be sufficient. That is to say, in this more general setting the entire moduli space of the polarized $K3$~surfaces cannot necessarily be obtained from a global anti-canonical section within the associated family of weak Fano threefolds. Therefore, Corti et al. introduce the notion of semi-Fano threefolds that furnish a subclass of weak Fano threefolds for which Beauville's theorem is still applicable \cite{MR3369307,MR3109862}. For the rather technical definition of semi-Fano threefolds we refer to ref.~\cite{MR3109862}, and instead we present a characterizing criterion for toric semi-Fano threefolds momentarily.

An important and large class of building blocks $(Z,S)$ is constructed from toric weak Fano threefolds~$P_\Sigma$\label{gl:tFano} described in terms of a three-dimensional toric fan~$\Sigma$. We describe the toric fan~$\Sigma$ of a toric weak Fano threefold~$P_\Sigma$ in terms of a three-dimensional reflexive lattice polytope $\Delta$ spanned by the one-dimensional cones of $\Sigma$, together with a triangulation, which encodes the higher-dimensional cones of the fan~$\Sigma$.\footnote{See, for instance, refs.~\cite{MR1234037,MR2810322} for an introduction to toric geometry.} By the classification of Kreuzer and Skarke \cite{Kreuzer:1998vb,Kreuzer:2002uu} there are $4\,319$ three-dimensional reflexive polytopes, which often admit several triangulations, i.e., typically of the order of ten to a few hundred triangulations. In this work we focus on the class of toric semi-Fano threefolds~$P_\Sigma$ in order to follow the outlined recipe to construct explicit matching pairs for twisted connected sum $G_2$-manifolds. In the toric setting the semi-Fano threefolds~$P_\Sigma$ are characterized by those reflexive polytopes $\Delta$ that do not have any interior points inside co-dimension one faces~\cite{MR3109862}. Note that there are $899$ three-dimensional reflexive polytopes of the semi-Fano type \cite{MR3109862}.

The toric approach to semi-Fano threefolds provides a powerful combinatorial machinery to explicitly carry out computations. For instance, a general global section~$s_\Delta$ of the anti-canonical line bundle $-K_{P_\Sigma}$ is readily described by
\begin{equation} \label{eq:sDeltas}
  s_\Delta\,=\,\sum_{\nu_i\in \Delta} s_i  \prod_{\nu^*_k\in\Delta^*} x_k^{\langle \nu_i,\nu^*_k\rangle+1}  \ ,
\end{equation}
with the points $\nu_i$ and $\nu^*_k$ of the lattice polytope~$\Delta$ and its dual lattice polytope~$\Delta^*$, the duality lattice pairing $\langle \cdot , \cdot \rangle$, the toric homogenous coordinates $x_k$, and the coordinates $s_i$ on the space of global anti-canonical sections. Furthermore, for the discussed three-dimensional semi-Fano lattice polytopes, a choice of generic sections $s_0$ and $s_1$ yields a smooth reduced curve $\mathcal{C}=\{ s_0=0\} \cap \{s_1=0 \} \subset P_\Sigma$. This curve has a smooth K3 surface $S=\left\{ \alpha_0 s_0 + \alpha_1 s_1 =0\right\}\subset P_\Sigma$ such that the blow-up \eqref{eq:ZBlowup} together with the proper transform $S$ yield indeed a well-defined building block $(Z,S)$, which we refer to as the toric semi-Fano building block $(Z,S)$ in the following. Such building blocks exhibit the required technical properties that the anti-canonical class $-K_Z$ is primitive, that $H^3(Z,\mathbb{Z})$ is torsion-free (because $H^3(P_\Sigma,\mathbb{Z})$ is torsion-free), and that $H^2(S,\mathbb{Z})$ embeds primitively into $H^2(X,\mathbb{Z})$ with $X=Z \setminus S$, cf. ref.~\cite{MR3109862}.

\subsection{Construction of $G_2$-manifolds}  \label{sec:gluing}
To construct explicit examples of twisted connected sum $G_2$-manifolds, Corti et al. introduce the method of orthogonal gluing in ref.~\cite{MR3369307}, which is a particular recipe to fulfill the matching condition for suitable pairs of building blocks $(Z_\LR,S_\LR)$.

In this work we apply the orthogonal gluing method mainly to Fano building blocks and toric semi-Fano building blocks to algorithmically find novel twisted connected sum $G_2$-manifolds. In order to specify a particular semi-Fano threefold, in the following we use the Mori--Mukai classification for Fano threefold varieties \cite{MR641971} and the Kasprzyk classification for reflexive polytopes with terminal singularities for certain toric semi-Fano threefolds \cite{MR2221794,MR2221794db}. We label the corresponding semi-Fano threefolds by their respective reference numbers MM\#${}_\rho$ or/and K\# in these classifications,\label{gl:MMKlist} where the subscript in the Mori--Mukai list denotes the rank $\rho$ of the Picard lattice of the Fano threefold.

Applying the method of orthogonal gluing to a pair of building blocks $(Z_\LR,S_\LR)$ of two semi-Fano threefolds $P_\LR$, we use the following three-step algorithm \cite{MR3369307}:
\begin{itemize}
\item \emph{Construction of the orthogonal pushout lattice $W$:}
To achieve the matching condition~\eqref{eq:HyperKaehler} first for a pair of polarized K3~surfaces $S_\LR$ via orthogonal gluing, choose a negative definite lattice $R$ embedded primitively into both Picard lattices $N_\LR$ of the polarized K3~surfaces~$S_\LR$. Then the lattice $W$ is constructed as
\begin{equation} \label{eq:defW}
   W=N_L+N_R \ , \qquad R = N_L \cap N_R \ ,
\end{equation}
such that
\begin{equation}
  N_L^\perp \subset N_R \ , \qquad N_R^\perp \subset N_L \ ,
\end{equation}   
with the orthogonal lattices $N_\LR^\perp$ defined in eq.~\eqref{eq:Nperp}. The lattice~$W$ is called the orthogonal pushout of $N_\LR$ with respect to $R$ and is also denoted by \cite{MR3369307}
\begin{equation} \label{eq:Wperp}
  W  = N_L \perp_R N_R \ . 
\end{equation}
Note that the pushout lattice $W$ is unique but in general need not exist because the non-degenerate lattice pairing $\left\langle \cdot , \cdot \right\rangle_W: W \times W \to \mathbb{Z}$ induced from the pairings $\left\langle \cdot , \cdot \right\rangle_{N_\LR}: N_\LR \times N_\LR \to \mathbb{Z}$ is not necessarily well-defined. That is to say, the induced pairing
\begin{equation} \label{eq:indpairing}
   \left\langle e_L + e_R , f_L + f_R \right\rangle_W =  \left\langle e_L  , f_L  \right\rangle_{N_L} + \left\langle e_R  , f_R  \right\rangle_{N_R}
\end{equation}   
must be integral for any pair of lattice points $(e_L + e_R,f_L+f_R)$ in $W$, which can be represented --- not necessarily uniquely --- by $e_\LR,f_\LR \in N_\LR$. Furthermore, we require that the intersections $W_\LR = N_\LR \cap N_{R/L}^\perp$\label{gl:WLR} with the (generic) K\"ahler cones~$\mathcal{K}(P_\LR)$ of (the deformation families of)  $P_\LR$ are non-empty, i.e.,
\begin{equation} \label{eq:IntKaehler}
   \mathcal{K}(P_\LR) \cap W_\LR \ne \emptyset \ .
\end{equation}
\item \emph{Primitive embedding of pushout lattice $W$ into K3 lattice $L$:}
The matching condition~\eqref{eq:HyperKaehler} for a suitable pair of polarized K3~surfaces~$S_\LR$ in their moduli spaces is achieved if we embed primitively the pushout lattice $W$ into the K3~lattice $L$. The existence of such an embedding can often be deduced from results by Nikulin~\cite{MR525944}. In particular, such a primitive embedding is guaranteed to exist if the following rank condition is fulfilled \cite{MR3369307}
\begin{equation}
   \operatorname{rk} N_{L} + \operatorname{rk} N_{R} \,\leq \, 11\ .
\end{equation}
\item \emph{Lift matching condition of K3~surfaces $S_\LR$ to building blocks $(Z_\LR,S_\LR)$:}
Finally, we must ensure that the matching condition~\eqref{eq:HyperKaehler} for the polarized K3~surfaces~$S_\LR$ can be achieved within the moduli space of the building blocks $(Z_\LR,S_\LR)$. Corti et al. show in Proposition~6.18 of ref.~\cite{MR3369307} that the imposed assumptions on the orthogonal pushout $W$ --- namely that $(Z_\LR,S_\LR)$ are building blocks of semi-Fano threefolds, that the lattice $R$ is negative definite, that the intersections~\eqref{eq:IntKaehler} are non-empty, and that $W$ embeds primitively into the K3~lattice $L$ --- are sufficient to ensure that the matching conditions of the polarized K3~surfaces~$S_\LR$ can indeed be lifted to the moduli spaces of the building blocks $(Z_\LR,S_\LR)$.
\end{itemize}

Let us now determine the cohomology groups~\eqref{eq:cohTwistedSum} of the $G_2$-manifolds $Y$ obtained from orthogonal gluing. First we observe that $\operatorname{rk} N_L \cap T_R$ and $\operatorname{rk} N_R \cap T_L$ equal $\operatorname{rk} W_L$ and $\operatorname{rk} W_R$, respectively, while $N_L+N_R$ becomes the orthogonal pushout $W$ with $\operatorname{rk} W = \operatorname{rk} W_L + \operatorname{rk} W_R + \operatorname{rk} R$. Therefore, we readily deduce for the Betti numbers $b_2(Y)=\dim H^2(Y)$ and $b_3(Y)=\dim H^3(Y)$ for the $G_2$-manifolds $Y$ or the orthogonal gluing type
\begin{equation} \label{eq:b2b3}
\begin{aligned}
   b_2(Y) \,&=\,  \operatorname{rk} R + \dim k_L + \dim k_R   \ , \\
   b_3(Y) \,&=\,  b_3(Z_L) + b_3(Z_R) + \dim k_L + \dim k_R - \operatorname{rk} R + 23  \ .
\end{aligned}
\end{equation}
Here $b_3(Z_\LR)$ are the three-form Betti numbers of the threefolds $Z_\LR$ and $\dim k_\LR$ are the dimensions of the kernels $k_\LR$ defined below eq.~\eqref{eq:cohTwistedSum}.

Recall that in the Kovalev limit the kernels $k_\LR$ describe the $\mathcal{N}=2$ gauge theory sectors, respectively, whereas the rank of the intersection lattice $R$ in the orthogonal pushout $W$ coincides with the rank of the gauge group of the $\mathcal{N}=4$ gauge theory sector, cf. Table~\ref{tab:locSpec}.  A particular simple choice of orthogonal gluing is achieved if the intersection lattice $R$ has rank zero, i.e., $N_L\cap N_R = \left\{ 0 \right\}$. This special case of orthogonal gluing is referred to as perpendicular gluing with its trivial orthogonal pushout $W$ denoted by \cite{MR3369307}
\begin{equation} \label{eq:Worth}
  W = N_L \perp N_R \ . 
\end{equation}
As a consequence, the $\mathcal{N}=4$ gauge theory sector in the Kovalev limit is absent if the twisted connected $G_2$-manifold is obtained via perpendicular gluing.

\subsection{Examples of $G_2$-manifolds from orthogonal gluing}  \label{sec:examples}
In this section we study concrete examples of twisted connected sum $G_{2}$-manifolds obtained via orthogonal gluing. In particular, we focus on examples with non-trivial intersection lattices~$R$ in the orthogonal pushout $W$. 

As explained in Section~\ref{sec:spectrumandcohomology}, in the Kovalev limit such examples yield $\mathcal{N}=4$ gauge theory sectors with the Abelian gauge group $U(1)^{\operatorname{rk} R}$. Enhancement to a non-Abelian gauge group $G$ of rank $r$ would occur if the intersection lattice $R=N_L\cap N_R$ had a sublattice $G(-1)$ or rank $r$ with the pairing given by minus the Cartan matrix of the Lie algebra of $G$ \cite{Witten:1995ex}. Then we could blow-down the mutual $r$ rational curves of both polarized K3~surfaces $S_\LR$ because --- by the definition of the intersection lattice $R$ --- $G(-1)$ resides in the intersection of both Picard lattices $N_\LR$. In this way we would arrive at singular polarized K3~surfaces~$S_\LR$ resulting in the enhanced $\mathcal{N}=4$ gauge theory sector with non-Abelian gauge group $G \times U(1)^{\operatorname{rk} R-r}$. However, using the method of orthogonal gluing, such a gauge theory enhancement is not possible because the orthogonal complement to $R$ in the polarized K3~surfaces $S_\LR$ is required to contain an ample class, which --- due to the ampleness ---  would always have a non-zero intersection with any rational curve. It would nevertheless be interesting to see if non-Abelian gauge groups are possible in the $\mathcal{N}=4$ gauge theory sector by generalizing the orthogonal gluing construction.

\paragraph{Orthogonal gluing of rank two semi-Fano threefolds:}
Let us consider orthogonal gluings among building blocks of semi-Fano threefolds with Picard number two. The only non-trivial orthogonal gluings among such building blocks --- that is to say apart from perpendicular gluings --- have an intersection lattice~$R$ of rank one. In ref.~\cite{Crowley:2014ij} Crowley and Nordstr\"om classify, for rank two Fano threefold building blocks, all possible non-trivial orthogonal gluings to twisted connected sum $G_2$-manifolds --- except for one missing pair.\footnote{In Table~2 and Table~4 of ref.~\cite{Crowley:2014ij} the authors list all rank two Fano building blocks and the resulting $G_2$-manifolds, respectively. However, the classification of $G_2$-manifolds in Table~4 misses the orthogonal gluing between the building blocks MM5${}_2$ and MM25${}_2$ of Table~2. Therefore, Theorem 5.10 of ref.~\cite{Crowley:2014ij} should enumerate nineteen instead of eighteen pairs of twisted connected sum $G_2$-manifolds.} As a warm-up we want to extend this classification by including all building blocks arising from toric semi-Fano threefolds with Picard number two.

It turns out that there is actually a unique toric semi-Fano threefold $P_\Sigma$ of Picard number two that is not Fano and is given by the projective bundle \cite{MR3109862}
\begin{equation} \label{eq:ProjBundle}
   \mathbb{P}(\mathcal{O}\oplus \mathcal{O}(-1) \oplus \mathcal{O}(-1)) \to \mathbb{P}^1 \ . 
\end{equation} 
Its toric realization of $P_\Sigma$ arises from the reflexive lattice polytope $\Delta$ and its dual reflexive polytope $\Delta^*$ spanned by the lattice points $\nu_1, \ldots, \nu_5$ and the dual lattice points $\nu_1^*, \ldots, \nu_5^*$ given by:
\begin{equation} \label{eq:K32Mcone}
\vcenter{
\halign{$#\quad$&$#=($&\hfil$#,$&\hfil$#,$&\hfil$#)$&\qquad$#\quad$&$#=($&\hfil$#,$&\hfil$#,$&\hfil$#)$\cr
\Delta:&\nu_1&-1&-1&0&\Delta^*:&\nu_1^*&-1&-1&1\cr
&\nu_2&1&0&0&&\nu_2^*&-1&2&-2\cr
&\nu_3&0&1&0&&\nu_3^*&-1&2&1\cr
&\nu_4&1&1&1&&\nu_4^*&2&-1&-2\cr
&\nu_5&0&0&-1&&\nu_5^*&2&-1&1\cr
}} 
\end{equation}
The reflexive lattice polytope $\Delta$ appears as number K32 in the Kasprzyk classification \cite{MR2221794,MR2221794db}. It admits two simplicial triangulations both realizing the projective bundle~\eqref{eq:ProjBundle}. The toric variety $P_\Sigma$ associated to the fan $\Sigma$ of one of these triangulations gives rise to the Mori cone spanned by the curves $C_B\simeq \mathbb{P}^1$ and the curve $C_F \simeq \mathbb{P}^1 \subset \mathbb{P}^2_F$ in a projective fiber $\mathbb{P}^2_F$, such that these curves have the following intersection numbers with the toric divisors $D_i$ associated to the vertices $\nu_i$:\footnote{For the other triangulation the Mori cone takes the form:
$$
\vcenter{
\halign{\strut$#$&\hfil$\ #$&\hfil$\ #$&\hfil$\ #$&\hfil$\ #$&\hfil$\ #$\quad\cr
& \ D_1 & \ D_2 & \ D_3 &\ D_4 &\ D_5 \cr
\noalign{\hrule}
C_F: & 1 & 0 & 0 & 1 & 1 \cr 
C_B: & \phantom{-}0 & \phantom{-}1 & \phantom{-}1 & -1 & -1 \cr
}}
$$}
\begin{equation} \label{eq:MoriCone}
\vcenter{
\halign{\strut$#$&\hfil$\ #$&\hfil$\ #$&\hfil$\ #$&\hfil$\ #$&\hfil$\ #$\quad\cr
& \ D_1 & \ D_2 & \ D_3 &\ D_4 &\ D_5 \cr
\noalign{\hrule}
C_F: & 1 & 1 & 1 & 0 & 0 \cr 
C_B: & \phantom{-}0 & -1 & -1 & \phantom{-}1 & \phantom{-}1 \cr
}}
\end{equation}
As a consequence, we find among the toric divisors the linear equivalences $D_2 \sim D_3$, $D_4\sim D_5$, $D_2 \sim D_1 - D_4$ and the K\"ahler cone $\mathcal{K}(P_\Sigma)$ spanned by
\begin{equation} \label{eq:Kcone}
  \mathcal{K}(P_\Sigma) \,=\, \langle\!\langle D_1 , D_4 \rangle\!\rangle \ .
\end{equation}
Furthermore, note that $-K_{P_\Sigma} = \sum_i D_i \sim 3 D_1$ and $-K_{P_\Sigma}^3 = 54$. The intersection matrix $\kappa_{P_\Sigma}$ of the generators $D_1$ and $D_4$ with the anti-canonical class $-K_{P_\Sigma}$ reads
\begin{equation}
   \kappa_{P_\Sigma} \,=\, \begin{pmatrix} 6 & 3 \\ 3 & 0 \end{pmatrix} \ ,
\end{equation}
and has discriminant $\Delta^{\kappa}=-9$. The intersection matrix $\kappa_{P_\Sigma}$ furnishes the intersection pairing of the Picard lattice of the anti-canonical K3 surface in $P_\Sigma$.

According to the described algorithm in Section~\ref{sec:gluing}, for a pair of Picard lattices $(N_L, N_R)$  of rank two to yield a non-trivial orthogonal pushout $W$, the rank one sublattices $W_\LR$ must be  generated by ample classes in $\mathcal{K}(P_\LR)$ with the intersection lattice~$R$ orthogonal to both $W_\LR$. Thus, we need to construct two ample classes $A_\LR$ together with orthogonal lattice vectors $e_\LR$ in $N_\LR$ with $e_L^2 = e_R^2$, which generate the rank one intersection lattice~$R$. Then --- as Crowley and Nordstr\"om show in ref.~\cite{Crowley:2014ij} --- the induced lattice pairing $\left\langle \cdot , \cdot \right\rangle_W$ is a well-defined integral lattice pairing, if and only if 
\begin{equation} \label{eq:mcond}
  \frac{\Delta_L^{\kappa}\Delta_R^{\kappa}}{A_L^2 A_R^2}=k^2 \ , \quad \text{for}\ k\in\mathbb{Z} \ ,
\end{equation}
in terms of the discriminants $\Delta^\kappa_\LR$ and the ample classes $A_\LR$. Moreover, in order to fulfill this matching condition with a rank two Fano threefold $P_{R/L}$, Crowley and Nordstr\"om deduce an upper bound for the rank two semi-Fano threefold $P_\LR$ \cite{Crowley:2014ij} 
\begin{equation} \label{eq:inequ}
  \left| \frac{\Delta_\LR^{\kappa}}{A_\LR^2} \right| \le \frac85 \ .
\end{equation}

\begin{table}[t]
\hfil
\hbox{
\vbox{
\offinterlineskip
\halign{\vrule width1.2pt~#\hfil\vrule width0.8pt&~\hfil$#$\hfil~&~\hfil$#$\hfil~&~\hfil$#$\hfil~&~\hfil$#$\hfil~&~\hfil$#$\hfil~&~\hfil$#$\hfil~&~\hfil$#$\hfil~&~\hfil$#$\hfil~\vrule width1.2pt\cr
\noalign{\hrule height 1.2pt}
\strut No. & -K^3 & \kappa & \Delta^\kappa & A & e & A^2 & e^2 & b_3(Z)\cr
\noalign{\hrule height 1.2pt}
K32 (semi-Fano) & 54 & \begin{pmatrix} 6 & 3 \\  3 & 0 \end{pmatrix} & -9 & \begin{pmatrix}1\\1\end{pmatrix} & \begin{pmatrix}-1\\3\end{pmatrix} & 12 & -12 & 56\cr
MM5${}_2$ (Fano) &12 & \begin{pmatrix} 0 & 3 \\  3 & 6 \end{pmatrix} &-9& \begin{pmatrix}1\\1\end{pmatrix} & \begin{pmatrix}3\\-1\end{pmatrix}&12 &-12 &26 \cr
MM25${}_2$ (Fano) & 32 & \begin{pmatrix} 0 & 4 \\  4 & 4 \end{pmatrix} &-16& \begin{pmatrix}1\\1\end{pmatrix} & \begin{pmatrix}2\\-1\end{pmatrix} &12 &-12&36 \cr
\noalign{\hrule height 1.2pt}
}
}}
\hfil
\caption{The rows of the table list the data of the unique rank two $\rho=2$ toric semi-Fano threefolds together with all rank two Fano threefolds that admit an intersection lattice $R$ generated by a vector of length square $-12$. The columns show the reference numbers MM\#$_\rho$ in the Mori--Mukai~\cite{MR641971} classification or K\# in the Kasprzyk~\cite{MR2221794db} classification, the triple intersection of the anti-canonical divisor $-K$, the intersection matrix $\kappa$ of the Picard lattice of the anti-canonical K3 surface, the discriminant $\Delta^\kappa$ of the intersection matrix $\kappa$, the chosen ample class $A$ in the basis of the intersection matrix $\kappa$, the orthogonal complement $e$ to the ample class $A$, the length squares of the classes $A$ and $e$, and the three-form Betti number $b_3(Z)$ of the associated building block $Z$.}
\label{tab:g2:f22}
\end{table}

To find a matching rank two Fano building block for non-trivial orthogonal gluing with the unique rank two toric semi-Fano threefold~\eqref{eq:ProjBundle}, we first choose an ample class $A$, which according to eq.~\eqref{eq:Kcone} is given by $A =n D_1 + m D_4$ with $A^2 = 6 n (n+m)$. Thus, in order to conform with the inequality~\eqref{eq:inequ} for $\Delta^\kappa=-9$, the only possible ample class is $A=D_1+D_4$ with $A^2 = 12$. For this class the orthogonal complement $R$ is generated by $e=-  D_1 + 3 D_4$ with $e^2 = -12$. Table~\ref{tab:g2:f22} summarizes the data of this toric semi-Fano threefold together with the corresponding data for the building blocks of rank two Fano threefolds with compatible rank one intersection lattices generated by vectors of length square $-12$. The latter entries are taken from the Crowley--Nordstr\"om classification \cite{Crowley:2014ij}.

For the entries in Table~\ref{tab:g2:f22},  condition~\eqref{eq:mcond} tells us the two possible gluings with rank one intersection lattices, namely\footnote{We would like to thank Johannes Nordstr\"om for pointing out a numerical error in the third Betti numbers in eq.~\eqref{eq:Orth1} in an earlier version of this work.}
\begin{equation} \label{eq:Orth1}
\begin{aligned}
    W^\text{K32}_{\text{MM25}_2} &= N_\text{K32} \perp_e N_{\text{MM25}_2} \, :   &&b_2(Y^\text{K32}_{\text{MM25}_2})=1 \ ,  &&b_3(Y^\text{K32}_{\text{MM25}_2})=114 \ , \\
    W^{\text{MM5}_2}_{\text{MM25}_2} &= N_{\text{MM5}_2} \perp_e N_{\text{MM25}_2} \, :  &&b_2(Y^{\text{MM5}_2}_{\text{MM25}_2})=1 \ , &&b_3(Y^{\text{MM5}_2}_{\text{MM25}_2})=84 \ . 
\end{aligned}
\end{equation} 
Here $Y^{\cdots}_{\cdots}$\label{gl:Ydot} denotes the twisted connected $G_2$-manifolds obtained from the orthogonal pushout $W^{\cdots}_{\cdots}$ with their Betti numbers computed by eq.~\eqref{eq:b2b3}. The first $G_2$-manifold realizes the only possible combination with the rank two toric semi-Fano threefold~K32, whereas the second $G_2$-manifold realizes the non-trivial orthogonal gluing among the rank two Fano threefolds that have been overseen in ref.~\cite{Crowley:2014ij}.

\paragraph{Orthogonal gluing of higher rank semi-Fano threefolds:}
As our next illustrating examples, we consider orthogonal gluings along a rank one intersection lattice with the rank three Fano threefold $P_L=\mathbb P^1\times \mathbb P^1\times \mathbb P^1$, which has the reference numbers MM27${}_3$ and K62 in the Mori--Mukai and Kasprzyk classifications, respectively  \cite{MR641971,MR2221794,MR2221794db}. Let $H_i$, $i=1,2,3$, be the hyperplane classes of the respective $\mathbb{P}^1$-factors of this Fano threefold, which generate the three-dimensional K\"ahler cone
\begin{equation} \label{eq:3dKcone}
   \mathcal{K}(P_\Sigma) \,=\, \langle\!\langle H_1 , H_2 , H_3 \rangle\!\rangle \ .
\end{equation}
The ample anti-canonical class becomes $-K_{P_L}=2 H_1 + 2 H_2 + 2 H_3$, and the intersection matrix $\kappa_{P_L}$ of the K\"ahler cone generators with the anti-canonical divisor $-K_{P_L}$ reads
\begin{equation} \label{eq:kp1p1p1}
    \kappa_{P_L}\,=\,\begin{pmatrix} 0 & 2 & 2 \\  2 & 0 & 2\\  2 & 2 & 0 \end{pmatrix} \ .
\end{equation}

Now, we focus on orthogonal gluing with the rank one intersection lattice generated by a vector $e$ of length square $e^2=-4$. Note that this length square realizes the maximal negative value, as the pairing $\kappa_{P_L}$ is even and vectors with length square $e^2=-2$ correspond to rational curves with a positive intersection number with any ample class $A$, which is in violation with the orthogonal gluing assumption~\eqref{eq:IntKaehler}. 

The intersection pairing~\eqref{eq:kp1p1p1} corresponds to the ternary quadratic form $q(x,y,z)=4(xy+yz+zx)$, which allows us to parametrize, with the help of ref.~\cite{MR1758806}, --- up to trivial permutations of the K\"ahler cone generators $H_1$, $H_2$, $H_3$ --- all vectors $e$ with $e^2=-4$ by
\begin{equation}
  e = (d_1 - k) H_1 + (d_2 - k) H_2 + k H_3 \ ,
\end{equation}
where the integers $k, d_1, d_2$ obey
\begin{equation} \label{eq:kd1d2rel}
   k^2 -d_1 d_2 \,=\, 1 \ , \qquad 0 \le d_1 < k \le d_2 \ .
\end{equation}  
In order to fulfill condition~\eqref{eq:IntKaehler}, we need to check that the orthogonal complement contains an ample class $A=a_1 H_1 + a_2 H_2 + a_3 H_3$ given in terms of positive integers $a_1$, $a_2$, $a_3$, i.e.,
\begin{equation}
  0\,=\, A.e \,=\, a_1 d_2 + a_2 d_1 + a_3( d_1+d_2 - 2 k) \ .
\end{equation}  
As the sum of the first two terms are always positive, this orthogonality condition can only be met if
\begin{equation}
   d_1 + d_2 < 2 k \ \Leftrightarrow \ (d_1 + d_2)^2 < 4 k^2 \ \Leftrightarrow (d_2 -d_1)^2 < 4 \ ,
\end{equation}    
where we used the relations~\eqref{eq:kd1d2rel}, which furthermore implies that $d_2 = d_1 + 1$ and hence $k=1, d_1=0, d_2=1$, corresponding to the vector 
\begin{equation} \label{eq:ep1p1p1}
   e = - H_1 + H_3 \ .
\end{equation}
For this vector $e$, the ample class $A = H_1 + H_2 + H_3$ is indeed orthogonal.

Therefore, for a left building block $(Z_L,S_L)$ of the Fano threefold $\mathbb{P}^1\times\mathbb{P}^1\times\mathbb{P}^1$ with a rank one intersection lattice~$R$ generated by a vector $e$ with $e^2=-4$, --- up to trivial relabelling of the K\"ahler cone generators --- the vector~\eqref{eq:ep1p1p1} is the only possibility. Its orthogonal complement $W_L$ is then generated by
\begin{equation}
     W_L \,=\, \mathbb{Z} \, w_1 + \mathbb{Z} \, w_2 \quad \text{with} \quad
     w_1 \,=\, H_1 + H_3 \ , \ w_2 = H_2 \ , 
\end{equation}
such that the Picard lattice $N_L$ in terms of $(w_1, w_2, e)$ reads
\begin{equation}
   N_L \,=\, \mathbb{Z} \, w_1 + \mathbb{Z} \, w_2 + \mathbb{Z} \, e + \frac12 \mathbb{Z}\, (w_1 + e) \ .
\end{equation}  

\begin{table}[t]
\hfil
\hbox{
\vbox{
\offinterlineskip
\halign{\vrule width1.2pt~#\hfil\vrule width0.8pt&~\hfil$#$\hfil~&~\hfil$#$\hfil~&~\hfil$#$\hfil~&~\hfil$#$\hfil~&~\hfil$#$\hfil~&~\hfil$#$\hfil~&~\hfil$#$\hfil~&~\hfil$#$\hfil~\vrule width1.2pt\cr
\noalign{\hrule height 1.2pt}
\strut No. & -K^3 & \kappa & \Delta^\kappa & A & e & A^2 & e^2 & b_3(Z)\cr
\noalign{\hrule height 1.2pt}
MM6${}_2$ &12 & \begin{pmatrix} 2 & 4 \\  4 & 2 \end{pmatrix} &-12& \begin{pmatrix}1\\1\end{pmatrix} & \begin{pmatrix}1\\-1\end{pmatrix}&12 &-4 &32 \cr
MM12${}_2$ & 20 & \begin{pmatrix} 4 & 6 \\  6 & 4 \end{pmatrix} &-20& \begin{pmatrix}1\\1\end{pmatrix} & \begin{pmatrix}1\\-1\end{pmatrix} &20 &-4 &28 \cr
MM21${}_2$ & 28 & \begin{pmatrix} 6 & 8 \\  8 & 6 \end{pmatrix} &-28& \begin{pmatrix}1\\1\end{pmatrix} & \begin{pmatrix}1\\-1\end{pmatrix} &28 &-4 &30 \cr
MM32${}_2$ & 48 & \begin{pmatrix} 2 & 4 \\  4 & 2 \end{pmatrix} &-12& \begin{pmatrix}1\\1\end{pmatrix} & \begin{pmatrix}1\\-1\end{pmatrix} &12 &-4 &50 \cr
\noalign{\hrule height 1.2pt}
}
}}
\hfil
\caption{The rows of the table list the data of all rank two Fano threefolds admitting an intersection lattice $R$ generated by a vector of length square $-4$. The columns show the reference number in the Mori--Mukai classification, the triple intersection of the anti-canonical divisor $-K$, the intersection matrix $\kappa$ of the Picard lattice of the anti-canonical K3 surface, the discriminant $\Delta^\kappa$ of the intersection matrix $\kappa$, the chosen ample class $A$ in the basis of the intersection matrix $\kappa$, the orthogonal complement $e$ to the ample class $A$, the length squares of the classes $A$ and $e$, and the three-form Betti number $b_3(Z)$ of the associated building block $Z$.}
\label{tab:Zrank2e2=-4Blocks}
\end{table}

So as to orthogonally glue this left Picard lattice $N_L$ along $e$ with a Picard lattice $N_R$ of a right rank two Fano building block $(Z_R, S_R)$, we find in the Crowley--Nordstr\"om classification~\cite{Crowley:2014ij} that the rank two Fano threefolds with Mori--Mukai reference numbers MM6${}_2$, MM12${}_2$, MM21${}_2$, and MM32${}_2$ give rise to compatible intersection lattices. For convenience, these particular building blocks together with some geometric data are summarized in Table~\ref{tab:Zrank2e2=-4Blocks}, and we readily see that for all these examples the rank two Picard lattices $N_R$ are generated in the orthogonal basis $(A,e)$ by
\begin{equation}
   N_R \,=\, \mathbb{Z} \, A + \mathbb{Z} \, e + \frac12 \mathbb{Z}\, (A + e) \ ,
\end{equation}
with $e^2 = -4$ and the length square of the ample class $A$ as listed in Table~\ref{tab:Zrank2e2=-4Blocks}.

\begin{table}[t]
\hfil
\hbox{\footnotesize{
\vbox{
\offinterlineskip
\halign{\vrule width1.2pt~#\hfil\vrule width0.8pt&~\hfil$#$\hfil~&~\hfil$#$\hfil~&~\hfil$#$\hfil~&~\hfil$#$\hfil~&~\hfil$#$\hfil~&~\hfil$#$\hfil~\vrule width1.2pt\cr
\noalign{\hrule height 1.2pt}
\strut No. &  \operatorname{rk} N & -K^3 & \kappa & e & e^2 & b_3(Z)\cr
\noalign{\hrule height 1.2pt}
MM27${}_3$, K62 (Fano) & 3 & 48 & \begin{pmatrix} 0 & 2 &2 \\  2 & 0 &2 \\ 2&2&0 \end{pmatrix} & \begin{pmatrix}1\\0\\-1\end{pmatrix} & -4 &  50 \cr
MM25${}_3$, K68 (Fano) & 3 & 44 & \begin{pmatrix} 0 & 2 &1 \\  2 & 0 &3 \\ 1&3&-2 \end{pmatrix} & \begin{pmatrix}-1\\1\\0\end{pmatrix} & -4 &  46 \cr
MM31${}_3$, K105 (Fano) & 3 & 52 & \begin{pmatrix} 0 & 2 &1 \\  2 & 0 &3 \\ 1&3&-2 \end{pmatrix} & \begin{pmatrix}-1\\1\\0\end{pmatrix} & -4 &  54 \cr
K124 (semi-Fano) & 3 & 48 & \begin{pmatrix} 2 & 4&2 \\  4 & 2 &2 \\ 2&2&0 \end{pmatrix} & \begin{pmatrix}-1\\1\\0\end{pmatrix} & -4 &  50 \cr
\noalign{\hrule}
MM12${}_4$, K218 (Fano) & 4 & 46 & \begin{pmatrix} 2 & 4&2&0 \\  4 & 2 &2&2 \\ 2&2&0&1\\0&2&1&-2 \end{pmatrix} & \begin{pmatrix} 1\\-1\\0\\0\end{pmatrix} & -4 &  48 \cr
MM10${}_4$, K266 (Fano) & 4 & 42 & \begin{pmatrix} 0 & 2&2&0 \\  2 & 0 &2&0 \\ 2&2&0&1\\0&0&1&-2 \end{pmatrix} & \begin{pmatrix} 1\\-1\\0\\0\end{pmatrix} & -4 &  44 \cr
K221 (semi-Fano) & 4 & 38 & \begin{pmatrix} -2 & 2&0&0 \\  2 & 2 &1&1 \\ 0&1&-2&1\\0&1&1&-2  \end{pmatrix} & \begin{pmatrix} 0\\-1\\1\\1\end{pmatrix} & -4 &  40 \cr
K232 (semi-Fano) & 4 & 40 & \begin{pmatrix} 0 & 2&2&0 \\  2 & 0 &2&0 \\ 2&2&0&0\\0&0&0&-2 \end{pmatrix} & \begin{pmatrix} -1\\0\\1\\0\end{pmatrix} & -4 &  42 \cr
K233 (semi-Fano) & 4 & 38 & \begin{pmatrix} -2 & 0&1&0 \\  0 & -2 &1&2 \\ 1&1&0&2\\0&2&2&-2 \end{pmatrix} & \begin{pmatrix} -1\\1\\0\\0\end{pmatrix} & -4 &  40 \cr
K247 (semi-Fano) & 4 & 44 & \begin{pmatrix} 4 & 3&3&2 \\  3 & 0 &2&0 \\ 3&2&0&0\\2&0&0&-2 \end{pmatrix} & \begin{pmatrix}0\\-1\\1\\0\end{pmatrix} & -4 &  46 \cr
K257 (semi-Fano) & 4 & 46 & \begin{pmatrix} 0& 2&0&3 \\  2 & 0 &0&3 \\ 0&0&-2&1\\3&3&1&6 \end{pmatrix} & \begin{pmatrix}-1\\1\\0\\0\end{pmatrix} & -4 &  48 \cr
\noalign{\hrule height 1.2pt}
}}}}
\hfil
\caption{The rows of the table list the data of all rank three and four (resolved) toric terminal Fano threefolds for an intersection lattice $R$ 
generated by a vector of length square $-4$. The columns show the reference number in the 
Mori--Mukai~\cite{MR641971} and/or Kasprzyk~\cite{MR2221794db} classification, the rank of the Picard lattice ---note that in the semi-Fano cases this rank as 
reported in~\cite{MR2221794db} is smaller since it refers to the singular variety--- the triple intersection of the anti-canonical divisor $-K$, the 
intersection matrix $\kappa$ of the Picard lattice of the anti-canonical K3 surface, the generator $e$ of the lattice $R$ and its length square, and the three-form Betti number $b_3(Z)$ 
of the associated building block $Z$.}
\label{tab:34Blocks}
\end{table}

Next we can construct the orthogonal pushout $W=N_L \perp_e N_R$, which in the basis $(w_1,w_2,e,A)$ takes the form
\begin{equation}
  W \,=\, \mathbb{Z} \, w_1 + \mathbb{Z} \, w_2 + \mathbb{Z} \, e + \mathbb{Z} \, A + \frac12 \mathbb{Z}\, (w_1 + e) +  \frac12 \mathbb{Z}\, (A + e) \ .
\end{equation}
This orthogonal pushout is well-defined since the potentially non-integral intersection pairing $\left\langle \frac12 (w_1 + e), \frac12 (A+e) \right\rangle_W = -1$ is indeed integral. For the integral generators $(\frac12 (w_1 + e),w_2,\frac12 (A+e),e)$, the intersection pairing $\kappa_W$ of the pushout $W$ becomes
\begin{equation} \label{eq:kappaW}
  \kappa_W \,=\, \begin{pmatrix} 0 & 2 & -1 & -2 \\ 2 & 0 & 0 & 0 \\  -1 & 0 & \tfrac14A^2-1 & -2 \\ -2 & 0 & -2 & -4 \end{pmatrix} \ , \qquad
  \det \kappa_W \,=\, 4 A^2 \ ,
\end{equation}
where --- according to Table~\ref{tab:Zrank2e2=-4Blocks} --- the entry $\frac14 A^2 -1$ is integral and even. As a result we obtain, for the orthogonal pushouts~$W$ of the rank three Fano threefold $\mathbb{P}^1\times\mathbb{P}^1\times\mathbb{P}^1$ with reference number MM27${}_3$ and the rank two Fano threefolds listed in Table~\ref{tab:Zrank2e2=-4Blocks}, the twisted connected sum $G_2$-manifolds $Y^\text{MM27${}_3$}_{\cdots}$
\begin{equation}\label{eq:Yrank2rank3}
\begin{aligned}
    W^\text{MM27${}_3$}_\text{MM6${}_2$} &= N_\text{MM27${}_3$} \perp_e N_\text{MM6${}_2$} \, :  &&b_2(Y^\text{MM27${}_3$}_\text{MM6${}_2$})=1 \ ,  &&b_3(Y^\text{MM27${}_3$}_\text{MM6${}_2$})=104 \ , \\
    W^\text{MM27${}_3$}_\text{MM12${}_2$} &= N_\text{MM27${}_3$} \perp_e N_\text{MM12${}_2$} \, :  &&b_2(Y^\text{MM27${}_3$}_\text{MM12${}_2$})=1 \ , &&b_3(Y^\text{MM27${}_3$}_\text{MM12${}_2$})=100 \ , \\ 
    W^\text{MM27${}_3$}_\text{MM21${}_2$} &= N_\text{MM27${}_3$} \perp_e N_\text{MM21${}_2$} \, :  &&b_2(Y^\text{MM27${}_3$}_\text{MM21${}_2$})=1 \ , &&b_3(Y^\text{MM27${}_3$}_\text{MM21${}_2$})=102 \ , \\
    W^\text{MM27${}_3$}_\text{MM32${}_2$} &= N_\text{MM27${}_3$} \perp_e N_\text{MM32${}_2$} \, :  &&b_2(Y^\text{MM27${}_3$}_\text{MM32${}_2$})=1 \ , &&b_3(Y^\text{MM27${}_3$}_\text{MM32${}_2$})=122 \ .
\end{aligned}
\end{equation} 

Analogously, we can construct twisted connected sum $G_2$-manifolds via orthogonal gluing along rank one intersection lattices for semi-Fano threefolds with higher rank Picard lattices. In Table~\ref{tab:34Blocks} we collect all (resolved) toric terminal Fano threefolds of Picard rank three and four that allow for a rank one intersection lattice generated by a vector $e$ of length square $e^2=-4$. The geometries of these threefolds are again specified by their reference number MM\#${}_\rho$ and/or K\# as arising in the Mori--Mukai and/or Kasprzyk classifications \cite{MR641971,MR2221794,MR2221794db}. The resulting twisted connected sum $G_2$-manifolds~$Y^{\cdots}_{\cdots}$ obtained from orthogonal gluing along the rank one intersection lattice~$R$ all have the two-form Betti number $b_2(Y^{\cdots}_{\cdots})=1$ and their three-form Betti numbers $b_3(Y^{\cdots}_{\cdots})$ are listed in Table~\ref{tab:G2examples}. These Betti numbers are easily calculated with relations~\eqref{eq:b2b3}. 
\clearpage 

\begin{table}[t]
\hfil
\hbox{\footnotesize{
\vbox{
\offinterlineskip
\halign{\vrule width1.2pt~\strut#\hfil\vrule width0.8pt&~\hfil#\hfil~&~\hfil#\hfil~&~\hfil#\hfil~&~\hfil#\hfil~\vrule&~\hfil#\hfil~&~\hfil#\hfil~&~\hfil#\hfil~&~\hfil#\hfil~&~\hfil#\hfil~&~\hfil#\hfil~&~\hfil#\hfil~\vrule width1.2pt\cr
\noalign{\hrule height 1.2pt}
$b_3(Y^{\cdots}_{\cdots})$&MM27${}_3$&MM25${}_3$&MM31${}_3$&K124&MM12${}_4$&MM10${}_4$&K221&K232&K233&K247&K257\cr
\noalign{\hrule height 1.2pt}
MM27${}_3$ & 122 & 118 & 126 & 122 & 120 & 116 & 112 & 114 & 112 & 118 & 120 \cr
MM25${}_3$ & 118 & 114 & 122 & 118 & 116 & 112 & 108 & 110 & 108 & 114 & 116  \cr
MM31${}_3$ & 126 & 122 & 130 & 126 & 124 & 120 & 116 & 118 & 116 & 122 & 124 \cr
K124 & 122 & 118 & 126 & 122 & 120 & 116 & 112 & 114 & 112 & 118 & 120 \cr
\noalign{\hrule}
MM12${}_4$ & 120 & 116 & 124 & 120 & 118 & 114 & 110 & 112 & 110 & 116 & 118  \cr
MM10${}_4$ & 116 & 112 & 120 & 116 & 114 & 110 & 106 & 108 & 106 & 112 & 114 \cr
K221 & 112 & 108 & 116 & 112 & 110 & 106 & 102 & 104 & 102 & 108 & 110 \cr
K232 & 114 & 110 & 118 & 114 & 112 & 108 & 104 & 106 & 104 & 110 & 112 \cr
K233 & 112 & 108 & 116 & 112 & 110 & 106 & 102 & 104 & 102 & 108 & 110 \cr
K247 & 118 & 114 & 122 & 118 & 116 & 112 & 108 & 110 & 108 & 114 & 116 \cr
K257 & 120 & 116 & 124 & 120 & 118 & 114 & 110 & 112 & 110 & 116 & 118 \cr
\noalign{\hrule height 1.2pt}
}}}}
\hfil
\caption{This table shows the three-form Betti numbers $b_3(Y^{\cdots}_{\cdots})$ of the twisted connected sum $G_2$-manifolds~$Y^{\cdots}_{\cdots}$ arising from the orthogonal pushout $N_{\cdots} \perp_e N_{\cdots}$ along the rank one intersection lattice with $e^2=-4$ from all pairs of building blocks collected in Table~\ref{tab:34Blocks}. By construction of gluing along a rank one intersection lattice, all these examples have the two-form Betti numbers $b_2(Y^{\cdots}_{\cdots})=1$. The reference numbers MM\#$_\rho$ or K\# for the rows and columns label the building blocks, and the lines in the table divides between the examples with rank three and rank four Picard lattices.}
\label{tab:G2examples}
\end{table}

\paragraph{Orthogonal gluing along rank two intersection lattice:}
A systematic analysis of orthogonal pushouts for higher rank intersection lattices $R$ is beyond the scope of this work. Instead, we present a particular example with a rank two intersection lattice $R$ with two orthogonal generators $e_1$ and $e_2$ both of length square $-4$. Certainly we do not expect that orthogonality and the maximal negative value for the length squares are necessary conditions to find a higher rank example. However, imposing these two conditions certainly simplifies the construction of a matching pair. 

Our example is based upon gluing a pair of building blocks $(Z_\LR,S_\LR)$ both obtained from the rank five Fano threefold $P_\LR=\mathbb{P}^1 \times dP_3$, where $dP_3$ denotes the del~Pezzo surface of degree six, which is the blow-up of $\mathbb{P}^2$ along three non-collinear points $p_1, p_2, p_3$. This rank five Fano threefold has the Mori--Mukai reference number~MM3${}_5$ and --- as it is toric --- the Kasprzky reference number~K324. 

First, we collect some basic properties of the del Pezzo surface $dP_3$.\label{gl:dP3} Let $E_1$, $E_2$, $E_3$ be the three exceptional divisors from the blow-ups at the points $p_1,p_2,p_3$, and let $H$ be the proper transform of the hyperplane class of $\mathbb{P}^2$. These divisors span the Picard lattice of $dP_3$ and their intersection numbers read
\begin{equation}
  \dPint{E_i}{E_j} \,=\, - \delta_{ij} \ , \qquad \dPint{H}{H} \,=\, 1 \ , \qquad \dPint{H}{E_i} \,=\, 0 \ .
\end{equation}
The ample anti-canonical divisor reads $-K_{dP_3}=3H - E_1 - E_2 - E_3$.

Let us further define the two divisors
\begin{equation}
  e_1 \,=\, E_1 + E_2 + E_3 - H \ , \qquad e_2 \,=\, E_1 - E_2 \ ,
\end{equation}
which are both differences of rational curves on $dP_3$. The most important point is that the defined divisors $e_1, e_2$ of length square $-2$ are both mutually orthogonal and orthogonal to the class $-K_{dP_3}$ in the K\"ahler cone $\mathcal{K}(dP_3)$, i.e., 
\begin{equation} \label{eq:perp}
  \dPint{e_1}{e_2} = \dPint{e_1}{K_{dP_3}}=\dPint{e_2}{K_{dP_3}}=0 \ , \quad \dPint{e_1}{e_1}=\dPint{e_2}{e_2} = -2 \ .
\end{equation}

Now we return to the rank five Fano threefold $\mathbb{P}^1\times dP_3$. With the hyperplane divisor $h$ of $\mathbb{P}^1$ and the described divisors of $dP_3$, the anti-canonical divisor becomes
\begin{equation}
  - K_{\mathbb{P}^1\times dP_3} \,=\, 2 h -K_{dP_3} \,=\, 2 h + 3H - E_1 - E_2 - E_3 \ ,
\end{equation}
Furthermore, the Picard lattice $N$ of the polarized K3 surface $S$ on $\mathbb{P}^1\times dP_3$ is generated by the divisors $h, H, E_1, E_2, E_3$ together with the intersection pairing
\begin{equation} \label{eq:P1dP3pairing}
   \langle h,h \rangle_N \,=\,0 \ , \qquad  \langle h,D \rangle_N \,=\,-\dPint{K_{dP_3}}{D} \ , \qquad \langle D,F \rangle_N \,=\,2 \dPint{D}{F} \ ,
\end{equation}
where $D$ and $F$ are some divisors on $dP_3$. 

For the orthogonal pushout we generate the rank two lattice $R$ with the two del~Pezzo divisors $e_1$ and $e_2$ as
\begin{equation}
   R \,=\, \mathbb{Z} e_1 + \mathbb{Z} e_2 \ , \qquad \langle e_i , e_j \rangle_N \,=\, -4 \delta_{ij} \ ,
\end{equation}
where eqs.~\eqref{eq:perp} and \eqref{eq:P1dP3pairing} determine the intersection pairing on $R$. Moreover, the orthogonal complement $W$ of $R$ becomes
\begin{equation}
   W\,=\,\mathbb{Z} w_1 + \mathbb{Z} w_2 + \mathbb{Z} w_3 \quad\text{with}\quad w_1 \,=\, h -K_{dP_3} \ , \quad  w_2 \,=\, H - E_3 \ , \quad w_3 \,=\, h \ ,
\end{equation}   
where, in particular, the ample generator $w_1$ is in the K\"ahler cone~$\mathcal{K}(\mathbb{P}^1\times dP_3)$. As a result for the rank five Picard lattice $N$ of the polarized K3~surface $S$ in $\mathbb{P}^1\times dP_3$ we arrive with $(w_1,w_2,w_3,e_1,e_2)$ at
\begin{equation} \label{eq:NP1dP3}
   N \,=\, \left(\mathbb{Z} w_1 + \mathbb{Z} w_2 + \mathbb{Z} w_3\right) + \left(\mathbb{Z} e_1 + \mathbb{Z} e_2\right) 
   + \frac12 \left( \mathbb{Z} (w_1+e_1) + \mathbb{Z} (w_1+w_2+e_2) \right) \ . 
\end{equation}
Now taking the decomposition~\eqref{eq:NP1dP3} of the Picard lattice for both the left and the right Picard lattice, i.e., $N_L = N_R = N$, we consider the orthogonal pushout $W = N_L \perp_R N_R$, which in the basis $(w_1^L,w_2^L,w_3^L,w_1^R,w_2^R,w_3^R,e_1,e_2)$ takes the form
\begin{multline}
  W \,=\, \left(\mathbb{Z} w^L_1 + \mathbb{Z} w^L_2 + \mathbb{Z} w^L_3\right) + \left(\mathbb{Z} w^R_1 + \mathbb{Z} w^R_2 + \mathbb{Z} w^R_3\right) \\
   + \left(\mathbb{Z} e_1 + \mathbb{Z} e_2\right) + \frac12 \left( \mathbb{Z} (w_1^L+e_1) +  \mathbb{Z} (w_1^R+e_1) \right) \\
   + \frac12 \left(\mathbb{Z} (w_1^L+w_2^L+e_2) + \mathbb{Z} (w_1^R+w_2^R+e_2) \right) \ . 
\end{multline}
This orthogonal pushout is well-defined because the potentially non-integral intersections $\langle \frac12(w_1^L+e_1), \frac12(w_1^R+e_1) \rangle=\langle \frac12(w_1^L+w_2^L+ e_2), \frac12(w_1^R+w_2^R+e_2) \rangle_W=-1$ and $\langle \frac12(w_1^L+e_1), \frac12(w_1^R+w_2^R+e_2) \rangle_W = \langle \frac12(w_1^R+e_1), \frac12(w_1^L+w_2^L+e_2) \rangle_W = 0$ are integral. 

As a result we obtain, from this orthogonal pushout along the rank two lattice $R$, the twisted connected $G_2$-manifold $Y^{\text{MM3}_5}_{\text{MM3}_5}$ with the Betti numbers 
\begin{equation} \label{eq:Yintrank2}
  W^\text{MM3${}_5$}_\text{MM3${}_5$} = N_\text{MM3${}_5$} \perp_R N_\text{MM3${}_5$} \, : \quad 
  b_2(Y^{\text{MM3}_5}_{\text{MM3}_5}) =2 \ , \ b_3(Y^{\text{MM3}_5}_{\text{MM3}_5}) = 97 \ .
\end{equation} 
Here we use that $b_3(Z_\LR)=\langle K_{\mathbb{P}^1\times dP_3},K_{\mathbb{P}^1\times dP_3} \rangle_N + 2 = 6\dPint{K_{dP_3}}{K_{dP_3}}+2=38$ because $b_3(\mathbb{P}^1 \times dP_3)=0$, cf. ref.~\cite{MR641971}.

\section{$\mathcal{N}=2$ gauge sectors on twisted connected sums}  \label{sec:N=2sectors}
Compared to the Abelian $\mathcal{N}=4$ gauge theory sectors studied in the previous section, the structure of the $\mathcal{N}=2$ gauge theory sectors turns out to be much richer.  The building blocks $(Z_\LR,S_\LR)$ of the twisted connected sum $G_2$-manifolds admit enhancement to $\mathcal{N}=2$ non-Abelian gauge theory sectors with an interesting branch structure that is geometrically accessible in terms of extremal transitions in the asymptotically cylindrical Calabi--Yau threefolds $X_\LR$.

In order to see what kind of features we can expect by degenerating the building blocks  $(Z_\LR,S_\LR)$, we recall from ref.~\cite{Acharya:2001gy} that there is a simple hierarchy of real codimension four, six and seven singularities in $G_2$-manifolds, which respectively lead to non-Abelian gauge groups, non-trivial matter representations, and chirality of the charged $\mathcal{N}=1$ matter spectrum. While our setup admits non-Abelian gauge groups with non-trivial matter representations, we should not expect singularities inducing chirality as the trivial $S^1$ fibration in the non-compact seven-manifolds ~$Y_{L/R}$ prevents the appearance of codimension seven singularities.

The picture proposed in ref.~\cite{Acharya:2001gy} uses the heterotic/M-theory duality \cite{Witten:1995ex}, the Strominger--Yau--Zaslow fibration of Calabi--Yau manifolds \cite{Strominger:1996it}, and the fact that $G_2$-manifolds 
can be locally constructed as degenerating $S^1$ fibrations over Calabi-Yau threefolds~\cite{Atiyah:2000zz,Aganagic:2001ug}, where the $S^1$ can be identified in a hyper K\"ahler quotient construction starting 
in eight dimensions~\cite{Atiyah:2001qf,Acharya:2001gy}. Namely, consider the heterotic string compactification on the Calabi--Yau threefold~$W$. We further assume that the threefold~$W$ admits a geometric mirror threefold such that it has a Strominger--Yau--Zaslow Lagrangian $T^3$ fibration over a (real) three-dimensional Lagrangian cycle $Q$.\footnote{Further suggestions on the role of mirror symmetry in the context of $G_2$-manifolds have been proposed in refs.~\cite{Gukov:2002jv,Braun:2017ryx}.} In the best known examples --- such as hypersurfaces in toric varieties --- it has the topology of a three-sphere.\footnote{It is possible to obtain a Lens space for the Lagrangian base $Q$, for instance by dividing out a freely acting finite group on a suitable Calabi--Yau threefold, see e.g., ref.~\cite{Gopakumar:1997dv}.}  In the limit where the volume of the base $Q$ is large compared to the volume of the Lagrangian fibers $T^3$, the essential idea is now to adiabatically extend the duality between the heterotic string on $T^3$ and M-theory on $K3$ over the entire base $Q$. The M-theory geometries defined in this way realize the same fibration structure~\eqref{eq:K3fib} as appearing in the twisted connected sum $G_2$-manifolds. Whenever the heterotic string has a non-Abelian ADE type gauge group $G$, the dual K3~fibers develop the corresponding ADE singularity extending over the entire real three-dimensional base $Q$.  The proposed construction can be viewed as an $\mathcal{N}=1$ version of the $\mathcal{N}=2$ heterotic/type~II duality between the heterotic string on $K3 \times T^2$ and type~IIA string on the dual K3 fibered Calabi--Yau threefolds, as proposed in refs.~\cite{Kachru:1995wm,Klemm:1995tj}. In the context of twisted connected sum $G_2$-manifolds a possibility to arrive at non-Abelian ADE type gauge theories has been contemplated in ref.~\cite{Halverson:2015vta}.

While the K3 fibration described in ref.~\cite{Acharya:2001gy} complies with the K3 fibration~\eqref{eq:K3fib} of the twisted connected sum construction, we should stress that the non-Abelian gauge theory enhancement encountered in this work arises from singularities along a three-cycle $S^1\times \mathcal{C}$, where the curve $\mathcal{C}$ of genus $g$ resides in K3 fibers along a circle $S^1$ in the base $Q$. Thus, compared to ref.~\cite{Acharya:2001gy} the non-Abelian gauge group still emerges from a real codimension four singularity, however, along different types of three-cycles.
In the Kovalev limit, the three-cycle  $S^1\times\mathcal{C}$ resides in one of the seven-manifold $Y_\LR = X_\LR \times S^1_\LR$ such that the curve $\mathcal{C}$ realizes an ADE singularity in one of the asymptotically cylindrical Calabi--Yau threefolds $X_\LR$. Therefore, the non-Abelian gauge theory enhancement discussed here directly relates to non-Abelian gauge groups from curves of ADE singularities in Calabi--Yau threefolds in the context of type~IIA strings \cite{Klemm:1996kv,Katz:1996ht}. Specifically, in this setting an ADE singularity along a curve $\mathcal{C}$ yields a four-dimensional $\mathcal{N}=2$ gauge theory with the associated gauge group $G$ together with $g$ hypermultiplets in the adjoint representation. More general matter representations occur at points along $\mathcal{C}$ where the ADE singularity further enhances, i.e., along real codimension six singularities. For instance, at the intersection point of two curves $\mathcal{C}$ and $\mathcal{C}'$ of ADE singularities we encounter matter in the bi-fundamental representation of the two associated gauge groups $G$ and $G'$ \cite{Katz:1996xe}. In the following we find that the described $\mathcal{N}=2$ gauge theory spectra can indeed be realized within the $\mathcal{N}=2$ gauge theory sectors of the building blocks $(Z_\LR,S_\LR)$. Remarkably, even the phase structure of the four-dimensional $\mathcal{N}=2$ gauge theory sectors --- connecting topologically distinct Calabi--Yau threefolds via extremal transitions --- carries over to the four-dimensional $\mathcal{N}=1$ M-theory compactifications on twisted connected sum $G_2$-manifolds, where now the gauge theory branches relate topologically distinct $G_2$-manifolds.

The picture presented in refs.~\cite{Witten:2001uq,Witten:1995ex} finally explains chirality of non-Abelian matter as a local effect that occurs in codimension seven. Since the twisted connected sum breaks supersymmetry from $\mathcal{N}=2$ to $\mathcal{N}=1$ non-locally via the twisted gluing recipe, the last step --- namely to construct chiral charged matter --- is more subtle to achieve. We discuss some ideas in the conclusions.

\subsection{Phases of $\mathcal{N}=2$ Abelian gauge theory sectors} \label{sec:AbelianPhases}
Let us now focus on twisted connected sum $G_2$-manifolds with non-trivial $\mathcal{N}=2$ gauge theory sectors in the Kovalev limit. According to Table~\ref{tab:locSpec} this amounts to constructing building blocks with $(Z_\LR,S_\LR)$ with non-trivial kernels $k_\LR$ as defined below eq.~\eqref{eq:cohTwistedSum}. This can be achieved with the proposal by Kovalev and Lee \cite{MR2823130}, generalizing the construction of asymptotically cylindrical Calabi--Yau threefolds outlined in Section~\ref{sec:ACYex}. In a particular example, the possibility to realize $\mathcal{N}=2$ Abelian gauge theory enhancement appeared in ref.~\cite{Halverson:2014tya}.

For the semi-Fano threefold $P$ we pick again two global sections $s_0$ and $s_1$ of the anti-canonical divisor $-K_P$. However, instead of choosing a generic section $s_0$, we assume that the global section $s_0$ factors into a product
\begin{equation} \label{eq:sfactors}
   s_0 \,=\, s_{0,1} \cdots s_{0,n} \ ,
\end{equation}    
such that $s_{0,i}$ are global sections of line bundles $\mathcal{L}_i$ with $-K_P = \mathcal{L}_1 \otimes \ldots \otimes \mathcal{L}_n$. As a consequence, the curve $\mathcal{C}_\text{sing}=\{ s_0=0\} \cap \{s_1=0 \}$ becomes reducible and decomposes into
\begin{equation} \label{eq:Ccomp}
  \mathcal{C}_\text{sing} \,=\, \sum_{i=1}^n \mathcal{C}_i \ , \qquad \mathcal{C}_i = \{ s_{0,i}=0\} \cap\{ s_1 = 0 \} \ ,
\end{equation}
where we assume that the individual curves $\mathcal{C}_i$ are smooth and reduced. Following Kovalev and Lee \cite{MR2823130}, we construct the building block $(Z^\sharp, S)$ associated to $P$ by the sequence of blow-ups $\pi_{\{\mathcal{C}_1,\ldots,\mathcal{C}_n\}}: Z^\sharp \to P$ along the individual curves $\mathcal{C}_i$ according to
\begin{equation} \label{eq:BlowUpSequence}
   Z^\sharp \,=\, \operatorname{Bl}_{\{\mathcal{C}_1,\ldots,\mathcal{C}_n\}}\!P \,=\, 
   \operatorname{Bl}_{\mathcal{C}_n}\!\operatorname{Bl}_{\mathcal{C}_{n-1}} \cdots  \operatorname{Bl}_{\mathcal{C}_{1}}\!P \ .
\end{equation}
Since the curves $\mathcal{C}_i$ and the semi-Fano threefold $P$ are smooth, the blow-up $Z^\sharp$ is smooth as well. As before, the K3 surface~$S$ arises as the proper transform of a smooth anti-canonical divisor $S^\sharp = \{ \alpha_0s_0 + \alpha_1 s_1=0 \} \subset P$ for some $[\alpha_0 : \alpha_1] \in \mathbb{P}^1$. By blowing up a semi-Fano threefold~$P$, the resulting dimension of the kernel $k$ --- defined below eq.~\eqref{eq:cohTwistedSum} --- is then given by \cite{MR3109862}
\begin{equation} \label{eq:dimk}
  \dim k \,=\, n - 1  \ .
\end{equation}
Furthermore, the three-form Betti number $b_3(Z^\sharp)$ of the blown-up threefold $Z^\sharp$ becomes
\begin{equation} \label{eq:dimb3}
  b_3(Z^\sharp) \,=\, b_3(P) + 2 \sum_{i=1}^n g(\mathcal{C}_i) \ ,
\end{equation}
in terms of the three-form Betti number $b_3(P)$ of the semi-Fano threefold $P$ and the genera $g(\mathcal{C}_i)$ of the smooth curve components $\mathcal{C}_i$. As all these curves $\mathcal{C}_i$ lie in the K3 fiber $S$, the genus $g(\mathcal{C}_i)$ is readily computed by the adjunction formula 
\begin{equation} \label{eq:genusC}
   g(\mathcal{C}_i) \,=\, \frac12 \mathcal{C}_i.\mathcal{C}_i +1 \ ,
\end{equation}
with the self-intersections $\mathcal{C}_i.\mathcal{C}_i$ in $S$.    

Corti et al. show in refs.~\cite{MR3369307,MR3109862} that, with such a building block $(Z^\sharp,S^\sharp)$, the orthogonal gluing recipe of Section~\ref{sec:gluing} can still be carried out in the same way. In particular, the lift from a matching pair of K3~surfaces $S_\LR$ to their asymptotically cylindrical Calabi--Yau threefolds $X_\LR$ still exists if such generalized building blocks are involved in the orthogonal gluing procedure.

Thus, we observe that, from a single semi-Fano threefold $P$, often several building blocks can be constructed depending on the properties of the curve $\mathcal{C}=\{ s_0=0\} \cap\{ s_1=0 \}$. Namely, for smooth irreducible curves $\mathcal{C}^\flat$, we obtain a smooth building block $(Z^\flat,S^\flat)$ with vanishing kernel $k$, whereas for reducible curves $\mathcal{C}_\text{sing}$ with smooth components $\mathcal{C}_i$ we arrive, with the sequence of blow-ups~\eqref{eq:BlowUpSequence}, at a smooth building block $(Z^\sharp,S^\sharp)$ with non-vanishing kernel $k$. The former building block does not contribute with any vector multiplets to the $\mathcal{N}=2$ gauge theory sector, while the latter building block contributes with Abelian vector multiplets to the $\mathcal{N}=2$ gauge theory sector. In the sequel we argue that these different possibilities realize distinct branches of the $\mathcal{N}=2$ gauge theory sectors.

\begin{table}[t]
\hfil
\hbox{
\vbox{
\offinterlineskip
\halign{\vrule height2.7ex depth1.2ex width1.2pt\hfil#\hfil\vrule width0.8pt&\hfil#\hfil&\vrule\hfil#\hfil&#\vrule width0.8pt&\hfil#\hfil&\vrule\hfil#\hfil&#\vrule width1.2pt\cr 
\noalign{\hrule height 1.2pt}
\,Multiplicity\,&\multispan2\hfil$\mathcal{N}=2$ multiplets\hfil&&\multispan2\hfil$\mathcal{N}=1$ multiplets\hfil&\cr
&$U(1)^{n-1}$ charges&\,multiplet\,&&$U(1)^{n-1}$ charges&\,multiplet\,&\cr
\noalign{\hrule height 1.2pt}
$n-1$&$(0,0,\ldots,0)$&vector&&$(0,\ldots,0)$&vector&\cr 
&&&&$(0,\ldots,0)$&chiral&\cr
\noalign{\hrule height 0.8pt}
$\chi_{ij}$&\,{\footnotesize{$(0,\ldots,+1_i,\ldots,+1_j,\ldots,0)$}}\,&hyper&&\,{\footnotesize{$(0,\ldots,+1_i,\ldots,+1_j,\ldots,0)$}}\,&chiral&\cr 
{\footnotesize{$1\le i < j < n$}}&&&&\,{\footnotesize{$(0,\ldots,-1_i,\ldots,-1_j,\ldots,0)$}}\,&chiral&\cr
\noalign{\hrule}
$\chi_{in}$&$(0,\ldots,+1_i,\ldots,0)$&hyper&&$(0,\ldots,+1_i,\ldots,0)$&chiral&\cr
$1\le i < n$&&&&$(0,\ldots,-1_i,\ldots,0)$&chiral&\cr
\noalign{\hrule height 1.2pt}
}
}}
\hfil
\caption{The table shows the spectrum of the Abelian $\mathcal{N}=2$ gauge theory sector arising from the conifold singularities in the building block $(Z_\text{sing},S)$. Listed are the four-dimensional $\mathcal{N}=2$ multiplets and their decomposition into the four-dimensional $\mathcal{N}=1$ multiplets together with their multiplicities $\chi_{ij}$. The subscripts of the entries of the $U(1)$ charges indicate their position in the charged vector.}
\label{tab:SpecAbelian}
\end{table}

To arrive at this gauge theory interpretation, let us consider a semi-Fano threefold $P$ with a curve $\mathcal{C}_\text{sing}$ of the reducible type \eqref{eq:Ccomp} with the factorized global anti-canonical section~\eqref{eq:sfactors}. Performing a blow-up along this reducible curve yields the fibration $\pi: Z_\text{sing} \to \mathbb{P}^1$ with
\begin{equation} \label{eq:ZsingAb}
   Z_\text{sing} \,=\, \operatorname{Bl}_{\mathcal{C}_\text{sing}}\!P \,=\, \left\{ (x,z) \in P \times \mathbb{P}^1 \,\middle|\, z_0 s_{0,1} \cdots s_{0,n} + z_1 s_1 = 0 \right\} \ .
\end{equation}
In the vicinity of the fiber $\pi^{-1}([1,0])$, the threefold $Z_\text{sing}$ becomes singular because in the patch of the affine coordinate $t = \frac{z_1}{z_0}$ we get
\begin{equation} \label{eq:sing1}
    s_{0,1} \cdots s_{0,n} + t s_1 = 0 \ .
\end{equation}
Thus --- assuming transverse intersections among the smooth curves $\mathcal{C}_i$ --- there are conifold singularities at the discrete intersection loci $\mathcal{I}_{ij} \,=\,\{ t =0\}\cap\{ s_1=0\} \cap \{ s_{0,i}=0\} \cap \{ s_{0,j}=0 \}$ for $1\le i \le j \le n$ with $\chi_{ij} = | \mathcal{I}_{ij} |$ intersection points. These numbers are given by
\begin{equation} \label{eq:intchi}
   \chi_{ij} \,=\, \mathcal{C}_i.\mathcal{C}_j \ ,
\end{equation}
in terms of the intersection numbers of the reduced curves $\mathcal{C}_i$ and $\mathcal{C}_j$ within the K3 surface~$S$. This singularity structure prevails in the asymptotically cylindrical Calabi--Yau threefold $X_\text{sing} = Z_\text{sing} \setminus S$ since the asymptotic fiber $S=\pi^{-1}([\alpha_0,\alpha_1])$ (for $\alpha_1\ne 0$) is disjoint from the singular fiber $\pi^{-1}([1,0])$.

In the vicinity of the singular fiber $\pi^{-1}([1,0]) \subset X_\text{sing}$, we interpret the dimensional reduction of M-theory on the local seven-dimensional singular space $S^1 \times X_\text{sing}$ as the dimensional reduction of type IIA~string theory on the asymptotically cylindrical Calabi--Yau threefold $X_\text{sing}$, where the $S^1$~factor corresponds to the M-theory circle of type~IIA string theory. In this IIA~picture, refs.~\cite{Strominger:1995cz,Greene:1995hu} establish that the conifold singularities~\eqref{eq:sing1} yield an Abelian $\mathcal{N}=2$ gauge theory with charged matter multiplets.\footnote{Starting from the Fano threefold~$\mathbb{P}^3$, it has also been proposed in ref.~\cite{Halverson:2014tya} that the singular building block $(Z_\text{sing},S)$ with conifold singularities realizes an Abelian gauge theory with charged matter.} Namely, to each curve $\mathcal{C}_i$ we assign an Abelian group factor $U(1)_i$ such that the total Abelian gauge group of rank $n-1$ becomes
\begin{equation} \label{eq:GAb}
   U(1)^{n-1} \,\simeq\, \frac{ U(1)_1 \times \ldots \times U(1)_n}{U(1)_\text{Diag}} \ ,
\end{equation} 
where $U(1)_\text{Diag}$ is the diagonal subgroup of $U(1)_1\times\ldots\times U(1)_n$. Thus, in the low-energy effective theory, we obtain $(n-1)$ four-dimensional $\mathcal{N}=2$ $U(1)$ vector multiplets, which decomposes into $(n-1)$ four-dimensional $\mathcal{N}=1$ $U(1)$ vector multiplets and $(n-1)$ four-dimensional $\mathcal{N}=1$ neutral chiral multiplets. Furthermore, to each intersection point in $\mathcal{I}_{ij}$ one assigns a four-dimensional $\mathcal{N}=2$ hypermultiplet of charge $(+1,+1)$ with respect to the $U(1)_i \times U(1)_j$ group factor. Then each of these $\mathcal{N}=2$ hypermultiplet of charge $(+1,+1)$ decomposes into two four-dimensional $\mathcal{N}=1$ chiral multiplets of charge $(+1,+1)$ and $(-1,-1)$, respectively. We summarize the resulting spectrum in Table~\ref{tab:SpecAbelian}.

Alternatively, we can dimensionally reduce M-theory on the local Calabi--Yau fourfold~$T^2 \times X_\text{sing}$ to three space-time dimensions. Then the conifold points in $X_\text{sing}$ become genus one curves of conifold singularities. This analysis has been carried out in ref.~\cite{Intriligator:2012ue}, and one arrives at three-dimensional $\mathcal{N}=4$ gauge theory sectors, which agree with the four-dimensional $\mathcal{N}=2$ spectrum in Table~\ref{tab:SpecAbelian} upon further dimensional reduction on a circle $S^1$.\footnote{Note that eight supercharges correspond to $\mathcal{N}=4$ supersymmetry in three dimensions and to $\mathcal{N}=2$ supersymmetry in four dimensions.} This further justifies that the local dimensional reduction of type~IIA theory on $X_\text{sing}$ correctly describes the gauge theory of M-theory on $S^1\times X_\text{sing}$ without requiring that the $S^1$ factor realizes the M-theory circle for the dual type~IIA description. 

The described four-dimensional $\mathcal{N}=2$ Abelian gauge theory now predicts a Higgs branch $H^\flat$ and a Coulomb branch $C^\sharp$. On the one hand, generic non-vanishing expectation values of the charged hypermultiplets break the $U(1)^{n-1}$ gauge theory entirely and parametrize the Higgs branch $H^\flat$ of the gauge theory. As a consequence $(n-1)$ charged $\mathcal{N}=2$ hypermultiplets play the role of $\mathcal{N}=2$ Goldstone multiplets that combine with the $(n-1)$ short massless $\mathcal{N}=2$ vector multiplets to $(n-1)$ long massive $\mathcal{N}=2$ vector multiplets. As a result --- according to the spectrum in Table~\ref{tab:SpecAbelian} --- we arrive at the Higgs branch $H^{\flat}$ of complex dimension $h^\flat$ \cite{Greene:1995hu}
\begin{equation} \label{eq:dimhAb}
    h^\flat \,=\, \dim_\mathbb{C} H^{\flat} \,=\,2 \left(\sum_{1\le i < j \le n} \chi_{ij} \right) - 2 (n-1) \ .
\end{equation}
Here the factor two takes into account that each hypermultiplet contains two complex scalar fields. This complex dimension readily describes the Higgs branch as parametrized by the expectation values of the corresponding charged $\mathcal{N}=1$ chiral multiplets. On the other hand, the expectation values of the neutral complex scalar fields in the $\mathcal{N}=2$ vector multiplets furnish the coordinates on the Coulomb branch $C^\sharp$ such that its complex dimension $c^\sharp$ reads
\begin{equation} \label{eq:dimcAb}
  c^\sharp \,=\, \dim_\mathbb{C} C^{\sharp} \,=\, n-1 \ .
\end{equation}  
In the $\mathcal{N}=1$ language, the Coulomb branch moduli space is parametrized by the expectation value of neutral $\mathcal{N}=1$ chiral multiplets.

In the geometry, the Higgs branch $H^\flat$ arises from deforming the conifold singularities in $X_\text{sing}$ to the deformed asymptotically cylindrical Calabi--Yau threefold $X^\flat$ \cite{Greene:1995hu}. On the level of the semi-Fano threefold $P$, this amounts to deforming the reducible curve $\mathcal{C}_\text{sing}$ in eq.~\eqref{eq:Ccomp} to the smooth reduced curve $\mathcal{C}^\flat$ such that the building block $(Z_\text{sing},S)$ deforms to the building block $(Z^\flat,S^\flat)$. According to eqs.~\eqref{eq:dimk} and \eqref{eq:dimb3}, this yields for the kernel $k^\flat$ and the three-form Betti number $b_3(Z^\flat)$
\begin{equation} \label{eq:kbflat}
  \dim k^\flat \,=\, 0 \ , \qquad b_3(Z^\flat) \,=\, b_3(P) + \mathcal{C}^\flat.\mathcal{C}^\flat +2 \ .
\end{equation}
Furthermore, the resolution of the conifold singularities in $X_\text{sing}$ geometrically yields the Coulomb branch $C^\sharp$ of the gauge theory \cite{Greene:1995hu}, which again on the level of the semi-Fano threefold $P$ realizes the sequential blow-ups \eqref{eq:BlowUpSequence} along the components~$\mathcal{C}_i$ of $\mathcal{C}_\text{sing}$ to the building block $(Z^\sharp,S^\sharp)$.\footnote{The sequential resolution~\eqref{eq:BlowUpSequence} in the Coulomb branch depends on the order of the performed blow-ups. Changing the order geometrically realizes bi-rational transformations among the resulting asymptotically cylindrical Calabi--Yau threefolds~$X^\sharp$, as discussed in ref.~\cite{Halverson:2014tya}.} With eqs.~\eqref{eq:dimk} and \eqref{eq:dimb3}, the dimension of its kernel $k^\sharp$ and the Betti number $b_3(Z^\sharp)$ becomes
\begin{equation} \label{eq:kbsharp}
  \dim k^\sharp \,=\, n-1 \ , \qquad b_3(Z^\sharp) \,=\, b_3(P) + 2n + \sum_{i=1}^n \mathcal{C}_i.\mathcal{C}_i \ .
\end{equation}
Let us now consider two twisted connected sum $G_2$-manifolds $Y^\flat$ and $Y^\sharp$ respectively constructed via orthogonal gluing of the left building blocks $(Z^\flat,S^\flat)$ and $(Z^\sharp,S^\sharp)$ with another right building block $(Z_R,S_R)$. Then we can use eq.~\eqref{eq:b2b3} to deduce from eqs.~\eqref{eq:kbflat} and \eqref{eq:kbsharp} the Betti numbers
\begin{equation} \label{eq:Yflatb2b3}
  b_2(Y^\flat) \,=\, \delta^{(2)}_R \ , \quad
  b_3(Y^\flat) \,=\,  b_3(P) + \mathcal{C}^\flat.\mathcal{C}^\flat + 25 + \delta^{(3)}_R \ ,
\end{equation}  
and
\begin{equation}
  b_2(Y^\sharp) \,=\,  (n - 1)+ \delta^{(2)}_R \ , \quad
  b_3(Y^\sharp) \,=\,  b_3(P) + \left(\sum_{i=1}^n \mathcal{C}_i.\mathcal{C}_i\right) + 3n + 22 + \delta^{(3)}_R \ ,
\end{equation}  
with the contributions from the right building block $(Z_R,S_R)$
\begin{equation} \label{eq:defdelta}
    \delta^{(2)}_R \,=\,  \dim k_R +  \operatorname{rk} R\ , \qquad  \delta^{(3)}_R \,=\,  b_3(Z_R) + \dim k_R - \operatorname{rk} R \ .
\end{equation}
The relations allow us to further define what we call the reduced Betti numbers $b^\flat_\ell$ and $b^\sharp_\ell$, $\ell=1,2$, given by
\begin{equation} \label{eq:redBetti}
  b_\ell^\flat \,=\, b_\ell(Y^\flat) - \delta^{(\ell)}_R \ , \qquad b_\ell^\sharp \,=\, b_\ell(Y^\sharp) - \delta^{(\ell)}_R \ , \qquad \ell=1,2 \ .
\end{equation}
Using the equivalence $\mathcal{C}^\flat \sim \mathcal{C}_1 + \ldots +\mathcal{C}_n$ on the semi-Fano threefold $P$ and the definition~\eqref{eq:intchi} of the multiplicities $\chi_{ij}$, we finally arrive at
\begin{equation} \label{eq:DeltaBetti}
\begin{aligned}
  b_2(Y^\flat) \,&=\, b_2(Y^\sharp) - (n-1) \ , \\
  b_3(Y^\flat) \,&=\, b_3(Y^\sharp) + 2 \left(\sum_{1\le i < j \le n} \chi_{ij} \right) - 3 (n -1) \ .
\end{aligned}  
\end{equation}  

The non-trivial result is now that the derived change in Betti numbers~\eqref{eq:DeltaBetti} between such twisted connected sum $G_2$-manifolds is in perfect agreement with the  phase structure of the proposed $U(1)^{n-1}$ gauge theory. The change in the Betti number $b_2$ geometrically realizes the difference of massless four-dimensional $\mathcal{N}=1$ vector multiplets, whereas the change of the  Betti number $b_3$ geometrically realizes the difference of four-dimensional $\mathcal{N}=1$ chiral multiplets. This is in agreement with the gauge theory expectation. Passing from the Coulomb branch $C^\sharp$ to the Higgs branch $H^\flat$ via the Higgs mechanism reduces the vector bosons by the rank $(n-1)$ of the gauge group. Furthermore, the difference in the four-dimensional $\mathcal{N}=1$ chiral multiplets agrees with the change in dimension of the moduli space of these gauge theory phases, i.e., 
\begin{equation}
    b_3(Y^\flat) -  b_3(Y^\sharp) \,=\, b_3^\flat - b_3^\sharp \,=\, h^\flat - c^\sharp \ .
\end{equation}

In a similar fashion, it is straightforward to establish the correspondence between the gauge theory and the geometry for mixed Higgs--Coulomb branches, where the gauge group $U(1)^{n-1}$ is broken to a subgroup $U(1)^{k-1}$ with $1<k<n$. The geometries of such mixed phases are obtained by partially resolving and partially deforming the conifold singularities in the asymptotically cylindrical Calabi--Yau threefold $X_\text{sing}$. We illustrate the analysis of mixed Higgs--Coulomb branches with an explicit example in Section~\ref{sec:GaugeExample}.

\subsection{Phases of $\mathcal{N}=2$ non-Abelian gauge theory sectors} \label{sec:NonAbelianPhases}
Let us now turn to the enhancement to non-Abelian $\mathcal{N}=2$ gauge theory sectors in the context of twisted connected sum $G_2$-manifolds, indicated as a possibility in ref.~\cite{Halverson:2015vta}. Let us assume that the anti-canonical line bundle $-K_P$ of the semi-Fano threefold $P$ factors as
\begin{equation}
    -K_P \,=\, \tilde{\mathcal{L}}_1^{\otimes k_1} \otimes \ldots \otimes \tilde{\mathcal{L}}_s^{\otimes k_s} \quad \text{with} \quad n = k_1 + \ldots + k_s \ ,
\end{equation}
where $\tilde{\mathcal{L}}_i$ are line bundles with global sections $\tilde s_{0,i}$. Then the global section $s_0$ of $-K_P$ can further degenerate to $s_0=\tilde s_{0,1}^{k_1} \cdots \tilde s_{0,s}^{k_s}$ and the singular building block~\eqref{eq:ZsingAb} reads
\begin{equation} \label{eq:ZsingNonAb}
   Z_\text{sing} \,=\, \left\{ (x,z) \in P \times \mathbb{P}^1 \,\middle|\, z_0 \tilde s_{0,1}^{k_1} \cdots \tilde s_{0,s}^{k_s} + z_1 s_1 = 0 \right\} \ ,
\end{equation}
with the singular equation in the affine coordinate $t=\frac{z_1}{z_0}$ given by
\begin{equation} \label{eq:sing2}
    \tilde s_{0,1}^{k_1} \cdots \tilde s_{0,s}^{k_s} + t s_1 = 0 \ .
\end{equation}
As before, we assume that all curves $\tilde{\mathcal{C}}_i = \{ \tilde s_{0,i}=0 \} \cap \{ s_1 = 0 \}$ are smooth. In the vicinity of the singular fiber $\pi^{-1}([1,0])\subset Z_\text{sing}$, the singular building block $(Z_\text{sing},S)$ develops $A_{k_i-1}$-singularities along those curves $\tilde{\mathcal{C}}_i$ with $k_i>1$. 

To arrive at the gauge theory description, we again analyze the local M-theory geometry on $S^1\times X_\text{sing}$ in terms of its dual type~IIA picture on the asymptotically cylindrical Calabi-Yau threefold $X_\text{sing}$. Refs.~\cite{Klemm:1996kv,Katz:1996ht} establish that type~IIA string theory on Calabi--Yau threefolds with a genus $g$ curve of $A_{k-1}$ singularities develops a $\mathcal{N}=2$ $SU(k)$ gauge theory with $g$ four-dimensional $\mathcal{N}=2$ hypermultiplets in the adjoint representation of $SU(N)$. Furthermore, according to ref.~\cite{Katz:1996xe}, each intersection point of two such curves of $A_{k_1-1}$ and $A_{k_2-1}$ singularities contributes a four-dimensional $\mathcal{N}=2$ hypermultiplet in the bi-fundamental representation $(\mathbf{k_1},\mathbf{k_2})$ of $SU(k_1)\times SU(k_2)$. 

\begin{table}[t]
\hfil
\hbox{
\vbox{
\offinterlineskip
\halign{\vrule height2.7ex depth1.2ex width1.2pt\hfil#\hfil\vrule width0.8pt&\hfil#\hfil&\vrule\hfil#\hfil&#\vrule width0.8pt&\hfil#\hfil&\vrule\hfil#\hfil&#\vrule width1.2pt\cr 
\noalign{\hrule height 1.2pt}
\,Multiplicity\,&\multispan2\hfil$\mathcal{N}=2$ multiplets\hfil&&\multispan2\hfil$\mathcal{N}=1$ multiplets\hfil&\cr
&\,$G$ reps.\,&\,multiplet\,&&\,$G$ reps.\,&\,multiplet\,&\cr
\noalign{\hrule height 1.2pt}
$s-1$&$\mathbf{1}$&\,$U(1)$ vector\,&&$\mathbf{1}$&\,$U(1)$ vector\,&\cr 
&&&&$\mathbf{1}$&chiral&\cr
\noalign{\hrule}
$i=1,\ldots,s$&$\mathbf{adj}_{SU(k_i)}$&\,$SU(k_i)$ vector\,&&$\mathbf{adj}_{SU(k_i)}$&\,$SU(k_i)$ vector\,&\cr 
&&&&$\mathbf{adj}_{SU(k_i)}$&chiral&\cr
\noalign{\hrule height 0.8pt}
$g(\tilde{\mathcal{C}}_i)$&\,$\mathbf{adj}_{SU(k_i)}$\,&hyper&&\,$\mathbf{adj}_{SU(k_i)}$\,&chiral&\cr 
{\footnotesize{$1\le i \le s$}}&&&&\,$\mathbf{adj}_{SU(k_i)}$\,&chiral&\cr
\noalign{\hrule}
$\tilde\chi_{ij}$&\,$(\mathbf{k_i},\mathbf{k_j})_{(+1_i,+1_j)}$\,&hyper&&\,$(\mathbf{k_i},\mathbf{k_j})_{(+1_i,+1_j)}$\,&chiral&\cr 
{\footnotesize{$1\le i < j < s$}}&&&&\,$(\mathbf{\bar k_i},\mathbf{\bar k_j})_{(-1_i,-1_j)}\,$\,&chiral&\cr
\noalign{\hrule}
$\tilde\chi_{is}$&\,$(\mathbf{k_i},\mathbf{k_s})_{(+1_i)}$\,&hyper&&\,$(\mathbf{k_i},\mathbf{k_s})_{(+1_i)}$\,&chiral&\cr 
$1\le i < s$&&&&\,$(\mathbf{\bar k_i},\mathbf{\bar k_s})_{(-1_i)}\,$\,&chiral&\cr
\noalign{\hrule height 1.2pt}
}
}}
\hfil
\caption{The table shows the spectrum of the $\mathcal{N}=2$ gauge theory sector with gauge group $G=SU(k_1) \times \ldots \times SU(k_s) \times U(1)^{s-1}$ as arising from the non-Abelian building blocks $(Z_\text{sing},S)$. It lists both the four-dimensional $\mathcal{N}=2$ and the four-dimensional $\mathcal{N}=1$ multiplet structure. The adjoint matter is determined by the genus $g(\tilde{\mathcal{C}}_i)$ of the curves $\tilde{\mathcal{C}}_i$, whereas the bi-fundamental matter is determined by their intersection numbers $\tilde\chi_{ij}$ within the K3~surface $S$.}
\label{tab:SpecNonAbelian}
\end{table}

Therefore --- putting all these ingredients together and including the $U(1)$ gauge theory factors of the previously discussed Abelian gauge theory sectors --- we propose the following non-Abelian gauge theory description for M-theory on the local singular seven space $S^1\times X_\text{sing}$. Firstly, the singularities along the curves $\tilde{\mathcal{C}}_i$ determine the gauge group 
\begin{equation} \label{eq:GnonAb}
   G\,=\,SU(k_1) \times \ldots \times SU(k_s) \times U(1)^{s-1} \simeq \frac{ U(k_1) \times \ldots \times U(k_s) }{ U(1)_\text{Diag}} \ ,
\end{equation}  
where any $SU(1)$ factors must be dropped out and $U(1)_\text{Diag}$ is the diagonal subgroup of $U(k_1) \times \ldots \times U(k_s)$. Secondly, for any $i$ with $k_i>0$, there are $g(\tilde{\mathcal{C}}_i)$ four-dimensional $\mathcal{N}=2$ hypermultiplets in the adjoint representation of $SU(k_i)$. Thirdly, we have $\tilde \chi_{ij}$ four-dimensional $\mathcal{N}=2$ hypermultiplets in the bi-fundamental representation $(\mathbf{k_i},\mathbf{k_j})_{(+1_i,+1_j)}$ of the gauge group factors $SU(k_i)\times SU(k_j)$, where the subscripts indicate the $U(1)$-charges with respect to the diagonal $U(1)_i$ and $U(1)_j$ subgroups of the respective unitary groups $U(k_i)$ and $U(k_j)$ in the relation~\eqref{eq:GnonAb}. The multiplicities $\tilde\chi_{ij}$ are again determined by the intersection numbers of the curves $\tilde{\mathcal{C}}_{i}$ and $\tilde{\mathcal{C}}_j$ in the K3 fiber $S$. The resulting gauge theory spectrum is summarized in Table~\ref{tab:SpecNonAbelian}.

From the described spectrum and the results of ref.~\cite{Katz:1996ht}, we are now ready to analyze the branches of the $\mathcal{N}=2$ gauge theory sectors. First, we determine the complex dimension $h^\flat$ of the Higgs branch
\begin{equation} \label{eq:HnonAb}
   h^\flat \,=\, \dim_\mathbb{C} H^\flat \,=\, 2 \left( \sum_{i=1}^s (g(\tilde{\mathcal{C}}_i) - 1) (k_i^2 -1) \right) + 2 \left( \sum_{1\le i< j \le s} \tilde\chi_{ij} k_i k_j  \right) 
   - 2 ( s-1) \ .
\end{equation}
Here, the first term captures the $2(k_i^2 -1)$ complex degrees of freedom of the four-dimensional $\mathcal{N}=2$ hypermultiplets in the corresponding adjoint representations of the $SU(k_i)$ gauge group factors --- reduced by one adjoint $\mathcal{N}=2$ Goldstone hypermultiplet rendering the four-dimensional $\mathcal{N}=2$ $SU(k_i)$ vector multiplet massive. The second term realizes the complex degrees of freedom of the four-dimensional $\mathcal{N}=2$ matter hypermultiplets in the bi-fundamental representations of the associated special unitary gauge groups and charged with respect to the appropriate $U(1)$ factors. The last term subtracts from the second term the $\mathcal{N}=2$ Goldstone hypermultiplets for higgsing the $(s-1)$ four-dimensional $\mathcal{N}=2$ $U(1)$ vector multiplets.

Next, we derive the complex dimension of the Coulomb branch $C^\flat$, in which the maximal Abelian subgroup $U(1)^{n-1}$ remains unbroken. It is parametrized by the expectation value of all four-dimensional $\mathcal{N}=2$ hypermultiplet components that are neutral with respect to this unbroken maximal Abelian subgroup. Therefore, the complex dimension $c^\sharp$ of the Coulomb branch becomes
\begin{equation} \label{eq:CnonAb}
  c^\sharp \,=\,  \dim_\mathbb{C} C^\sharp \,=\, 2 \left( \sum_{i=1}^s g(\tilde{\mathcal{C}}_i) (k_i -1) \right) + (n-1) \ .
\end{equation}
The first term counts the traceless neutral diagonal degrees of freedom of the four-dimensional $\mathcal{N}=2$ matter hypermultiplets in the adjoint representation, while the second term adds the contributions of the complex scalar fields in the four-dimensional unbroken Abelian $\mathcal{N}=2$ vector multiplets.

The next task is to compute the Betti numbers of the twisted connected sum $G_2$-manifolds $Y^\flat$ and $Y^\sharp$, which geometrically realize the Higgs and Coulomb branch by orthogonal gluing of the building blocks $(Z^\flat,S^\flat)$ and $(Z^\sharp,S^\sharp)$ to a common right building block $(Z_R,S_R)$. We construct the building block $(Z^\flat,S^\flat)$ by blowing-up the semi-Fano threefold~$P$ along the smooth irreducible curve $\mathcal{C}^\flat$, which --- as in the Higgs branch of the Abelian gauge theories --- arises from a generic deformation of the section $s_0$ of the anti-canonical line bundle $-K_P$. Then, relations~\eqref{eq:Yflatb2b3} determine again the two-form and three-form Betti numbers of the $G_2$-manifold $Y^\flat$. The smooth Coulomb branch building block $(Z^\sharp,S^\sharp)$ results from the sequence of $n=k_1+ \ldots + k_s$ blow-ups
\begin{equation}
  Z^\sharp \,=\, \operatorname{Bl}_{\{\tilde{\mathcal{C}}_1^{k_1},\ldots,\tilde{\mathcal{C}}_s^{k_s}\}}\!P \ ,
\end{equation}
where each individual curve~$\tilde{\mathcal{C}}_i$ is resolved $k_i$ times such that $\dim k^\sharp = n -1$. Therefore, using eqs.~\eqref{eq:b2b3}, \eqref{eq:dimb3} and \eqref{eq:genusC}, we arrive at the Betti numbers for the smooth $G_2$-manifold $Y^\sharp$
\begin{equation}
  b_2(Y^\sharp) \,=\,(n - 1)+ \delta^{(2)}_R \ , \quad 
  b_3(Y^\sharp) \,=\,  b_3(P) + \left(\sum_{i=1}^s k_i\,\tilde{\mathcal{C}}_i.\tilde{\mathcal{C}}_i\right)  + 3n + 22 + \delta^{(3)}_R\ ,
\end{equation}  
with the definitions~\eqref{eq:defdelta}. Using the equivalence relation $\mathcal{C}^\flat \sim k_1 \tilde{\mathcal{C}}_1 + \ldots + k_s \tilde{\mathcal{C}}_s$, we calculate the change of Betti numbers
\begin{equation} \label{eq:DeltaBetti2}
\begin{aligned}
  b_2(Y^\flat) \,&=\, b_2(Y^\sharp) - (n-1) \ , \\
  b_3(Y^\flat) \,&=\, b_3(Y^\sharp) +\left( \sum_{i=1}^s \tilde\chi_{ii} k_i(k_i-1) \right)+  2 \left(\sum_{1\le i < j \le s} \tilde\chi_{ij}k_i k_j \right) - 3 (n -1) 
\end{aligned}  
\end{equation} 
in terms of the intersection numbers $\tilde\chi_{ij}=\tilde{\mathcal{C}}_i.\tilde{\mathcal{C}}_j$ on the K3~surface~$S$.

As for the Abelian gauge theory sectors, the computed change of Betti numbers is also in accord with the phase structure of the proposed non-Abelian gauge theory description. Namely, the change of the two-form Betti number conforms with the difference of the four-dimensional $\mathcal{N}=1$ vector multiplets in the Higgs and Coulomb branches, given by the rank of the non-Abelian gauge group~\eqref{eq:GnonAb}. The difference of four-dimensional $\mathcal{N}=1$ chiral multiplets is accurately predicted by the complex dimensions of the Higgs and Coulomb branches. That is to say that, with eqs.~\eqref{eq:genusC}, \eqref{eq:HnonAb} and \eqref{eq:CnonAb}, we find for the discussed non-Abelian gauge theories
\begin{equation}
    b_3(Y^\flat) -  b_3(Y^\sharp) \,=\, b_3^\flat - b_3^\sharp \,=\, \dim_\mathbb{C} H^\flat - \dim_\mathbb{C} C^\sharp \ .
\end{equation}

As for the previously discussed Abelian gauge theories, the established correspondence between $G_2$-manifolds and non-Abelian Higgs and Coulomb branches carries over to mixed Higgs--Coulomb branches as well, which we illustrate with an explicit example in Section~\ref{sec:GaugeExample}. The fact that the performed analysis of the non-Abelian gauge theory sectors closely parallels the study of the Abelian gauge theories does not come as a surprise, because the Abelian gauge group~\eqref{eq:GAb} arises from partially higgsing the non-Abelian gauge group~\eqref{eq:GnonAb} to its maximal Abelian subgroup. As a result, the topological data of the $G_2$-manifolds for the Higgs, Coulomb and mixed Higgs--Coulomb phases resulting from a given semi-Fano threefold $P$ are the same for both the discussed Abelian and non-Abelian gauge theory sectors.

\subsection{Examples of $G_2$-manifolds with $\mathcal{N}=2$ gauge theories} \label{sec:GaugeExample}
Following the general discussion of $\mathcal{N}=2$ gauge theory sectors in Section~\ref{sec:AbelianPhases} and Section~\ref{sec:NonAbelianPhases}, we now illustrate the emergence of $\mathcal{N}=2$ gauge theory sectors in twisted connected sum $G_2$-manifolds with a few explicit examples: 
\paragraph{$SU(4)$ gauge theory with adjoint matter from the Fano threefold $\mathbb{P}^3$:}
Consider the Fano threefold $\mathbb{P}^3$ with the anti-canonical divisor $-K_{\mathbb{P}^3} = 4 H$ in terms of the hyperplane class $H$. Let $\tilde s_{0,1}$ and $s_1$ be a (generic) global section of $H$ and $-K_{\mathbb{P}^3}$, respectively. Then we obtain, with eq.~\eqref{eq:ZsingNonAb}, the resolved building block $Z_\text{sing} \subset \mathbb{P}^3\times\mathbb{P}^1$ as the hypersurface equation
\begin{equation} \label{eq:HP3}
   \tilde s_{0,1}^4 + t s_1 \,=\, 0 \ ,
\end{equation}
with the affine coordinate $t$ of the factor $\mathbb{P}^1$. This equation exhibits an $A_3$ singularity along the curve $\tilde{\mathcal{C}}_1 = \{ \tilde s_{0,1} = 0 \} \cap \{ s_1 = 0 \} \cap \{ t = 0 \}$, which yields a $\mathcal{N}=2$ gauge theory sector with gauge group $SU(4)$. Note that for this particular example the deformed phases of the non-enhanced $\mathcal{N}=2$ Abelian gauge theory sector are discussed as well in ref.~\cite{Halverson:2014tya}.

We first note that the curves $\mathcal{C}^{(k)} = (-K_{\mathbb{P}^3}) \cap (k H)$ have the following intersection numbers on the K3~surface $S$ and --- according to eq.~\eqref{eq:genusC} --- genera
\begin{equation} \label{eq:P3CurveData}
   \mathcal{C}^{(k)}.\mathcal{C}^{(l)} \,=\, 4 k l \ , \qquad
   g(\mathcal{C}^{(k)}) \,=\, \frac12 \mathcal{C}^{(k)}.\mathcal{C}^{(k)}+1 \,=\, 2k^2 +1 \ .
\end{equation}   
Due to the equivalence $\tilde{\mathcal{C}}_1 \sim \mathcal{C}^{(k)}$, we arrive at $g(\tilde{\mathcal{C}}_1) = 3$ four-dimensional $\mathcal{N}=2$ hypermultiplets in the adjoint representation of $SU(4)$. This spectrum predicts with eqs.~\eqref{eq:HnonAb} and \eqref{eq:GnonAb} the dimensions of the Higgs and Coulomb branches
\begin{equation} \label{eq:P3HC}
  \dim_\mathbb{C} H^\flat \,=\, 60 \ , \qquad \dim_\mathbb{C} C^\sharp \,=\, 21 \ , \qquad \dim_\mathbb{C} H^\flat - \dim_\mathbb{C} C^\sharp \,=\, 39 \ .
\end{equation}
As proposed in Section~\ref{sec:NonAbelianPhases}, by sequentially blowing-up $\mathbb{P}^3$ four times along the curve~$\tilde{\mathcal{C}}_1$, we arrive at the building block $(Z^\sharp,S^\sharp)$ with
\begin{equation}
   \dim k^\sharp \,=\, 3 \ , \qquad b_3(Z^\sharp) \,=\, 4 \cdot 2   g(\tilde{\mathcal{C}}_1) \,=\, 24 \ .
\end{equation}   
Deforming the hypersurface equation~\eqref{eq:HP3} to $s_0+t s_1=0$ with a generic section of $-K_{\mathbb{P}^3}$, we resolve along the reduced smooth curve $\mathcal{C}^\flat \subset\mathbb{P}^3$ with $\mathcal{C}^\flat \sim \mathcal{C}^{(4)}$ in order to determine the building block $(Z^\flat,S^\flat)$ of the Higgs branch~$H^\flat$ with
\begin{equation}
  \dim k^\flat \,=\, 0 \ , \qquad b_3(Z^\flat) \,=\, 2 g(\mathcal{C}^\flat) \,=\, 66 \ .
\end{equation}

\begin{table}[t]
\hfil
\hbox{\footnotesize{
\vbox{
\offinterlineskip
\halign{\vrule height2.7ex depth1.2ex width1.2pt\strut\,$#$\hfil\,\vrule width0.8pt&\,#\hfil\,\vrule&\,#\hfil\,\vrule width0.8pt&\,\hfil$#$\,\vrule&\,\hfil$#$\,\vrule width0.8pt&\,\hfil$#$\,\vrule&\,\hfil$#$\,\vrule width 0.8pt&\,\hfil$#$\hfil\,\vrule width1.2pt\cr
\noalign{\hrule height 1.2pt}
s_0\,\text{factors}&Gauge Group&$\mathcal{N}=2$ Hypermultiplet spectrum&h^\flat\hfil&c^\sharp\hfil&b_3^\flat\hfil&b_3^\sharp\hfil&k^\sharp\cr
\noalign{\hrule height 1.2pt}
1^4&$SU(4)$&$3\times\adj$&60&21&89&50&3\cr
\noalign{\hrule}
1^3\cdot1&$SU(3)\times U(1)$&$3\times\adj; 4\times\fun3_{+1}$&54&15&89&50&3\cr
\noalign{\hrule}
1^2\cdot1^2&$SU(2)^2\times U(1)$&$3\times(\adj,\fun1); 3\times(\fun1,\adj); 4\times(\fun2,\fun2)_{+1}$&54&15&89&50&3\cr
\noalign{\hrule}
1^2\cdot1\cdot1&$SU(2)\times U(1)^2$&$3\times\adj; 4\times\fun2_{(+1,+1)}; 4\times\fun2_{(+1,0)}; 4\times\fun2_{(0,+1)}$&48&9&89&50&3\cr
\noalign{\hrule}
1\cdot1\cdot1\cdot1&$U(1)^3$&$4\times(+1,+1,0); 4\times(+1,0,+1); 4\times(0,+1,+1);$&42&3&89&50&3\cr
&&$4\times(+1,0,0);4\times(0,+1,0);4\times(0,0,+1)$&&&&&\cr
\noalign{\hrule height 0.8pt}
2\cdot1^2&$SU(2)\times U(1)$&$3\times\adj; 8\times\fun2_{+1}$&42&8&89&55&2\cr
\noalign{\hrule}
2\cdot1\cdot1&$U(1)^2$&$4\times(+1,+1); 8\times(+1,0); 8\times(0,+1)$&36&2&89&55&2\cr
\noalign{\hrule height 0.8pt}
2^2&$SU(2)$&$9\times\adj$&48&19&89&60&1\cr
\noalign{\hrule}
2\cdot2&$U(1)$&$16\times(+1)$&30&1&89&60&1\cr
\noalign{\hrule height 0.8pt}
3\cdot1&$U(1)$&$12\times(+1)$&22&1&89&68&1\cr
\noalign{\hrule height 1.2pt}
}}}}
\hfil
\caption{Depicted in this table are the gauge theory branches of the $SU(4)$ gauge theory of the building blocks associated to the rank one Fano threefold $\mathbb{P}^3$. The columns list the factorization of the anti-canonical section $s_0$ with degrees and multiplicities, the gauge group of the gauge theory branch, the matter spectrum of $\mathcal{N}=2$ hypermultiplets with their non-Abelian representations together with the Abelian $U(1)$ charges, the complex dimensions $h^\flat$ and $c^\sharp$ of the Higgs and Coulomb branches, the reduced three-form Betti numbers $b_3^\flat$ and $b_3^\sharp$ of the twisted connected sum $G_2$-manifolds $Y^\flat$ and $Y^\sharp$, and the kernel $k^\sharp$ of the Coulomb phase building block $(Z^\sharp,S^\sharp)$.}
\label{tab:tabP3}
\end{table}

Finally, orthogonally gluing the building blocks $(Z^\flat,S^\flat)$ and $(Z^\sharp,S^\sharp)$ to a suitable right building block $(Z_R,S_R)$, we obtain, with eq.~\eqref{eq:b2b3}, the twisted connected sum $G_2$-manifolds $Y^\flat$ and $Y^\sharp$ with the reduced Betti numbers
\begin{equation}
\begin{aligned}
    b_2^\flat \,&=\, 0  \ ,\qquad & b_3^\flat \,&=\, 89 \ , \\
    b_2^\sharp \,&=\,3  \ ,\qquad & b_3^\sharp \,&=\, 50 \ , 
\end{aligned}
\end{equation}
which we defined in eq.~\eqref{eq:redBetti}. We observe that the differences $b_2^\sharp - b_2^\flat = 3$ and $b_3^\flat - b_3^\sharp = 39$ agree with the rank of the gauge group and the change in the dimensionality of the Higgs and Coulomb branches, respectively, which is in accord with the anticipated gauge theory description established in Section~\ref{sec:NonAbelianPhases}.

By partially deforming the first term $\tilde s_{0,1}^4$ in the hypersurface equation~\eqref{eq:HP3}, we can realize hypersurface singularities describing various Abelian and non-Abelian subgroups of $SU(4)$. Such partial deformations geometrically realize mixed Higgs--Coulomb branches of the $SU(4)$ gauge theory. We collect the geometry and phase structure of these mixed Higgs--Coulomb branches in Table~\ref{tab:tabP3}, where the entries of this table are determined by eqs.~\eqref{eq:dimk}, \eqref{eq:dimb3}, \eqref{eq:HnonAb}, \eqref{eq:CnonAb}, and \eqref{eq:P3CurveData}. Note that --- depending on the breaking pattern of $SU(4)$ arising from partially higgsing --- the dimensions of Higgs and Coulomb branches vary because only the charged matter spectrum of the unbroken gauge group plays a role for the Higgs and Coulomb branches in this gauge theory sector. For all entries in Table~\ref{tab:tabP3} we find that
\begin{equation}
    b_3^\flat - b_3^\sharp = h^\flat - c^\sharp \ , \qquad \dim k^\sharp = \operatorname{rk} G \ .
\end{equation}
This agreement confirms nicely the correspondence between gauge theory branches and phases of twisted connected sum $G_2$-manifolds.

\begin{table}[t]
\hfil
\hbox{\footnotesize{
\vbox{
\offinterlineskip
\halign{\vrule height2.7ex depth1.2ex width1.2pt\strut\,#\hfil\,\vrule width0.8pt&\,#\hfil\,\vrule&\,#\hfil\,\vrule width0.8pt&\,\hfil$#$\,\vrule&\,\hfil$#$\,\vrule width0.8pt&\,\hfil$#$\,\vrule&\,\hfil$#$\,\vrule width 0.8pt&\,\hfil$#$\hfil\,\vrule width1.2pt\cr
\noalign{\hrule height 1.2pt}
$s_0$ factors&Gauge Group&$\mathcal{N}=2$ Hypermultiplet spectrum&h^\flat\hfil&c^\sharp\hfil&b_3^\flat\hfil&b_3^\sharp\hfil&k^\sharp\cr
\noalign{\hrule height 1.2pt}
$(1,0)^2(0,1)^2$&$SU(2)\times SU(2)\times U(1)$&$2\times(\adj,\fun1); 2\times (\fun1,\adj); 4\times(\fun2,\fun2)_{+1}$&42&11&50&19&3\cr
\noalign{\hrule}
$(1,0)^2(0,1)(0,1)$&$SU(2)\times U(1)^2$&$2\times\adj; 4\times\fun2_{(1,0)}; 4\times\fun2_{(0,1)}; 2\times\fun1_{(1,1)}$&38&7&50&19&3\cr
\noalign{\hrule}
$(1,0)(1,0)$&$U(1)^3$&$2\times(1,1,0);4\times(1,0,1);4\times(0,1,1);$&34&3&50&19&3\cr
$\qquad\cdot(0,1)(0,1)$&&$4\times(1,0,0);4\times(0,1,0);2\times(0,0,1)$&&&&&\cr
\noalign{\hrule height 0.8pt}
$(2,0)(0,1)^2$&$SU(2)\times U(1)$&$2\times\adj; 8\times\fun2_{+1}$&36&6&50&20&2\cr
\noalign{\hrule}
$(2,0)(0,1)(0,1)$&$U(1)^2$&$8\times(1,0); 8\times(0,1); 2\times(1,1)$&32&2&50&20&2\cr
\noalign{\hrule height 0.8pt}
$(1,1)^2$&$SU(2)$&$7\times\adj$&36&15&50&29&1\cr
\noalign{\hrule}
$(1,1)(1,1)$&$U(1)$&$12\times(+1)$&22&1&50&29&1\cr
\noalign{\hrule height 0.8pt}
$(2,0)(0,2)$&$U(1)$&$16\times(+1)$&30&1&50&21&1\cr
\noalign{\hrule height 0.8pt}
$(2,1)(0,1)$&$U(1)$&$10\times(+1)$&18&1&50&33&1\cr
\noalign{\hrule height 1.2pt}
}}}}
\hfil
\caption{The table shows the branches of the $SU(2)\times SU(2)\times U(1)$ gauge theory associated to the Fano threefold $W_6$ with Mori--Mukai label MM48${}_2$~\cite{MR641971}. Listed are the factors of the anti-canonical section $s_0$ with bi-degrees and multiplicities, the unbroken gauge subgroup, the $\mathcal{N}=2$ matter hypermultiplets, the complex dimensions $h^\flat$ and $c^\sharp$ of the Higgs and Coulomb branches, the reduced three-form Betti numbers $b_3^\flat$ and $b_3^\sharp$ of the twisted connected $G_2$-manifolds $Y^\flat$ and $Y^\sharp$, and the kernel $k^\sharp$ of the Coulomb phase.}
\label{tab:tabW6}
\end{table}

\paragraph{$SU(2)\times SU(2)\times U(1)$ gauge theory from the Fano threefold MM48${}_2$:}
The rank two Fano threefold $W_6$ with reference number MM48${}_2$ is a hypersurface of bidegree $(1,1)$ in $\mathbb{P}^2\times\mathbb{P}^2$ with $b_3(W_6)=0$ \cite{MR641971}. Let $H_1$ and $H_2$ be the hyperplane classes of $\mathbb{P}^2\times\mathbb{P}^2$. Then, by adjunction, the anti-canonical divisor of $W_6$ reads $-K_{W_6} = 2 H_1 + 2 H_2$. Furthermore, the self-intersection numbers of the curves $\mathcal{C}^{(k,l)} = (-K_{W_6}) \cap (k H_1 + l H_2)$ in the anti-canonical divisor $-K_{W_6}$ and hence their genera are
\begin{equation} \label{eq:W6CurveData}
   \mathcal{C}^{(k_1,l_1)}.\mathcal{C}^{(k_2,l_2)} \,=\, 2 (k_1k_2+l_1l_2+2 k_1 l_2 + 2 l_1 k_2 )   \ , \qquad
   g(\mathcal{C}^{(k,l)}) \,=\,  k^2 + l^2 + 4 kl +1   \ .
\end{equation}
With generic global sections $\tilde s_{0,1}$, $\tilde s_{0,2}$ and $s_1$ of $H_1$, $H_2$ and $-K_{W_6}$, the equation for the singular building block $Z_\text{sing} \subset W_6 \times \mathbb{P}^1$ becomes, with the affine coordinate $t$ of $\mathbb{P}^1$,
\begin{equation} \label{eq:HW6}
   \tilde s_{0,1}^2\tilde s_{0,2}^2 + t s_1 \,=\, 0 \ .
\end{equation}
Thus, we find $A_1$ singularities along the two curves $\tilde{\mathcal{C}}_i = \{ \tilde s_{0,i} = 0 \} \cap \{ s_1 = 0 \} \cap \{ t = 0 \}$ with $i=1,2$. Thus --- following the general discussion in Section~\ref{sec:NonAbelianPhases} --- we find a $SU(2)\times SU(2) \times U(1)$ gauge theory both with adjoint matter and with bi-fundamental matter from the intersection points $\tilde{\mathcal{C}}_1 \cap \tilde{\mathcal{C}}_2$. Due to the equivalences $\tilde{\mathcal{C}}_1\sim  \mathcal{C}^{(1,0)}$ and $\tilde{\mathcal{C}}_1\sim  \mathcal{C}^{(0,1)}$ and relations~\eqref{eq:W6CurveData}, we arrive at the four-dimensional $\mathcal{N}=2$ hypermultiplet matter spectrum
\begin{equation}
  2 \times (\adj,\fun1)\ ; \quad 2 \times (\fun1,\adj)\ ; \quad 4 \times (\fun2,\fun2)_{+1} \ .
\end{equation}
The resulting correspondence between the gauge theory branches and the phase structure of the twisted connected sum $G_2$-manifold is summarized in Table~\ref{tab:tabW6}, where the entries are computed with the formulas~\eqref{eq:dimk}, \eqref{eq:dimb3}, \eqref{eq:HnonAb}, \eqref{eq:CnonAb}, and \eqref{eq:W6CurveData}. 

\paragraph{Further examples from toric semi-Fano threefolds:} 
Our last class of examples concerns $\mathcal{N}=2$ gauge theory sectors from toric semi-Fano threefolds $P_\Sigma$, where the fan~$\Sigma$ is obtained from a triangulation of a three-dimensional reflexive lattice polytope~$\Delta$. In this toric setup, the anti-canonical divisor reads
\begin{equation}
   - K_{P_\Sigma} \,=\, D_1 + \ldots + D_n \ ,
\end{equation}
where the toric divisors $D_i$ correspond to the one-dimensional cones of $\Sigma$, that is to say to the rays of the lattice polytope $\Delta$.  For smooth toric varieties~$P_\Sigma$, the toric divisors $D_i$ are smooth and intersect transversely \cite{MR2810322}. As the anti-canonical divisor $-K_P$ is base point free, we can apply Bertini's theorem --- see, e.g., ref.~\cite{MR1288523} --- to argue that we can find a smooth global section~$s_1$ of the anti-canonical divisor~$-K_{P_\Sigma}$ and further generic global sections $s_{0,i}$ of $D_i$ such that the curves $\mathcal{C}_i=\{ s_{0,i}=0 \} \cap \{ s_1 = 0 \}$ are smooth and mutually intersect transversely. Hence, the toric semi-Fano threefold $P_\Sigma$ realizes indeed a $U(1)^{n-1}$ gauge theory sector. The four-dimensional matter spectrum is then given by Table~\ref{tab:SpecAbelian}, where the multiplicities $\chi_{ij}$ are the toric triple intersection numbers 
\begin{equation}  \label{eq:tintnumbers}
  \chi_{ij} \,=\, -K_{P_\Sigma}.D_i.D_j \ .
\end{equation}
As proposed in Section~\ref{sec:AbelianPhases}, we construct the building blocks $(Z^\sharp,S^\sharp)$ of the Coulomb branch $C^\sharp$ by the sequential blow-ups~\eqref{eq:BlowUpSequence} along the curves $\mathcal{C}_i$, while we determine the building block $(Z^\flat,S^\flat)$ of the Higgs branch $H^\flat$ by blowing a smooth curve $\mathcal{C}^\flat=\{ s_0 = 0 \} \cap \{ s_1 = 0 \}$ obtained by deforming the singular section $s_{0,1}\cdots s_{0,n}$ to a generic anti-canonical section $s_0$. Then we arrive at the twisted connected sum $G_2$-manifolds $Y^\sharp$ and $Y^\flat$ by orthogonally gluing these gauge theory building blocks with a right building block $(Z_R,S_R)$ in the usual way.\footnote{For toric semi-Fano threefolds $P_\Sigma$, some of the performed blow-ups discussed here and in the following can also be described with toric geometry techniques \cite{Braun:2016igl}. However, such a toric description is not advantageous to extract the relevant geometric data for us.}

\begin{table}[t]
\hfil
\hbox{\footnotesize{
\vbox{
\offinterlineskip
\halign{\vrule height2.5ex depth1.0ex width1.2pt\strut\,#\hfil\,\vrule width0.8pt&\,\hfil$#$\hfil\,\vrule width0.8pt&\,#\hfil\,\vrule&\,#\hfil\,\vrule width0.8pt&\,\hfil$#$\,\vrule&\,\hfil$#$\,\vrule width0.8pt&\,\hfil$#$\,\vrule&\,\hfil$#$\,\vrule width 0.8pt&\,\hfil$#$\hfil\,\vrule width1.2pt\cr
\noalign{\hrule height 1.2pt}
~No.&\rho&~Gauge Group&~$\mathcal{N}=2$ Hypermultiplet spectrum&h^\flat\hfil&c^\sharp\hfil&b_3^\flat\hfil&b_3^\sharp\hfil&k^\sharp\cr
\noalign{\hrule height 1.2pt}
K24,&2&$SU(3)\times SU(2)$&$2\times(\adj,\fun1); (\fun1,\adj); 3\times(\fun3,\fun2)_{+1}$&50&14&79&43&4\cr
MM34${}_2$&&\hfill$\times U(1)$&&&&&&\cr
\noalign{\hrule}
K32&2&$SU(3)^2\times U(1)$&$(\adj,\fun1); (\fun1,\adj); 3\times(\fun3,\fun3)_{+1}$&52&13&79&40&5\cr
\noalign{\hrule}
K35,&2&$SU(5)\times SU(2)$&$2\times(\adj,\fun1); (\fun5,\fun2)_{+1}$&60&22&87&49&6\cr
MM36${}_2$&&\hfill$\times U(1)$&&&&&&\cr
\noalign{\hrule}
K36,&2&$SU(4)\times SU(2)$&$2\times(\adj,\fun1); 2\times(\fun4,\fun2)_{+1}$&54&17&81&44&5\cr
MM35${}_2$&&\hfill$\times U(1)$&&&&&&\cr
\noalign{\hrule}
K37,&2&$SU(4)\times SU(3)$&$(\adj,\fun1); 3\times(\fun4,\fun3)_{+1}$&54&12&79&37&6\cr
MM33${}_2$&&\hfill$\times U(1)$&&&&&&\cr
\noalign{\hrule height 0.8pt}
K62,&3&$SU(2)^3\times U(1)^2$&$(\adj,\fun1^2); (\fun1,\adj,\fun1); (\fun1^2,\adj); 2\times(\fun2^2,\fun1)_{(1,1)}$&44&11&73&40&5\cr
MM27${}_3$&&&$2\times(\fun2,\fun1,\fun2)_{(1,0)}; 2\times(\fun1,\fun2^2)_{(0,1)}$&&&&&\cr
\noalign{\hrule}
K68,&3&$SU(3)\times SU(2)$&$(\adj,\fun1); 3\times(\fun3,\fun2)_{(1,1)}; 2\times(\fun3,\fun1)_{(1,0)}; (\fun1,\fun2)_{(0,1)}$&42&9&69&36&6\cr
MM25${}_3$&&\hfill$\times U(1)^2$&&&&&&\cr
\noalign{\hrule}
K105,&3&$SU(3)^2\times SU(2)$&$(\adj,\fun1^2); (\fun1,\adj,\fun1); 2\times(\fun3^2,\fun1)_{(1,1)}; (\fun3,\fun1,\fun2)_{(1,0)};$&50&15&77&42&7\cr
MM31${}_3$&&\hfill$\times U(1)^2$&$(\fun1,\fun3,\fun2)_{(0,1)}$&&&&&\cr
\noalign{\hrule}
K124&3&$SU(4)\times SU(2)^2$&$(\adj,1,1); 2\times(\fun4,\fun2,\fun1)_{(1,1)}; 2\times(\fun4,\fun1,\fun2)_{(1,0)}$&48&13&73&38&7\cr
&&\hfill$\times U(1)^2$&&&&&&\cr
\noalign{\hrule height 0.8pt}
K218,&4&$SU(4)\times SU(3)$&$(\adj,\fun1^3); (\fun4,\fun3,\fun1^2)_{(1,1,0)}; (\fun4,\fun1,\fun2,\fun1)_{(1,0,1)};$&46&16&71&41&10\cr
MM12${}_4$&&\hfil$\times SU(2)^2\times U(1)^3$&$(\fun4,\fun1^2,\fun2)_{(1,0,0)}; (\fun1,\fun3,\fun2,\fun1)_{(0,1,1)}; (\fun1,\fun3,\fun1,\fun2)_{(0,1,0)}$&&&&&\cr
\noalign{\hrule}
K266,&4&$SU(3)\times SU(2)^3$&$(\fun1,\adj,\fun1^2); (\fun3,\fun2,\fun1^2)_{(1,1,0)}; 2\times(\fun3,\fun1,\fun2,\fun1)_{(1,0,1)};$&42&10&67&35&8\cr
MM10${}_4$&&\hfill$\times U(1)^3$&$2\times(\fun3,\fun1^2,\fun2)_{(1,0,0)}; (\fun1,\fun2^2,\fun1)_{(0,1,1)}; (\fun1,\fun2,\fun1,\fun2)_{(0,1,0)}$&&&&&\cr
\noalign{\hrule}
K221&4&$SU(3)\times SU(2)^2$&$2\times(\fun3,\fun2,\fun1)_{(1,1,0)}; 3\times(\fun3,\fun1,\fun2)_{(1,0,1)}; (\fun3,\fun1^2)_{(1,0,0)}$&40&7&63&30&7\cr
&&\hfill$\times U(1)^3$&$2\times(\fun1,\fun2,\fun1)_{(0,1,0)}$&&&&&\cr
\noalign{\hrule}
K232&4&$SU(4)\times SU(2)^3$&$2\times(\fun4,\fun2,\fun1^2)_{(1,1,0)}; 2\times(\fun4,\fun1,\fun2,\fun1)_{(1,0,1)};$&42&9&65&32&9\cr
&&\hfill$\times U(1)^3$&$2\times(\fun4,\fun1^2,\fun2)_{(1,0,0)}$&&&&&\cr
\noalign{\hrule}
K233&4&$SU(3)\times SU(2)^2$&$3\times(\fun3,\fun2,\fun1)_{(1,1)}; 3\times(\fun3,\fun1,\fun2)_{(1,0)}$&40&6&63&29&6\cr
&&\hfill$\times U(1)^2$&&&&&&\cr
\noalign{\hrule}
K247&4&$SU(4)\times SU(3)^2$&$2\times(\fun4,\fun3,\fun1^2)_{(1,1,0)}; 2\times(\fun4,\fun1,\fun3,\fun1)_{(1,0,1)};$&46&11&69&34&11\cr
&&\hfill$\times SU(2)\times U(1)^3$&$(\fun1,\fun3,\fun1,\fun2)_{(0,1,0)}; (\fun1^2,\fun3,\fun4)_{(0,0,1)}$&&&&&\cr
\noalign{\hrule}
K257&4&$SU(5)\times SU(3)^2$&$2\times(\fun5,\fun3,\fun1^2)_{(1,1,0)}; 2\times(\fun5,\fun1,\fun3,\fun1)_{(1,0,1)};$&48&12&71&35&12\cr
&&\hfill$\times SU(2)\times U(1)^3$&$(\fun5,\fun1^2,\fun2)_{(1,0,0)}$&&&&&\cr
\noalign{\hrule height 1.2pt}
}}}}
\hfil
\caption{The table exhibits the $\mathcal{N}=2$ gauge theory sectors for some smooth toric semi-Fano threefolds $P_\Sigma$ of Picard rank two and higher. The columns display the number of the threefold $P_\Sigma$ in the 
Mori--Mukai~\cite{MR641971} and/or Kasprzyk~\cite{MR2221794db} classification, its Picard rank $\rho$, the maximally enhanced gauge group of maximal rank by factorizing the anti-canonical bundle, the $\mathcal{N}=2$ matter hypermultiplets, the complex dimensions $h^\flat$ and $c^\sharp$ of the Higgs and Coulomb branches, the reduced three-form Betti numbers $b_3^\flat$ and $b_3^\sharp$, and the kernel $k^\sharp$ of the Coulomb branch.}
\label{tab:tabToric}
\end{table}

Note that, due to linear equivalences among the toric divisors $D_i$~, the Abelian gauge theory can enhance to non-Abelian gauge groups as well. Namely, assume that the anti-canonical bundle $-K_{P_\Sigma}$ is linearly equivalent to
\begin{equation}
   - K_{P_\Sigma} \,\sim\, k_1 \tilde D_1  + \ldots + k_s \tilde D_s \ ,
\end{equation}
where, for some divisors, $\tilde D_\alpha \sim \sum_i a_{\alpha i} D_i$ with global sections $\tilde s_{0,\alpha}$. Furthermore, we require that the curves $\tilde{\mathcal{C}}_\alpha$ are smooth and mutually intersect transversely. Then, following Section~\ref{sec:NonAbelianPhases}, we arrive at the $\mathcal{N}=2$ gauge theory sector with gauge group
\begin{equation}
     G\,=\,SU(k_1) \times \ldots \times SU(k_s) \times U(1)^{s-1} \ .
\end{equation} 
Note that rank of the gauge group $\tilde n = k_1 + \ldots + k_s - 1$ is a priori not correlated with the number $n$ of toric divisors. Instead, it depends on the precise nature of the linear equivalences among the toric divisors $D_i$, $i=1,\ldots,n$, and the divisors $\tilde{D}_\alpha$, $\alpha=1,\ldots,s$.

Let us exemplify the study of four-dimensional $\mathcal{N}=2$ gauge theory sectors with the rank two toric semi-Fano threefold $P_\Sigma$ of reference number~K32 described in some detail in Section~\ref{sec:examples}. Using for this example the linear equivalences among the toric divisors $D_1, \ldots, D_5$ --- cf. below eq.~\eqref{eq:K32Mcone} --- we find for the anti-canonical divisor 
\begin{equation}
  -K_{P_\Sigma} \,=\, D_1 + \ldots + D_5  \,\sim\, 3 D_1 \,\sim\, 3 D_2 + 3 D_4 \ .
\end{equation}
With these linearly equivalent representations for $-K_{P_\Sigma}$, we arrive, for instance, at the gauge groups $U(1)^4$ of rank four, $SU(3)$ of rank three, or $SU(3)\times SU(3) \times U(1)$ of rank five. Note that the phases of the lower rank gauge groups $U(1)^4$ and $SU(3)$ enjoy again the interpretation as mixed Higgs--Coulomb branches of the $SU(3)\times SU(3)\times U(1)$ gauge theory of rank five which, by applying eq.~\eqref{eq:tintnumbers} and eq.~\eqref{eq:genusC}, yields the spectrum
\begin{equation}
  1 \times (\adj,\fun1)\ ; \quad 1 \times (\fun1,\adj)\ ; \quad 3 \times (\fun3,\fun3)_{+1} \ .
\end{equation}

In Table~\ref{tab:tabToric} we list the gauge theory sectors of a few toric semi-Fano threefolds $P_\Sigma$. This table does not display all mixed Higgs--Coulomb branches. Here, we focus on the resulting twisted connected sum $G_2$-manifolds $Y^\flat$ and $Y^\sharp$ associated to the Higgs $H^\flat$ and Coulomb branches $C^\sharp$ of the maximally enhanced gauge group of maximal rank, as obtained by the factorization of the anti-canonical bundle $-K_{P_\Sigma}$. 
\clearpage

\subsection{Transitions among twisted connected sum $G_2$-manifolds} \label{sec:Transitions}
The proposed correspondence between phases of twisted connected sum $G_2$-manifolds and gauge theory branches of the described $\mathcal{N}=2$ gauge theory sectors is essentially based upon the correspondence between extremal transitions in the asymptotically cylindrically Calabi--Yau threefolds $X_\LR$ and the Higgs--Coulomb phase structure of the associated $\mathcal{N}=2$ gauge theories. In the original type~IIA string theory setting the $\mathcal{N}=2$ matter spectrum arises from solitons of massless D2-branes wrapping the vanishing cycles of the singular Calabi--Yau threefolds at the transition point \cite{Strominger:1995cz,Greene:1995hu}, which become membranes in the discussed context of M-theory. However, while in the type~IIA setting these D2-branes furnish BPS states of the $\mathcal{N}=2$ algebra, the corresponding interpretation of membrane states becomes more subtle in the context of M-theory on twisted connected sum $G_2$-manifolds because the corresponding membrane states do not admit a BPS interpretation due to minimal four-dimensional $\mathcal{N}=1$ supersymmetry. Therefore, a natural question now is whether the described M-theory transitions are actually dynamically realized.

As discussed in Section~\ref{sec:Mcompactification}, the semi-classical moduli space $\mathcal{M}_\mathbb{C}$ of M-theory on $G_2$-manifolds has the geometric moduli space $\mathcal{M}$ of Ricci-flat $G_2$-manifolds as a real subspace. From the low-energy effective $\mathcal{N}=1$ supergravity point of view, this is a consequence of the semi-classical shift symmetries with respect to the real parts of the chiral fields~\eqref{eq:CStarget}. However, due to arguments about the absence of global continuous symmetries in consistent theories of gravity, see e.g., ref.~\cite{Banks:2010zn}, these shift symmetries should be broken non-perturbatively such that the flat directions of the chiral moduli fields are lifted. In the context of M-theory on $G_2$-manifolds, membrane instantons on suitable three-cycles generate non-perturbative superpotential terms that break these continuous shift symmetries \cite{Harvey:1999as}. As these non-perturbative corrections are exponentially suppressed in the volume of the wrapped three-cycles, the flat directions --- as described by the semi-classical moduli space $\mathcal{M}_\mathbb{C}$ --- are expected to be only realized in the large volume limit of the $G_2$-compactification. Hence, M-theory transitions among $G_2$-manifolds should only occur in the absence of such non-perturbative effects, as for instance in the case of the large volume limit.\footnote{In the presence of small non-perturbative obstructions we can still have quantum-mechanical transitions among four-dimensional vacua. Then the transition probability is governed by the tunneling rate through the barrier of the non-perturbative scalar potential.} 

If we now take both the large volume limit and the Kovalev limit simultaneously, gravity decouples, and we arrive at a genuine four-dimensional $\mathcal{N}=2$ gauge theory sector with eight unbroken supercharges. Then the lower energy dynamics is indeed described as in refs.~\cite{Strominger:1995cz,Greene:1995hu,Klemm:1996kv,Katz:1996ht}, and the gauge theory phases connect asymptotically cylindrical Calabi--Yau threefolds via extremal transitions. Thus, we claim that, in the large volume and in the large Kovalev limit, the transitions among the $\mathcal{N}=2$ gauge theory sectors geometrically realize the anticipated transitions among twisted connected sum $G_2$-manifolds. 

If we maintain the large volume limit but allow for finite Kovalevton, the situation becomes more subtle. While the massless spectrum is still $\mathcal{N}=2$, we expect that the appearance of further interaction terms breaks $\mathcal{N}=2$ supersymmetry to $\mathcal{N}=1$. Then the $\mathcal{N}=2$ gauge theory sector is partially broken to a $\mathcal{N}=1$ gauge theory, whose supersymmetry breaking couplings are governed by the scale of the Kovalevton. In this $\mathcal{N}=1$ language, the transition between non-Abelian $\mathcal{N}=2$ Higgs and Coulomb branches essentially describes an enhancement to an Abelian gauge symmetry within the $\mathcal{N}=1$ Higgs branches. Namely, in the $\mathcal{N}=1$ language, the $\mathcal{N}=2$ Coulomb phase corresponds to the partially higgsing of the non-Abelian group to its maximal Abelian subgroup. Thus, at low energies, the proposed (non-Abelian) $\mathcal{N}=2$ Higgs--Coulomb phase transition describes the Higgs mechanism of a weakly-coupled Abelian $\mathcal{N}=1$ gauge theory. These observations provide for some evidence that, in the large volume limit, the anticipated phase structure among the described twisted connected sum $G_2$-manifolds is still realized --- even for finite Kovalevton.

Geometrically, we therefore propose that in the M-theory moduli space~$\mathcal{M}_\mathbb{C}$ the presented transitions among twisted connected sum~$G_2$-manifolds are indeed unobstructed. That is to say, we conjecture that the construction of orthogonally gluing commutes with extremal transitions in the asymptotically cylindrical Calabi--Yau threefolds $X_\LR$. Furthermore, our proposal implies that the moduli space $\mathcal{M}$ of Ricci-flat $G_2$-metrics of the twisted connected sum type should exhibit a stratification structure as predicted by the phase structure of the analyzed $\mathcal{N}=2$ gauge theories sectors. In the context of Abelian gauge theory sectors our proposal conforms with a similar conjecture put forward in ref.~\cite{Halverson:2014tya}.

\section{Conclusions}  \label{sec:conclusions}

In this work we have studied the four-dimensional low-energy effective $\mathcal{N}=1$ supergravity 
action arising from M-theory compactified on $G_2$-manifolds of the Kovalev's twisted connected sum type. By suitably gluing a 
pair of non-compact asymptotically cylindrical Calabi--Yau threefolds times a circle $Y_\LR=X_\LR\times S^1$ in their 
asymptotic regions \cite{MR2024648}, this construction realizes a large class of 
examples for compact $G_2$-manifolds \cite{MR3369307,Halverson:2014tya,Braun:2016igl}, which are yet of a 
specific type realizing only a particular (non-trivial) homotopy invariant of $G_2$-manifolds \cite{MR3416118}. 

From the cohomology of such $G_2$-manifolds, we established that this class of M-theory compactifications yields two 
neutral universal $\mathcal{N}=1$ chiral moduli fields associated to the complexified overall volume 
modulus $\nu$ and the gluing modulus  --- called the Kovalevton $\kov$ ---  respectively. 
The latter parametrizes the Kovalev limit taken by ${\rm Re}(\kov)\rightarrow \infty$. 

The proper interpretation of the different contributions in \eqref{eq:split1} to the cohomology of the $G_2$-manifold $Y$ 
implies that there is a decomposition of the fields of the $\mathcal{N}=1$ effective supergravity theory on $Y$ into
$\mathcal{N}=1$ neutral chiral moduli multiplets, into two $\mathcal{N}=2$ gauge theory 
sectors coming from the two asymptotic regions $Y_\LR$, and into one $\mathcal{N}=4$ gauge sector that comes 
from the trivial $K3$ fibration with fibre $S$ in the gluing region $T^2\times S\times (0,1)$, cf.~Table 
\ref{tab:locSpec}. This decomposition into these gauge theory sectors becomes exact in a controllable way in the Kovalev limit and yields a scheme in which the four-dimensional low-energy effective theory can be approximated in terms of these 
sectors with small corrections. In particular, we worked out the dependence on these two universal chiral moduli fields
in the four-dimensional low-energy effective $\mathcal{N}=1$ supergravity action resulting from a compactification on 
a smooth twisted connected sum $G_2$-manifold $Y$. Moreover, the obtained two 
scales do also control the behavior of $M$-theory corrections. Let us now list mathematical, physical and 
eventually even phenomenological prospects of this decomposition, specific to $M$-theory compactifications
on twisted connected sum $G_2$-manifolds.

A first consequence is that we can identify Abelian and non-Abelian gauge theory enhancements with 
various matter content from singularities in the asymptotic cylindrical Calabi--Yau 
threefolds $X_\LR$ in codimension four and six that occur in the twisted connected sum~$Y$ away from 
the gluing region. These lead to transitions in the threefolds~$X_\LR$, whose deformations and resolutions can 
be  described by methods of algebraic geometry familiar in the context of $\mathcal{N}=2$ 
theories. The significant point established in Section \ref{sec:N=2sectors} is that these 
transitions commute with the Kovalev limit and the gluing construction. Namely, they  
connect $G_2$-manifolds whose change in the cohomology groups corresponds exactly to the change  
in the spectrum of $\mathcal{N}=1$ vector and chiral superfields as predicted by the transitions.
Concretely, starting with the equations that describe the blow-up of the 
anti-canonical divisor in semi-Fano threefolds and analyzing all their possible degenerations 
lead to a great variety of gauge groups and matter spectra as well as to many novel examples of 
twisted connected sum $G_2$-manifolds corresponding to the different branches of these 
gauge theories.

This suggests that, in a suitably compactified moduli space of the Ricci-flat $G_2$-metrics, there are many 
new types of singular loci through which it is possible to reach topological inequivalent 
$G_2$-manifolds. This question is a priori independent of the possible $\mathcal{N}=1$ non-perturbative
membrane instanton corrections that could lift the flat directions in the $\mathcal{N}=1$ scalar potential 
(which are protected in the pure $\mathcal{N}=2$ limit). Therefore, by taking the 
large volume limit and the Kovalev limit, these directions certainly remain flat. However, we argued in  
Section~\ref{sec:Transitions} that even for finite expectation values of the Kovalevton $\kov$, the
transitions should remain physical in the effective $\mathcal{N}=1$ theory, as long as non-perturbative
membrane instantons remain suppressed.
 
Another interesting physical consequence of the decomposition and the Kovalev limit 
is that the more advanced $\mathcal{N}=2$ techniques --- like calculating the 
exact gauge coupling and the exact BPS masses from the periods of the holomorphic 
three-form --- are approached in this limit and serve as a zeroth order 
approximation with inverse power laws or exponential corrections in the Kovalevton $\kov$
and the volume modulus~$\nu$, similarly as the calculations carried out in refs.~\cite{Atiyah:2000zz,Aganagic:2001ug}
in the context of local $G_2$-manifolds. Those corrections leading to holomorphic terms in the four-dimensional $\mathcal{N}=1$ effective theory
are expected to be accessible by techniques similar to the ones used to calculate four-dimensional
$\mathcal{N}=1$ F-terms in flux and/or brane compactifications of type~II theories.
Note, however, that these computations would require a detailed study of the relative 
Calabi--Yau three-form periods on the two non-compact Calabi--Yau threefolds --- for instance by using variation
of mixed Hodge structure techniques along the lines of refs.~\cite{Aganagic:2000gs,Aganagic:2001nx,Lerche:2002yw,
Jockers:2008pe,Alim:2009rf,Grimm:2009ef,Donagi:2012ts} --- and a 
moduli-dependent analysis of the matching conditions~\eqref{eq:HyperKaehler}.\footnote{A similar 
analysis of moduli dependent matching conditions is required for building blocks $(Z,S)$ 
arising from general weak Fano threefold for which Beauville's theorem~\cite{MR2112574} is not applicable.}

An attractive feature of the twisted connected sum compactification is that in 
the two individual $\mathcal{N} =2$ gauge theory sectors from $X_\LR$ we have algebraic methods
to geometrically engineer gauge groups, spectra and interactions. Already the few examples presented
in Table~\ref{tab:tabToric} yield small rank gauge groups such as the standard model group and
possible grand unification scenarios. The matter contents could in principle be broken into 
phenomenologically more suitable massless $\mathcal{N}=1$ chiral matter multiplets.
In fact, the computed $\mathcal{N}=2$ spectra can be broken to $\mathcal{N}=1$ multiplets by various
non-local effects. As discussed for finite Kovalevton $\kov$ the twisted gluing itself 
and non-perturbative effects --- such as membrane instantons --- introduce genuine $\mathcal{N}=1$
interaction terms. Adding a flux-induced superpotential~\eqref{eq:Superpotential} offers yet another
attractive mechanism to break the $\mathcal{N}=2$ spectra into $\mathcal{N}=1$ multiplets \cite{Gukov:1999ya},
potentially introducing chirality as well. Due to the absence of tadpole constraints for four-form fluxes on $G_2$-manifolds,
the local scenario for fluxes in type~II string theories proposed in ref.~\cite{Taylor:1999ii}
is readily realized on the level of the non-compact asymptotically cylindrical Calabi--Yau summands $X_\LR$. 
We expect that non-trivial background four-form fluxes provides for a much more intricate and genuine $\mathcal{N}=1$ gauge
theory branch structure, similarly as in refs.~\cite{Intriligator:2012ue,Jockers:2016bwi}.
All these effects come with different scales --- partially exponentially suppressed --- which exhibit potentially attractive hierarchies. A systematic analysis of phenomenologically attractive $\mathcal{N}=1$ or 
even $\mathcal{N}=0$ interaction terms in the context of M-theory on twisted connected sum $G_2$-manifolds
is beyond the scope of this work and will be addressed elsewhere.

While in type~II Calabi--Yau threefold compactifcations we arrive at four-dimen\-sional $\mathcal{N}=2$ effective supergravity theories with two
massless gravitinos realizing extended supersymmetry, breaking the $\mathcal{N}=2$ gravity multiplet down to the $\mathcal{N}=1$
is rather non-trivial, see for instance the discussion in ref.~\cite{Louis:2002vy}. In our case, however,
the obtained four-dimensional supergravity theory has already minimal supersymmetry. It is only the gauge theory
sectors in the Kovalev limit that approximately exhibit extended global supersymmetries. Therefore, introducing background
fluxes to break supersymmetry in the gauge theory sectors is much simpler than in the type~II Calabi--Yau threefold compactifications.
In particular, turning on background fluxes, resembles to a great extent the type~II scenario of ref.~\cite{Taylor:1999ii}, 
in which, however, gravity is decoupled.

Let us point out a further potentially phenomenologically attractive possibility. 
As explained in Section \ref{sec:spectrumandcohomology}, the separation of the two sectors 
$X_L$  and $X_R$ in Figure \ref{fig:twistedsum} is controlled by the real part of the 
Kovalevton $\kappa$. Together with the local construction of the spectra on $X_\LR$ described 
in Section \ref{sec:N=2sectors}, this offers the possibility to consider a hidden and a 
visible sector and to employ the mechanism of mediation of supersymmetry breaking only in 
the gravitational sector with a controllable scale set by Kovalevton $\kappa$. Or alternatively, as there 
is an anomaly inflow mechanism in the local theories~\cite{Witten:2001uq,Acharya:2001gy}, 
one could use the anomaly mediation of supersymmetry breaking as proposed in~\cite{Randall:1998uk}.

Finally, we comment on the possible relation of the twisted connected sum construction to other non-perturbative 
descriptions of $\mathcal{N}=1$ theories.  In lower dimensions the algebraic-geometrical approach towards the Ho\v{r}ava--Witten setup describes a 
duality to F-theory \cite{Aspinwall:1997eh}. Namely, certain Calabi--Yau compactifications of the heterotic string 
are dual to F-theory on elliptically fibered Calabi--Yau fourfolds in a particular stable degeneration limit 
\cite{Friedman:1997yq}. To obtain four-dimensional $\mathcal{N}=1$ supergravity theory this heterotic--F-theory correspondence 
is realized on the level of elliptically-fibered Calabi--Yau fourfolds. It is intriguing to observe that such Calabi--Yau fourfolds 
in the stable degeneration limit are obtained by gluing a pair of suitably chosen Fano fourfolds along their mutual 
anti-canonical Calabi--Yau threefold divisor \cite{MR1818182,Jockers:2009ti}. This construction of Calabi--Yau 
fourfolds in the stable degeneration limit shows a certain resemblance --- yet in one real dimension higher --- 
to twisted connected sum $G_2$-manifolds in the Kovalev limit. It would be interesting to see if such a speculation 
could be made precise, namely establishing a duality between M-theory on $G_2$-manifolds in the Kovalev
limit and F-theory on elliptically-fibered Calabi--Yau fourfolds in a certain degeneration limit.

\begin{center}
\textbf{Acknowledgements} 
\end{center}We would like to thank 
Dominic Joyce, Dave Morrison, Stefan Schreieder, and Eric Zaslow 
for discussions and correspondences. Thaisa C. da C. Guio would like to thank financial support from CNPq (Brazil) under grant number 205626/2014-9 and the Bonn-Cologne 
Graduate School of Physics and Astronomy (BCGS). Hung-Yu Yeh would like to thank the 
IMPRS scholarship and great research environment in the Max Planck Institute for Mathematics during the period of the PhD study.

\appendix
\section{Kaluza-Klein reduction of fermionic terms} \label{app:KKFermions}
In this appendix we present the Kaluza--Klein reduction of the fermionic terms obtained from the compactification of  the eleven-dimensional supergravity action \eqref{eq:11dSUGRA} on a seven-dimensional $G_2$-manifold. We analyze the four-dimensional fermionic spectrum, and explicitly derive the zero modes of the eleven-dimensional Rarita--Schwinger field compactified on the $G_2$-manifold. We further determine the holomorphic superpotential induced by internal four-form fluxes from certain fermionic interaction terms. Compared to the purely bosonic interactions with quadratic dependences on the superpotential, the fermionic interactions are linear in the superpotential, see for instance ref.~\cite{Wess:1992cp}. Thus, analyzing the fermionic interactions, as opposed to the purely bosonic ones, is more tractable for determining the superpotential.

\subsection{Definitions and useful relations}
Following the definitions and conventions of ref.~\cite{House:2004pm}, we first spell out the properties of the representation of the used eleven-, seven- and four-dimensional gamma matrices. The eleven-dimensional gamma matrices are represented by 32-dimensional matrices, which satisfy the usual Clifford algebra
\begin{equation} 
  \{\hat{\Gamma}_M,\hat{\Gamma}_N\} \,=\, 2g_{MN} \ ,
\end{equation}
with the eleven-dimensional Lorentzian metric $g_{MN}$. Furthermore, in the chosen 32-dimensional Majorana representation, the gamma matrices obey \cite{House:2004pm}
\begin{equation}
  \hat{\Gamma}_0\cdots\hat{\Gamma}_{10} \,=\, \mathbb{I} \ ,
\end{equation}
in terms of the 32-dimesional identity matrix $\mathbb{I}$. With the compactification ansatz $M^{1,10} = \mathbb{M}^{1,3} \times Y$, the eleven-dimensional gamma matrices split into two sets of commuting gamma matrices, i.e.,
\begin{equation} \label{SplitGammas}
	\hat{\Gamma}_M  = ( \hat{\Gamma}_{\mu} , \hat{\Gamma}_m)\ , \quad
	\hat{\Gamma}_{\mu} = \gamma_{\mu} \otimes \mathbb{I} \ , \quad
	\hat{\Gamma}_m  = \gamma \otimes \gamma_m \ ,
\end{equation}
where $\mathbb{I}$ is the seven-dimensional identity matrix, $\gamma_{\mu}$, $\mu=0,1,2,3$, are the four-dimensional imaginary gamma matrices, $\gamma_m$, $m=4,\ldots,10$, are purely imaginary seven-dimensional gamma matrices satisfying $\gamma_4\cdots\gamma_{10} = i$. Furthermore, we define $\gamma = (i/4!)\epsilon_{\mu\nu\rho\sx} \gamma^{\mu}\gamma^{\nu}\gamma^{\rho}\gamma^{\sx}$ as the four-dimensional chirality matrix, which is purely imaginary
and satisfies $\gamma^2=1$.

The four- and seven-dimensional gamma matrices satisfy the Clifford algebra in their corresponding dimensions
\begin{equation} \label{eq:Clifford4D7D}
	\{\gamma_{\mu}, \gamma_{\nu}\} \,=\, 2\eta_{\mu\nu}\ , \qquad \{\gamma_m, \gamma_n\} \,=\, 2g_{mn}\ .
\end{equation}
Here we use the Minkowski metric $\eta_{\mu\nu}$ with signature $(-1,+1,+1,+1)$ and $g_{mn}$ denotes the Riemannian metric of the seven-dimensional compactification space.

We define the anti-symmetrized product of eleven-dimensional gamma matrices as
\be \label{CompletelyAntiGammas}
	\hat{\Gamma}^{M_1 \cdots M_n} = \hat{\Gamma}^{[M_1} \cdots \hat{\Gamma}^{M_n]}\ ,
\ee
and we use the same notation for the anti-symmetrized products of four- and seven-dimensional gamma matrices, i.e., $\gamma^{\mu_1 \cdots \mu_n} = \gamma^{[\mu_1} \cdots \gamma^{\mu_n]}$ and $\gamma^{m_1 \cdots m_n} = \gamma^{[m_1} \cdots \gamma^{m_n]}$. For the decomposition of $\hat{\Gamma}^{MNP}$ into lower-dimensional gamma matrices we arrive at the useful relation
\be 	\label{Gamma3Antisymmetrized}
	\begin{split}
	\hat{\Gamma}^{MNP}  =~&(\gamma^{\mu\nu\rho}\otimes\mathbb{I}) + (\gamma^{\mu\nu-}\otimes\gamma^p) + (\gamma^{\mu-\rho}\otimes\gamma^n) + (\gamma^{-\nu\rho}\otimes\gamma^m)\\ 
	+ & \frac{1}{3} (\gamma^{\rho}\otimes\gamma^{mn} + \gamma^{\nu}\otimes\gamma^{pm} + \gamma^{\mu}\otimes\gamma^{np})\\
	+ & \gamma\otimes\gamma^{mnp}~,
	\end{split}
\ee
where the index `$-$' refers to the four-dimensional chirality matrix $\gamma$.

For the forthcoming zero mode analysis we record here a few useful identities. First of all, we record a few useful identities among products of anti-symmetrized gamma matrices, namely
\begin{equation} \label{eq:antigamma}
\begin{aligned}
  \gamma^{mnp}\gamma^q \,&=\, \gamma^{mnpq} + 3 g^{q[ m}\gamma^{np]} \ , \\
  \gamma^{mnp}\gamma^{qr} \,&=\, \gamma^{mnpqr} + 3 \left( g^{q[m}\gamma^{np]r} - g^{r[m}\gamma^{np]q}\right) + 6 g^{q[m}\gamma^ng^{p]r}  \ . 
\end{aligned}
\end{equation}
Furthermore, the $G_2$-structure $\varphi$ fulfills the contraction relations
\begin{equation} \label{eq:convarphi}
  \varphi_{mnp} \varphi^{npq} \,=\, 6\,\delta_n^q \ , \qquad
  \varphi_{mnp} \varphi^{pqr} \,=\, {\Phi_{mn}}^{qr} +  \delta_m^q\delta_n^r - \delta_m^r \delta_n^q  \ ,
\end{equation}
with the Hodge dual form $\Phi=*\varphi$, and the Fierz identity
\begin{equation} \label{eq:fierz}
  \gamma^{mn}\eta \,=\, -i \varphi^{mnp} \gamma_p \eta \ ,
\end{equation}
in terms of the covariantly constant spinor $\eta$. Finally, the Levi--Civita connection $\nabla$, the exterior derivative $d$ and its adjoint $d^\dagger$ fulfill the relations
\begin{equation} \label{eq:ExteriorDerivative}
\begin{aligned}
	(dA)_{n_1 \ldots n_{p+1}} & = (p+1) \nabla_{[n_1}A_{n_2 \ldots n_{p+1}]} \ ,\\
	(d^{\dagger}A)_{n_1 \ldots n_{p-1}} &= - \nabla^m A_{m n_1 \ldots n_{p-1}} \ , 
\end{aligned}
\end{equation}
for any $p$-form $A = \frac{1}{p!}A_{n_1 \ldots n_p} dy^{n_1} \wedge \ldots \wedge dy^{n_p}$.

\subsection{$G_2$-representations and the Rarita--Schwinger $G_2$-bundle}
In this section we present further details for the decomposition of the global section \eqref{eq:zeta1} of the Rarita--Schwinger bundle $T^{\ast}Y \otimes SY$ on $G_2$-manifolds, as introduced in Section~\ref{sec:Mcompactification}.

First of all, the differential $p$-forms on a manifold with $G$-structure fall into irreducible representations with respect to the structure group $G$. Specifically, for a seven-manifold with $G_2$-structure the spaces of differential $p$-forms $\Lambda^p$ decompose according to ref.~\cite{MR1424428}, namely
\begin{equation} \label{eq:pformsG2}
\begin{aligned}
	\Lambda^0 & = \Lambda^0_\mathbf1~,&&
	\Lambda^1 & = \Lambda^1_\mathbf7~,&&
	\Lambda^2 & = \Lambda^2_\mathbf7 \oplus \Lambda^2_\mathbf{14}~,&&
	\Lambda^3 & = \Lambda^3_\mathbf1 \oplus \Lambda^3_\mathbf7 \oplus \Lambda^3_{\mathbf{27}}~,\\
	\Lambda^7 & = \Lambda^7_\mathbf1~,&&
	\Lambda^6 & = \Lambda^6_\mathbf7~,&&
	\Lambda^5 & = \Lambda^5_\mathbf7 \oplus \Lambda^5_{\mathbf{14}}~,&&
	\Lambda^4 & = \Lambda^4_\mathbf1 \oplus \Lambda^4_\mathbf7 \oplus \Lambda^4_{\mathbf{27}}~,
\end{aligned}	
\end{equation}
where the summands $\Lambda^p_\mathbf{q}$ are $p$-forms transforming in the $q$-dimensional irreducible representations of the structure group $G_2$. As indicated in the arrangement of the form space $\Lambda^p_\mathbf{q}$ in eq.~\eqref{eq:pformsG2}, the Hodge star $\ast$ provides for an isometry between $\Lambda^p_\mathbf{q}$ and $\Lambda^{7-p}_\mathbf{q}$. The differential $p$-form spaces $\Lambda^p_\mathbf7$ are isomorphic to each other for $p=1,\ldots,6$. Moreover, $\Lambda^3_\mathbf1$ and $\Lambda^4_\mathbf1$ are generated by $\varphi$ and $\ast \varphi$, respectively. For a compact $G_2$-manifold $Y$ equipped with a torsion-free $G_2$-structure, the de Rham cohomologies $H^p(Y,\mathbb{R})$ have a similar decomposition into $H^p_\mathbf{q}(Y,\mathbb{R})$ with harmonic representatives, see, e.g., ref.~\cite{MR1424428},
\begin{equation}
	\begin{split}
	H^2 (Y,\mathbb{R}) & = H^2_\mathbf7(Y,\mathbb{R}) \oplus H^2_\mathbf{14}(Y,\mathbb{R})~,\\
	H^3 (Y,\mathbb{R}) & = H^3_\mathbf1(Y,\mathbb{R}) \oplus H^3_\mathbf7(Y,\mathbb{R}) \oplus H^3_\mathbf{27}(Y,\mathbb{R})~,\\
	H^4 (Y,\mathbb{R}) & = H^4_\mathbf1(Y,\mathbb{R}) \oplus H^4_\mathbf7(Y,\mathbb{R}) \oplus H^4_\mathbf{27}(Y,\mathbb{R})~,\\
	H^5 (Y,\mathbb{R}) & = H^5_\mathbf7(Y,\mathbb{R}) \oplus H^5_\mathbf{14}(Y,\mathbb{R})~.\\
	\end{split}	
\end{equation}
Notice that $H^3_\mathbf1(Y,\mathbb{R}) = \langle\!\langle [\varphi] \rangle\!\rangle$ and $H^4_\mathbf1(Y,\mathbb{R}) = \langle\!\langle [\ast \varphi] \rangle\!\rangle$. Moreover, $H^p_{\mathbf{q}}(Y, \mathbb{R}) \cong H^{7-p}_\mathbf{q}(Y, \mathbb{R})$, which implies for the Betti numbers $b^p_\mathbf{q}(Y) = b^{7-p}_\mathbf{q}(Y)$ and $b^3_\mathbf{1}(Y) = b^4_\mathbf{1}(Y)=1$. If the holonomy group is $G_2$ and not a subgroup thereof, we further have $H^p_\mathbf{7} = \{0\}$ for $p=1,\ldots,6$.

Let us now turn to the Rarita--Schwinger bundle on $G_2$-manifolds, which --- due to the covariantly constant spinor $\eta$ --- becomes reducible, namely  $T^{\ast}Y \otimes SY \cong T^{\ast}Y \otimes (T^{\ast}Y \oplus \mathbb{R})$. This allows us to make the following identification
\begin{equation}
	\begin{split}
	T^{\ast}Y \otimes SY \cong~& T^{\ast}Y \otimes (T^{\ast}Y \oplus \mathbb{R}) \\
	=~& (T^{\ast}Y \otimes T^{\ast}Y) \oplus T^{\ast}Y \\
	=~& \text{Sym}^2(T^{\ast}Y) \oplus \Lambda^2 T^{\ast}Y \oplus T^{\ast}Y~,
	\end{split}
\end{equation}
where $\text{Sym}^2(T^{\ast}Y)$ is the space of symmetric two-tensors on $Y$ and $\Lambda^2 T^{\ast}Y$ is the space of two-forms. Furthermore, it is shown in ref.~\cite{MR2559631} that $\text{Sym}^2(T^{\ast}Y) \cong \Lambda^3_\mathbf{1} \oplus \Lambda^3_\mathbf{27}$. Since the spaces $\Lambda^{2}_\mathbf{7}$ and $\Lambda^{3}_\mathbf{7}$ are isomorphic to the cotangent bundle $\Lambda^{1}_\mathbf{7}$, we arrive at
\begin{equation}
	T^{\ast}Y \otimes SY \cong \Lambda^3_\mathbf{1}  \oplus \Lambda^3_\mathbf{27} \oplus \Lambda^2_\mathbf{14} \oplus \Lambda^1_\mathbf7 \ .
\end{equation}
This decomposition of the Rarita--Schwinger $G_2$-bundle justifies the ansatz for the global Rarita--Schwinger section~\eqref{eq:zeta1} in Section~\ref{sec:Mcompactification}.

\subsection{The massless four-dimensional fermionic spectrum} \label{sec:ZeroModes}
We are now ready to determine the four-dimensional fermionic terms in the dimensional reduction of eleven-dimensional supergravity action on $G_2$-manifolds. We focus on the four-dimensional fermionic kinetic and mass terms.

Let us perform the dimensional reduction of the Rarita--Schwinger kinetic term for the gravitino $\hat{\Psi}$, which is given by the third term of the first line in \eqref{eq:11dSUGRA}. Inserting the expansion~\eqref{eq:ExpansionGravitino} for the gravitino $\hat\Psi$ and relation~\eqref{Gamma3Antisymmetrized} we obtain
\be  \label{FirstFermionicTerm}
	\begin{split}
	-\frac{1}{2\kappa^2_{11}} \int *_{11} i \, \bar{\hat\Psi}_M \hat\Gamma^{MNP} \hat{\nabla}_N \hat \Psi_P 
	=  & - \frac{i}{2\kappa^2_{11}} \int\limits_{\mathbb{M}^{1,3}} \ast_4 \bar{\psi}_{\mu} \gamma^{\mu\nu\rho}\nabla_{\nu} \psi^{\ast}_{\rho} \int\limits_{Y} \ast_7 \bar{\zeta}\zeta \\
	 & - \frac{i}{2\kappa^2_{11}} \int\limits_{\mathbb{M}^{1,3}} \ast_4 \bar{\psi}_{\mu} \gamma^{\mu - \rho} \psi^{\ast}_{\rho} \int\limits_{Y} \ast_7 \bar{\zeta} \gamma^n \nabla_n \zeta \\
	 & - \frac{i}{2\kappa^2_{11}} \int\limits_{\mathbb{M}^{1,3}} \ast_4 \frac{1}{3} \bar{\chi} \gamma^{\nu} \nabla_{\nu} \chi^{\ast} \int\limits_{Y} \ast_7 \bar{\zeta}^{(1)}_{m} \gamma^{pm} \zeta^{(1)}_{p} \\
	 & - \frac{i}{2\kappa^2_{11}} \int\limits_{\mathbb{M}^{1,3}} \ast_4 \bar{\chi} \gamma \chi^{\ast} \int\limits_{Y} \ast_7 \bar{\zeta}^{(1)}_{m} \gamma^{mnp} \nabla_n \zeta^{(1)}_{p} \\
	 & + \text{c.c.} \ .
	\end{split}
\ee
The resulting terms comprise the kinetic and mass terms for both the four-dimensional gravitinos $\psi_{\mu}$ --- the first and second line on the right hand side of eq.~\eqref{FirstFermionicTerm}, respectively --- and the four-dimensional fermions $\chi$ --- the third and fourth line on the right hand side of eq.~\eqref{FirstFermionicTerm}, respectively. It also gives rise to mixed terms between $\psi$ and $\chi$. However, since such mixed terms are not present in standard four-dimensional supergravity theories, they have been neglected in our analysis.\footnote{Actually, one should perform a redefinition of $\Psi_{\mu}$ with $\Psi_{\mu} \rightarrow \Psi'_{\mu} = \Psi_{\mu} + \hat{\Gamma}_{\mu}\hat{\Gamma}^m \Psi_m$ in order for such terms to cancel out. However, we do not consider this field redefinition as such a shift does not affect the gravitino mass \cite{House:2004pm,Gurrieri:2004dt}.}

Now, we turn to the discussion of the massless four-dimensional fermionic spectrum, which is obtained from the zero modes of the Dirac operator $\slashed{D} = \gamma^n \nabla_n$ and the Rarita--Schwinger operator $\slashed{D}^\text{RS} = \gamma^{mnp}\nabla_n$, i.e., $\slashed{D}\zeta=0$ and $\slashed{D}^\text{RS} \zeta^{(1)} = 0$.\label{gl:DRS}  

With the ansatz \eqref{eq:zeta} for the section $\zeta$ of the spin bundle~$SY$, we arrive at the zero modes equation
\begin{equation} \label{eq:DiracOpAction}
	\slashed{D} \zeta \,=\, (\nabla_n a_m) \gamma^n \gamma^m \eta + (\partial_n b)  \gamma^n \eta 
	 \,=\, \nabla_{[n} a_{m]}  \gamma^{nm}\eta + (\nabla^n a_n) \eta + (\partial_n b) \gamma^n \eta \,=\, 0\ ,
\end{equation}
which --- due to the linear independence of $\eta$, $\gamma^n \eta$, and $\gamma^{nm}\eta$ --- yields for the coefficient one-form $a(y)= a_{n}(y)dy^n$ and the function $b(y)$ together with eqs.~\eqref{eq:ExteriorDerivative}
\begin{equation}
   d a(y) \, =\, 0 \ , \qquad d^\dagger a(y) \,=\, 0  \ , \qquad d b(y) \,=\, 0 \ .
\end{equation}   
The first two equations imply that $a(y)$ must be a harmonic one-form, whereas the last equation determines the function $b(y)$ to be constant. As there are no harmonic one-forms on the $G_2$-manifold $Y$, the covariantly constant spinor $\eta$ furnishes the only zero mode in the spin bundle $SY$. This zero mode gives rise to the massless four-dimensional gravitino field $\psi_\mu$ and its conjugate $\psi_\mu^*$ of the four-dimensional massless $\mathcal{N}=1$ gravity multiplet listed in Table~\ref{tab:N1multi}.

Analogously, by acting with the Rarita--Schwinger operator $\slashed{D}^\text{RS}$ on the ansatz~\eqref{eq:zeta1} for $\zeta^{(1)}$ and using eqs.~\eqref{eq:antigamma}, \eqref{eq:convarphi} and \eqref{eq:fierz}, we arrive after a straightforward but somewhat tedious calculation at 
\begin{multline} \label{eq:RSzeromodes}
  \slashed{D}^\text{RS}\zeta^{(1)} \,=\, (\nabla_{[n} b_{m]}^7) \gamma^{mnp} dy_p \otimes \eta \\
  -\left(\nabla^n a^{14}_{nm} \right) dy^m \otimes \eta  +\frac32\left(\nabla_{[n} a^{14}_{pq]}\right)  \gamma^{mnpq} dy_m \otimes \eta  \\
  -\frac{3i}2 \left(\nabla^n a^{28}_{npq} \right) dy^p \otimes \gamma^q \eta +\frac{i}3 \left(\nabla_{[m} a^{28}_{npq]}\right) \gamma^{mnpqr} dy_r \otimes \eta  \\
  -\frac12\partial_p (\operatorname{tr}_g a^{28}_{(mn)}) dy_q  \otimes \gamma^{pq}\eta \ ,
\end{multline}
in terms of the singlet $\operatorname{tr}_g a^{28}_{(mn)}= a^{28}_{nm}g^{nm}$ and the three-form $a_{[nmp]}$ 
\begin{equation}
  a_{[mnp]}^{28} \,=\,g^{rs} a^{28}_{r[m} \varphi_{np]s} \ , \qquad 
  a^{28}_{(nm)} \,=\, \frac34  a_{[npq]}^{28}\varphi^{pqr}g_{rm} - \frac{1}{12} g_{nm} a_{[pqr]}^{28} \varphi^{pqr} \ .
\end{equation} 

Let us now analyze the zero modes of the Rarita--Schwinger section $\zeta^{(1)}$ from eq.~\eqref{eq:RSzeromodes}. The one-form $b(y)= b_{n}(y)dy^n$ does not contribute any zero modes, because with eq.~\eqref{eq:ExteriorDerivative} such a zero mode must be a closed one-form $db(y)=0$. Furthermore, due to $b_1(Y)=0$ it also must be exact $b(y)=df(y)$. However, an exact one-form $df(y)$ furnishes no physical degrees of freedom as it can be removed by a gauge transformation of the Rarita--Schwinger section, i.e., $\zeta^{(1)} \to \zeta^{(1)} - \nabla (f(y)\otimes \eta)$. For the remaining tensors we find that, with the help of eqs.~\eqref{eq:ExteriorDerivative}, the zero modes of the Rarita--Schwinger operator $\slashed{D}^\text{RS}$ are given by
\begin{equation}
\begin{aligned}
    &d a^{14}(y) \,=\, 0 \ , &&d^\dagger a^{14}(y)\,=\, 0 \ , \\
    &d a^{28}(y) \,=\, 0 \ ,\qquad &&d^\dagger a^{28}(y)\,=\, 0 \ ,
\end{aligned}    
\end{equation}
in terms of the two-form $a^{14}(y)= \frac12 a_{nm}^{14}(y)dy^n\wedge dy^m$ and the three-form $a^{28}(y)= \frac16 a_{nmp}^{28}(y)dy^n\wedge dy^m\wedge dy^p$. Thus, the zero modes are in one-to-one correspondence with harmonic two-forms $a^{14}(y)$ and three-forms $a^{28}(y)$ on the $G_2$-manifolds $Y$, where the harmonic property of $a^{28}(y)$ implies that the symmetric tensor $a_{(nm)}^{28}$ must be solutions to the Lichnerowicz Laplacian as well, cf. eq.~\eqref{eq:OpSymAntiSym}. Altogether, we can therefore deduce from the cohomology of the $G_2$-manifold $Y$ the fermionic zero modes listed in Table~\ref{tab:N1multi}.

\subsection{The flux-induced holomorphic superpotential} \label{app:super}
Let us now determine the holomorphic superpotential generated by a cohomologically non-trivial four-form background flux $G$ on the $G_2$-manifold $Y$, which is locally given by $d\hat C$. The superpotential can be read off from the four-dimensional gravitino mass term~\eqref{eq:gravimass}. Such a term arises from the dimensional reduction of the fourth term in the eleven-dimensional action~\eqref{eq:11dSUGRA}. That is to say, we find
\begin{multline}
  -\frac{1}{192\kappa_{11}^2} \int *_{11} \bar{\hat\Psi}_M \hat\Gamma^{MNPQRS}  \hat \Psi_N (d\hat C)_{[PQRS]} 
  \,\supset\,  -\frac{1}{192\kappa_{11}^2} \int *_{11} \bar{\hat\Psi}_{\mu} \hat\Gamma^{\mu\nu pqrs}  \hat \Psi_{\nu} (d\hat{C})_{[pqrs]} \\
   \,=\, -\frac{1}{192\kappa_{11}^2} \int *_{11}  (\bar{\psi}_{\mu} + \bar{\psi}^{\ast}_{\mu})\bar{\zeta} \gamma^{\mu\nu}\gamma^{pqrs} (\psi_{\nu} + \psi^{\ast}_{\nu})\zeta (d\hat{C})_{[pqrs]} \ .
\end{multline}
Since there are no harmonic one-forms on the $G_2$-manifold $Y$, we can identify the spinorial section~$\zeta$ with the unique covariant constant spinor $\eta$ on the $G_2$-manifold $Y$, cf. Section~\ref{sec:ZeroModes}. Furthermore, we notice that the covariantly constant three-form $\varphi$ and its Hodge dual four-form $\Phi = \ast\varphi$ are bilinear in $\eta$, namely $\varphi_{mnp} = i\bar{\eta}\gamma_{mnp}\eta$ and $\Phi_{[mnpq]} = (\ast \varphi)_{[mnpq]} = - \bar{\eta} \gamma_{mnpq} \eta$ such that
\begin{multline}
	-\frac{1}{192\kappa_{11}^2} \int *_{11} \bar{\hat\Psi}_M \hat\Gamma^{MNPQRS}  \hat \Psi_N (d\hat C)_{[PQRS]} \\
	\,\supset\, \frac{1}{192\kappa_{11}^2} \int *_{11} \bar{\psi}_{\mu}\gamma^{\mu\nu}\psi^{\ast}_{\nu} \Phi_{[pqrs]} (d\hat C)^{[pqrs]} + \text{c.c.}
\end{multline}
To arrive at the four-dimensional $\mathcal{N}=1$ supergravity action in the conventional Einstein frame, we employ the Weyl rescalings
\begin{equation}
	g_{\mu\nu}  \rightarrow \frac{g_{\mu\nu}}{\lambda_0(S^i)}\ , \qquad
	\gamma^{\mu}  \rightarrow \sqrt{\lambda_0(S^i)}\gamma^{\mu}\ ,\qquad
	\psi_{\mu}  \rightarrow \frac{\psi_{\mu}}{(\lambda_0(S^i))^{1/4}} \ .
\end{equation}	
Using the dimensionless volume factor defined in eq.~\eqref{eq:Vol} and $\kappa^2_{11} = V_{Y_0}\kappa^2_4$ in terms of the reference volume $V_{Y_0}=V_Y(S^i_0)$ defined in Section~\ref{sec:Mcompactification}, we obtain
\begin{multline}
  -\frac{1}{192\kappa_{11}^2} \int *_{11} \bar{\hat\Psi}_M \hat\Gamma^{MNPQRS}  \hat \Psi_N (d\hat C)_{[PQRS]} \\
  \,\supset\, \frac{1}{8\lambda^{7/2}_0 \kappa^2_4} \int_Y G \wedge \varphi \int_{\mathbb{M}^{1,3}} *_{4} \bar{\psi}_{\mu}\gamma^{\mu\nu}\psi^{\ast}_{\nu} + \text{c.c.}
\end{multline}
Therefore, with the K\"ahler potential $K=-3\text{ln}\lambda_0$ derived in Section~\ref{sec:Mcompactification}, the gravitino mass term~\eqref{eq:gravimass}, and the Weyl rescaled four-dimensional metric $g_{\mu\nu}$, we deduce the following superpotential contribution
\begin{equation}
	W \, \supset \,\frac{1}{4} \int_Y G \wedge \varphi \ .
\end{equation}
Note that the derived term is not holomorphic in the four-dimensional chiral coordinates since it couples to the three-form $\varphi$ and not its complexification. To arrive at the full superpotential we must render the moduli dependence holomorphic in terms of the replacement $\varphi \to \varphi + i \hat C$, which is in accord with the deduced chiral coordinates~\eqref{eq:CStarget}. This proposed replacement is in agreement with the domain wall tensions interpolating between distinct flux vacua \cite{Gukov:1999ya, Gukov:1999gr}. Thus, altogether we arrive at the flux-induced superpotential~\eqref{eq:Superpotential}.

\newpage
\bibliographystyle{amsmod}
\bibliography{G2notes}
\end{document}